\documentclass[twocolumn,twocolappendix]{aastex631}

\usepackage{multirow}
\usepackage{soul}

\begin{document}

\title{JWST NIRCam Reveals the Largest Known M-dwarf Debris Disk Around TWA 10 and New Scattered-Light Observations of the TWA 25 Debris Disk}

\correspondingauthor{Katie A. Crotts}
\email{kcrotts@stsci.edu}

\author[0000-0003-4909-256X]{Katie A. Crotts}
\affiliation{Space Telescope Science Institute (STScI), 3700 San Martin Drive, Baltimore, MD 21218, USA}

\author[0000-0001-5365-4815]{Aarynn L. Carter}
\affiliation{Space Telescope Science Institute (STScI), 3700 San Martin Drive, Baltimore, MD 21218, USA}

\author[0000-0003-4614-7035]{Beth Biller}
\affiliation{Scottish Universities Physics Alliance, Institute for Astronomy, University of Edinburgh, Blackford Hill, Edinburgh EH9 3HJ, UK} 
\affiliation{Centre for Exoplanet Science, University of Edinburgh, Edinburgh EH9 3HJ, UK}

\author[0000-0001-8568-6336]{Mark Booth}
\affiliation{UK Astronomy Technology Centre, Royal Observatory Edinburgh, Blackford Hill, Edinburgh EH9 3HJ, UK}

\author[0000-0001-5831-9530]{Rachel Bowens-Rubin}
\affiliation{Eureka Scientific Inc., 2542 Delmar Ave., Suite 100, Oakland, CA 94602, USA}

\author[0009-0000-0303-2145]{Rapha\"el Bendahan-West}
\affiliation{Department of Physics and Astronomy, University of Exeter, Stocker Road, Exeter EX4 4QL, UK}

\author[0000-0002-2683-2396]{Rodrigo Ferrer-Chavez}
\affiliation{Department of Physics and Astronomy, Northwestern University, 2145 Sheridan Rd, Evanston, IL 60208, USA}
\affiliation{Center for Interdisciplinary Exploration and Research in Astrophysics (CIERA), Northwestern University, 1800 Sherman Ave, Evanston, IL 60201, USA}

\author[0000-0002-6964-8732]{Kellen Lawson}
\affiliation{Center for Space Sciences and Technology, University of Maryland, Baltimore County, 1000 Hilltop Circle, Baltimore, MD 21250, USA}
\affiliation{Astrophysics Science Division, NASA-GSFC, 8800 Greenbelt Rd, Greenbelt, MD 20771, USA}
\affiliation{Center for Research and Exploration in Space Science and Technology, NASA-GSFC, 8800 Greenbelt Rd, Greenbelt, MD 20771, USA}

\author[0000-0002-8984-4319]{Briley L. Lewis}
\affiliation{Department of Physics, University of California, Santa Barbara, CA 93106, USA}

\author[0000-0002-5352-2924]{Sebastian Marino}
\affiliation{Department of Physics and Astronomy, University of Exeter, Stocker Road, Exeter EX4 4QL, UK}

\author[0000-0001-5653-5635]{Tim Pearce}
\affiliation{Department of Physics, University of Warwick, Gibbet Hill Road, Coventry CV4 7AL, UK}

\author[0000-0002-3191-8151]{Marshall Perrin}
\affiliation{Space Telescope Science Institute (STScI), 3700 San Martin Drive, Baltimore, MD 21218, USA}

\author[0000-0002-1652-420X]{Giovanni M. Strampelli}
\affiliation{Institute for Astronomy, University of Edinburgh, Blackford Hill, Edinburgh EH9 3HJ, UK}

\author[0000-0002-2428-9932]{Cl\'emence Fontanive}
\affiliation{Institute for Astronomy, University of Edinburgh, Blackford Hill, Edinburgh EH9 3HJ, UK}

\author[0009-0004-0868-1186]{Aiza Kenzhebekova}
\affiliation{Scottish Universities Physics Alliance, Institute for Astronomy, University of Edinburgh, Blackford Hill, Edinburgh EH9 3HJ, UK} 
\affiliation{Centre for Exoplanet Science, University of Edinburgh, Edinburgh EH9 3HJ, UK}

\author[0000-0002-1018-6203]{Patricia Luppe}
\affiliation{School of Physics, Trinity College Dublin, the University of Dublin, College Green, Dublin 2, Ireland}

\author[0000-0002-4388-6417]{Isabel Rebollido}
\affiliation{European Space Agency (ESA), European Space Astronomy Centre (ESAC), Camino Bajo del Castillo s/n, 28692 Villanueva de la Ca\~nada, Madrid, Spain}

\author[0000-0002-9962-132X]{Ben J. Sutlieff}
\affiliation{Institute for Astronomy, University of Edinburgh, Blackford Hill, Edinburgh EH9 3HJ, UK} 
\affiliation{Centre for Exoplanet Science, University of Edinburgh, Edinburgh EH9 3HJ, UK}

\author[0000-0002-7325-5990]{Ellis Bogat}
\affiliation{Laboratoire d'Astrophysique de Marseille, 38 Rue Fr\'{e}d\'{e}ric Joliot Curie, 13013 Marseille, France}

\author[0009-0001-0275-7811]{Evelyn L. Bruinsma}
\affiliation{Department of Physics \& Astronomy, Johns Hopkins University, 3400 N. Charles Street, Baltimore, MD 21218, USA}

\author[0000-0002-8382-0447]{Christine H. Chen}
\affiliation{Space Telescope Science Institute (STScI), 3700 San Martin Drive, Baltimore, MD 21218, USA}

\author[0000-0001-8627-0404]{Julien H. Girard}
\affiliation{Space Telescope Science Institute (STScI), 3700 San Martin Drive, Baltimore, MD 21218, USA}

\author[0000-0002-9803-8255]{Kielan Hoch}
\affiliation{Space Telescope Science Institute (STScI), 3700 San Martin Drive, Baltimore, MD 21218, USA}

\author[0009-0005-6943-6819]{Andrew D. James}
\affiliation{Department of Physics and Astronomy, University of Exeter, Stocker Road, Exeter EX4 4QL, UK}

\author[0009-0009-1074-3696]{Rohan Kane}
\affiliation{Space Telescope Science Institute (STScI), 3700 San Martin Drive, Baltimore, MD 21218, USA}

\author[0000-0002-0834-6140]{Jarron Leisenring}
\affiliation{Department of Astronomy / Steward Observatory, University of Arizona, Tucson, AZ 85721, USA}

\author[0000-0003-4203-9715]{Emily Rickman}
\affiliation{European Space Agency (ESA), ESA Office, Space Telescope Science Institute (STScI), 3700 San Martin Drive, Baltimore, MD 21218, USA}

\author[0000-0001-6098-3924]{Andy Skemer}
\affiliation{Department of Astronomy \& Astrophysics, University of California, Santa Cruz, 1156 High Street, Santa Cruz, CA 95064}

\author[0009-0008-2252-7969]{Klaus Subbotina Stephenson}
\affiliation{Department of Physics and Astronomy, University of Victoria, 3800 Finnerty Road, Victoria, BC, V8P 5C2, Canada}



\begin{abstract}

We present JWST NIRCam observations of two M-dwarf systems located in the TW Hydra association, TWA 10 and TWA 25. Both systems harbor detected debris disks in the F200W and F444W filters. Whereas the TWA 25 disk has been previously imaged, these observations represent the discovery and first images of the TWA 10 disk. In addition to planet searches within these systems, we also conduct an analysis of each debris disk, where the TWA 10 debris disk is characterized for the first time. We find that the TWA 10 debris disk is very large, with a radius of $\sim$191 au, significantly greater than other known M-dwarf debris disks. The TWA 25 disk hosts a sharp inner dust surface density power-law and a moderate brightness asymmetry present at 2 $\mu$m, suggesting potential sculpting from inner planets and potentially enhanced collisional activity. Finally, we find one potential companion candidate within the TWA 10 system and two within the TWA 25 system, although the measured F200W-F444W color suggests that these candidates are likely background objects. Both systems do not have measured IR-excesses in their SEDs, where radiative-transfer modeling suggests that these disks (and potentially more M-dwarf disks) were likely missed by previous disk detection surveys due to having low luminosity fractions.

\end{abstract}



\section{Introduction} \label{sec:intro}
Debris disks are optically-thin, gas-poor, dusty disks found around $\sim$20\% of main-sequence stars \citep{Matthews14}. They arise after the protoplanetary disk stage, i.e. after the majority of primordial gas and dust has dissipated, leaving behind formed exoplanets and planetesimals (\citealt{Wyatt08}, and references therein). Planetesimal belts, similar to the Asteroid and Kuiper belts in our own Solar System, can become dynamically excited or ``stirred" through interactions with planets in the system \citep{Mustill09}, their own self gravity \citep{Kenyon01}, or other more rare mechanisms such as stellar flybys \citep{Ida00,KB02}. This stirring leads to collisions between planetesimals, creating sub-micron to millimeter sized dust grains that can be observed through imaging at various wavelengths \citep{Wyatt08,Matthews14,Hughes18}.

Debris disks were first identified through infrared (IR) excess present in the stellar spectral energy distribution (SED), where the dust boosts the flux at mid-infrared (MIR) to far-infrared (FIR) wavelengths above what is expected for a stellar photosphere alone. Many surveys have been conducted to detect debris disks around nearby stars in the MIR-FIR with space-based instruments, including the \textit{Infrared Astronomical Satellite} (IRAS), \textit{Wide-Field Infrared Survey Explorer} (WISE), as well as the \textit{Herschel} and \textit{Spitzer} space telescopes. Observations with such instruments have been crucial in characterizing the SEDs of disk-hosting systems and determining the frequency of occurrence of debris disks around stars of various spectral types and ages (e.g. \citealt{Trilling08,Patel14,Montesinos16,Sibthorpe18}).

One issue with these detection surveys is the clear observational bias towards spectral type, such that there have been significantly less detections of IR-excesses around M-dwarfs compared to AFGK stars. Additional ground based imaging surveys targeted at M-dwarfs have also yielded few detections (e.g. \citealt{Cronin23}). It remains unclear whether or not the lack of M-dwarf disk detections is due to lower luminosity fractions (i.e. $L_{disk} / L_{*}$) (e.g. \citealt{ML14}) or potentially differing formation properties (e.g. lower initial disk masses \citealt{Gaidos17}). However, studies have found that M-dwarf disks may be just as common as AFGK disks and have evaded detection due to observational limitations \citep{Luppe20}. To date, only a small number of debris disks around M-dwarfs have been detected over multiple observations, including well known disks such as AU Mic \citep{Kalas04}, TWA 7 \citep{Choquet16}, and GSC 07396-00759 \citep{Sissa18}. However, since the launch of the James Webb Space Telescope (JWST), at least one new M-dwarf debris disk with no IR-excess detection has been discovered using the NIRCam instrument (i.e. TWA 20 \citealt{Palatnick25}). Thus, this discovery demonstrates that NIRCam's impressive sensitivity may provide the means to detect and spatially resolve faint M-dwarf debris disks for the first time. 

In this paper, we present debris disk observations around the TWA 10 and TWA 25 M-dwarfs using the NIRCam instrument on JWST. The properties of the systems can be found in Table \ref{tab:target_sum}. While the TWA 25 debris disk has been previously detected in scattered light with instruments on the Hubble Space Telescope (HST; \citealt{Choquet16,Ren23}) and the Very Large Telescope (VLT; \citealt{Engler25}), the TWA 10 debris disk is a new detection and is now the second M-dwarf debris disk to be discovered with JWST. Similarly to TWA 20, both the TWA 10 and 25 systems have no clear IR-excesses observed in their SEDs; these two systems have non-detections from \textit{Spitzer} and \textit{Herschel}. Here, we study the morphologies of these disks in our 2 $\mu$m and 4 $\mu$m observations, characterizing their geometrical, surface density, and scattering phase function (SPF) properties. Additionally, we search for potential planet candidates in both systems, as well as predict the maximum semi-major axis and minimum planet-mass for each system assuming a singular massive planet sculpting the disk inner-edge.

\begin{table}
	\centering
	\caption{\label{tab:target_sum}Target summary. Spectral types are taken from \citet{Torres06} and \citet{Herczeg14} for TWA 10 and TWA 25, respectively. Distance is from Gaia DR3 \citep{Gaia21} and system age is taken from \citet{Miret25}. The stellar luminosity is derived from the models fit to the photometry using the SED fitting code from \citet{Yelverton19}, which can be seen in Figure \ref{fig:seds}.}
	\begin{tabular}{ccccc}
	    \hline
	    \hline
		Name & SpT & d (pc) & Age (Myr) & $L_{*}$ (L$_{\odot}$) \\
		\hline
        TWA 10 & M2V & 57.5 & 9$^{+2}_{-1}$ & 0.106$\pm$0.001 \\
		  TWA 25 & M0.5 & 53.6 & 9$^{+2}_{-1}$ & 0.246$\pm$0.002 \\
		\hline
		\hline
	\end{tabular}
\end{table}

\section{Data Observations \& Reductions} \label{sec:obs}

\subsection{Observations}
Both targets were observed on June 16th of 2024 with JWST NIRCam in coronagraphy mode \citep{Girard22} as part of the General Observers (GO) 4050 program: ``Uncharted Worlds: Towards a Legacy of Direct Imaging of Sub-Jupiter Mass Exoplanets" (PI: A. Carter) \citep{CarterJWSTprop2023}. The goal of the program is to search for new planetary companions around nearby stars using NIRCam's high sensitivity. Each system was imaged simultaneously in the following filters; F200W ($\lambda_{\text{pivot}} \sim 2 \ \mu$m, $\Delta \lambda \sim 0.46 \ \mu$m) and F444W ($\lambda_{\text{pivot}} \sim 4.4 \ \mu$m, $\Delta \lambda \sim 1.0 \ \mu$m). The F200W observations were taken with a field-of-view (FOV) of $10\arcsec \times 10\arcsec$ and a pixel scale of 31 mas per pixel, while the F444W observations were taken with a FOV of $20\arcsec \times 20\arcsec$ and a pixel scale of 63 mas per pixel. Both observations were taken with the MASK335R focal plane mask with an inner working angle of $0\farcs63$. Table \ref{tab:data_sum} displays the summary of the data for TWA 10 and TWA 25. As part of the GO 4050 program, 7 dedicated reference stars were also observed (in addition to the 22 targets observed) utilizing a 9-POINT-CIRCLE dither pattern. 

\subsection{Data Reduction}
Both datasets were reduced using the codes \texttt{spaceKLIP} \citep{Kammerer22,Carter23} and \texttt{Winnie}, which is a \texttt{spaceKLIP} compatible code designed for JWST coronagraphic disk observations \citep{Lawson23}. The raw science and reference images are first reduced with the \texttt{spaceKLIP} reduction pipeline in three stages. In stage 1, the pipeline first applies detector-level corrections such as saturation, superbias, non-linearity, and dark current, followed by ramp-fitting of the raw, uncalibrated data. During this stage, we turn off the 1/f noise correction as removing this noise can lead to artifacts in images with disks. The data are then further processed in stage 2 of the pipeline, where additional correction and calibration are performed, including median background subtraction and identifying bad pixels. In the final stage, the data are bad-pixel repaired, re-centered to ensure that all reference and science images are aligned, and are padded with an 80-pixel border of nans to avoid any images being cut-off when rotated to North up.

Once the data are reduced, \texttt{Winnie} is used to perform Reference-star Differential Imaging (RDI) to produce a single PSF-subtracted image of the target. Similar to previous studies of debris disks observed with NIRCam (see \citealt{Crotts25,Palatnick25}), we use the entire GO 4050 library for our reference observations to improve our PSF subtraction. We additionally include the other target stars, minus those with high levels of noise or debris disks, as we find that this significantly improves the signal-to-noise (SNR) of our final reductions. The reference stars used (all M-dwarfs but one) are the following; TWA 23 A, TWA 26, TWA 29, TWA 30 A, TWA 33, TWA 35, TWA 36, TWA 43, TWA 45, TWA 46, TWA 47, CD-23 9765, HD 89063, HD 101581, IRAS 09432-4847, IRAS 13041-5653, J1058-2346, i-Vel, and V$^{*}$ Y Crt. 

\begin{figure*}
    \centering
    \includegraphics[width=\textwidth]{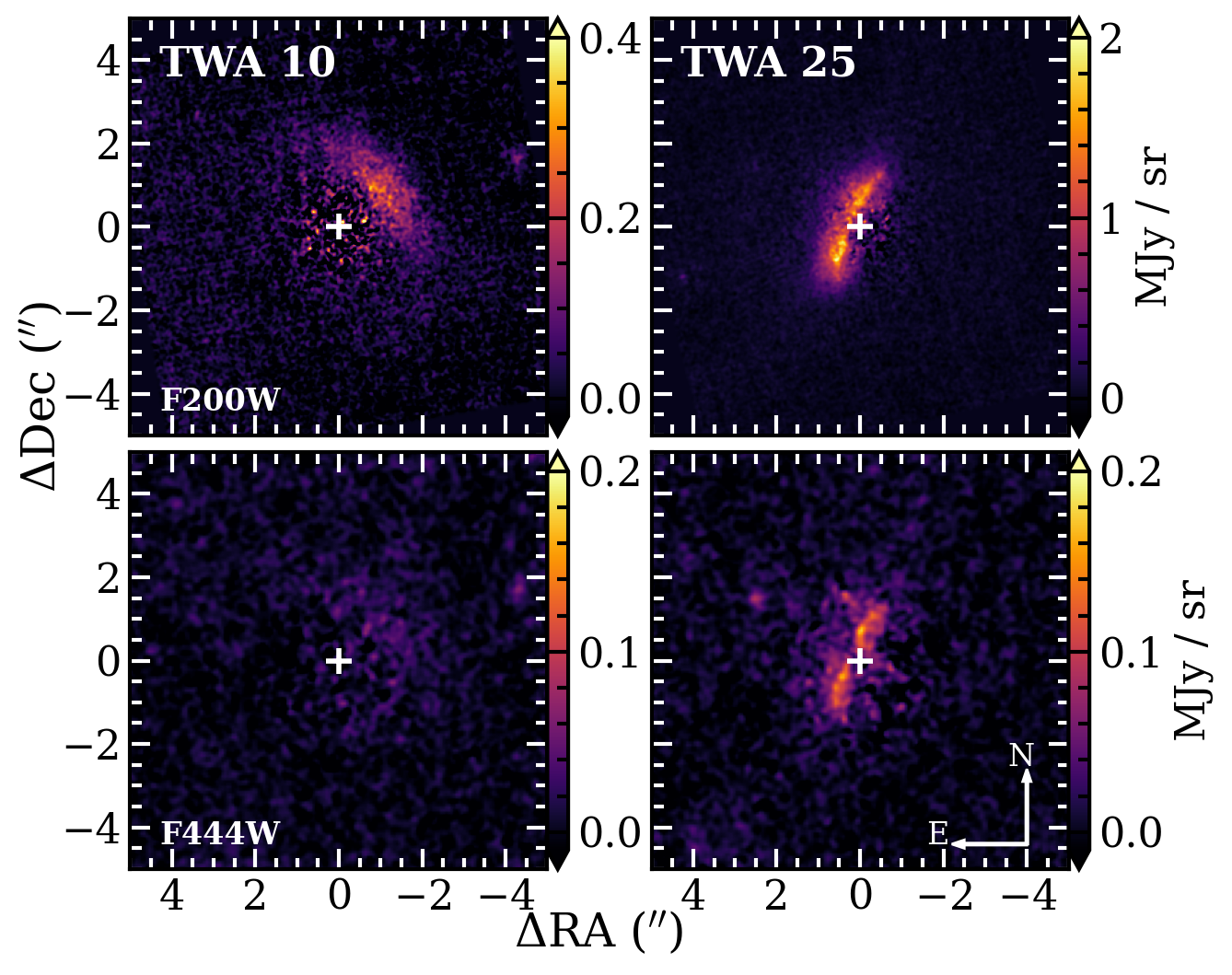}
    \caption{\label{fig:reduc} Final data reductions for (from left to right) TWA 10 and TWA 25 in the F200W (\textit{top}) and F444W (\textit{bottom}) filters. All images are rotated with North up, and the $+$ represents the star location.} 
\end{figure*}

Finally, we perform Model-Constrained RDI (MCRDI; \citealt{Lawson22}), a process which involves forward-modeling and using best-fit synthetic disk models to fine-tune RDI constraints. The reasoning for using MCRDI is to optimize our PSF-subtracted images and minimize the loss of disk signal. For examples and further discussion of RDI vs. MCRDI NIRCam reductions, we point the interested reader to \citet{Lawson23}. The forward modeling process and results will be further discussed in Section \ref{sec:mods}. Our final MCRDI images of TWA 10 and TWA 25 can be seen in Figure \ref{fig:reduc}.

\begin{table*}
	\centering
	\caption{\label{tab:data_sum}Summary of the data used in this paper. $N_{ints}$ is the total number of integrations, $N_{groups}$ is the number of groups per integration, $N_{frames}$ is the number of frames per group, t$_{int}$ is the effective integration time, and t$_{exp}$ is the total effective exposure time.}
	\begin{tabular*}{\textwidth}{c @{\extracolsep{\fill}} cccccc}
	    \hline
	    \hline
		Name & Filter & Date & Readout Pattern & $N_{ints}$/$N_{groups}$/$N_{frames}$ & t$_{int}$ (s) & t$_{exp}$ (s) \\
		\hline
		  TWA 10 & F200W/F444W & 2024 Jun 16 & DEEP8 & 5/18/8 & 372 & 1865  \\
        \hline
        TWA 25 & F200W/F444W & 2024 Jun 16 & DEEP8 & 5/18/8 & 372 & 1865  \\
		\hline
		\hline
	\end{tabular*}
\end{table*}

\section{Disk Forward Modeling} \label{sec:mods}

\subsection{Procedure}
In the previous section, we summarize the data reduction and PSF-subtraction process, where MCRDI is utilized to create the final images for both disks. Here, we further discuss the MCRDI process, which uses forward-modeling to optimize the PSF-subtracted image while simultaneously finding a best-fit model to the data. We use a simple ring-like disk morphology based on \citet{Augereau99} for our model, which assumes a 2-parameter Henyey-Greenstein (HG; \citealt{HG41}) SPF. The disk geometry and surface density of our models are defined with the following parameters: the fiducial disk radius ($r_{0}$), the ratio between the scale height and $r_{0}$ at $r_{0}$ ($h_{0}$), inclination ($\theta_{inc}$), position angle ($\theta_{PA}$), the radial density power-law exponent interior and exterior to $r_{0}$ ($\alpha_{in}$ and $\alpha_{out}$, respectively), and the vertical density exponent ($\gamma$). Additionally, the SPFs of the disk models are defined by two HG asymmetry parameters ($g_{1}$ and $g_{2}$, respectively) and the weight for the asymmetry parameter $g_{1}$ ($w_{1}$).

To find the best-fit model, we use a minimization routine via the code \textsc{lmfit.minimize} \citep{lmfit25}. For TWA 25, we utilize previous best-fit models of the disk found in \citet{Choquet16} to define the initial starting values for each parameter. In the case of TWA 10, because this is the first resolved detection of the disk, the initial geometrical parameters (e.g. $r_{0}$, $\theta_{inc}$, $\theta_{PA}$) are estimated by eye. We start the minimization routine by fitting the F200W data, which have stronger disk detections. Additionally, we use the ``Powell" method \citep{Powell65} in order to find the local minimum for each parameter in a computationally efficient manner. In each case, several hundred models are produced within a range of priors for each parameter, where an optimized model is then returned. The minimization routine is then run again with the best-fit parameters from the ``Powell" method, this time using the ``emcee" method \citep{emcee13}, which utilizes Markov Chain Monte Carlo (MCMC) to further explore the parameter space and calculate uncertainties for each parameter. For both disks, we run the MCMC for 5000 steps with 30 walkers and a 300 step burn-in, resulting in $\sim$100,000 models evaluated. Finally, we perform forward-modeling for the F444W observations using the results for the F200W. While the geometrical and surface density profiles should not significantly change between wavelengths, the SPF may differ. Therefore, to fit the F444W observations, we repeat the same procedure as before, keeping the geometrical and surface density parameters fixed and only allowing the SPF parameters ($g_{1}$, $g_{2}$, and $w_{1}$) to vary. However, we note that the SPF parameters are difficult to constrain, especially for the F444W, due to the low SNR of the observations and the removal of flux at the forward scattering peak in the case of TWA 25.

\begin{figure*}
    \centering
    \includegraphics[width=0.74\textwidth]{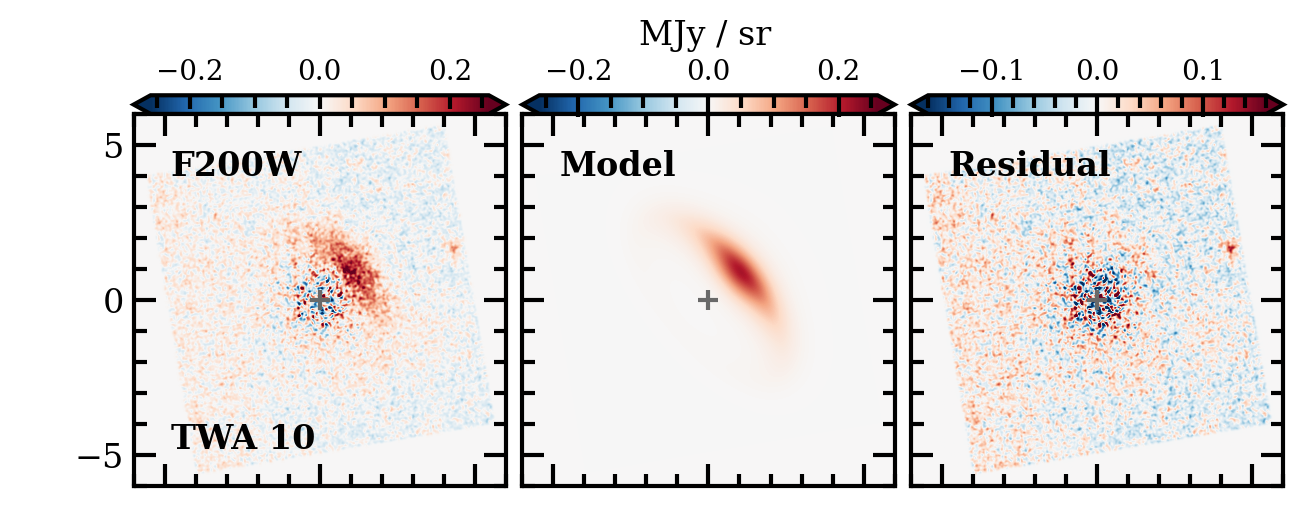}
    \includegraphics[width=0.74\textwidth]{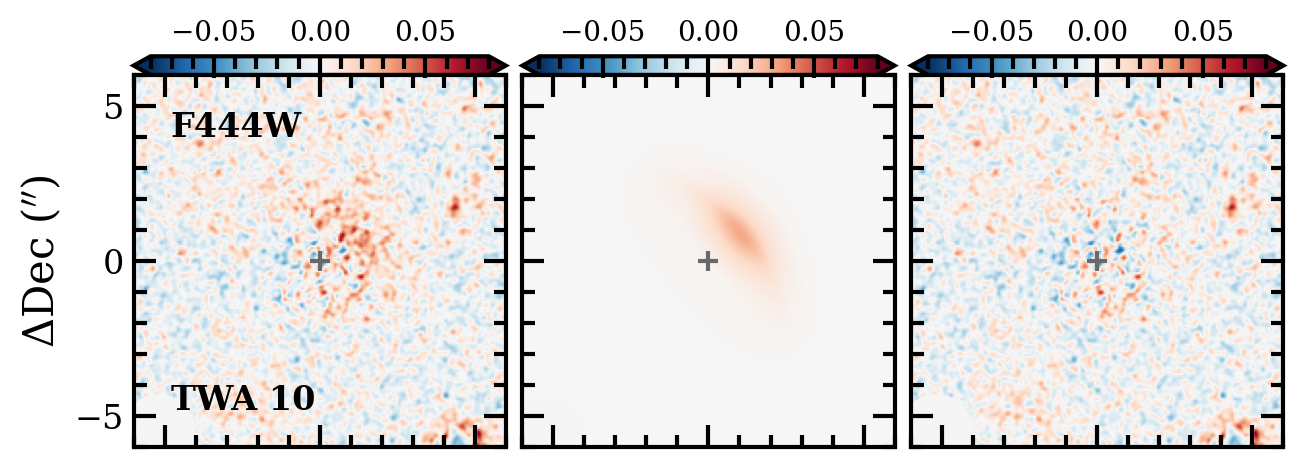}
    \includegraphics[width=0.74\textwidth]{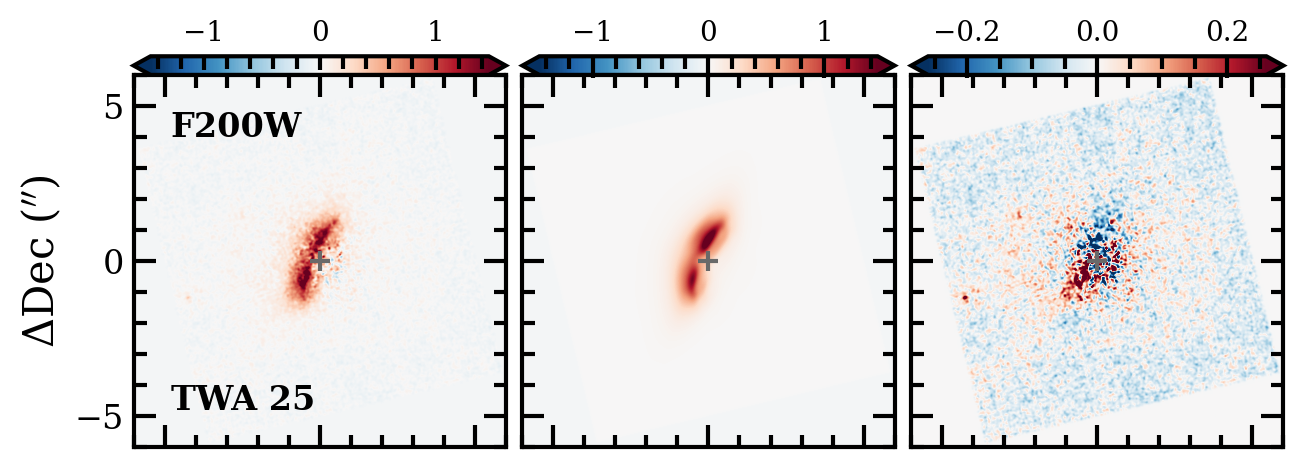}
    \includegraphics[width=0.74\textwidth]{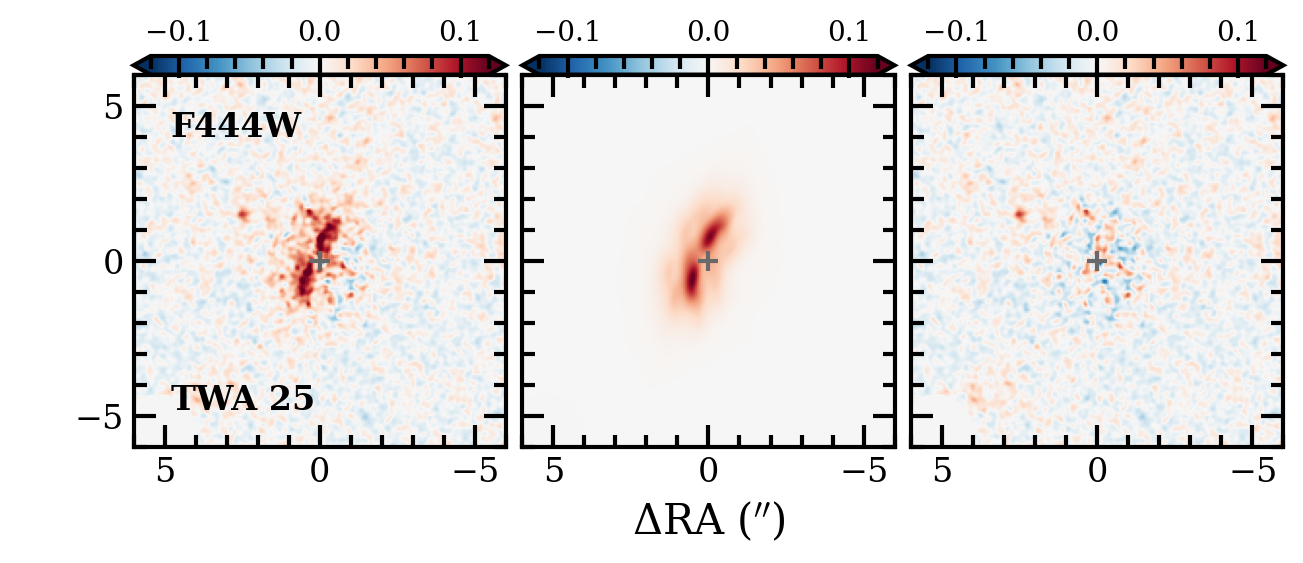}
    \caption{\label{fig:mod2micron} Forward modeling results for TWA 10 and TWA 25 in the F200W and F444W filters. The left panels are our MCRDI reductions, the middle panels are the best fitting emcee forward models, and right panels are the resulting residuals. All images are rotated with North up, and the $+$ represents the star location.} 
\end{figure*}

The results of the forward modeling or both the Powell and emcee methods can be found in Table \ref{tab:mod_res}, where each row shows the best fit values found for each parameter with $1\sigma$ uncertainties for the emcee method. Furthermore, Figure \ref{fig:mod2micron} shows our best-fit emcee models (middle panels) for the TWA 10 and TWA 25 disks, respectively, in both filters compared to our MCRDI reductions (left panels). The residuals from subtracting the model from the data are shown in the right panels. The best-fit disk models will be further discussed in Section \ref{sec:discuss}.

\begin{table*}
	\centering
	\caption{\label{tab:mod_res}List of best fitting model parameters for each disk using a minimization routine via the Powell and emcee methods. Parameters which were fixed during the minimization process are labeled.}
	\begin{tabular*}{\textwidth}{c @{\extracolsep{\fill}} ccccc}
	    \hline
	    \hline
		Name & Parameter & F200W Powell & F200W emcee &  F444W Powell & F444W emcee \\
		\hline
		TWA 10 & $r_{0}$ (au) & 157.13 & 191.15$\pm$6.53 & -- & -- \\
		   & $\theta_{inc}$ ($^{\circ}$) & 55.76 & 63.67$\pm$3.36 & -- & -- \\
              & $\theta_{PA}$ ($^{\circ}$) & 219.78 & 220.45$\pm$3.67 &  -- & -- \\
              & $h_{0}$ & 0.05 (fixed) & -- & -- &  -- \\
              & $\alpha_{in}$ & 4.78 & 4.53$\pm$3.36 & -- & -- \\
              & $\alpha_{out}$ & -9.86 & -8.45$\pm$2.56 & -- & -- \\
              & $\gamma$ & 1.15 & 2.52$\pm$1.29  & -- & -- \\
              & $g_{1}$ & 0.99 & 0.80$\pm$0.13  & 0.99 & 0.02$\pm$0.68 \\
              & $g_{2}$ & 0.99 & 0.57$\pm$0.65  & 0.98 & 0.02$\pm$0.68  \\
              & $wg_{1}$ & 0.55 & 0.80$\pm$0.18 & 0.51 & 0.77$\pm$0.17  \\
            \hline
		  TWA 25 & $r_{0}$ (au) & 62.42 & 63.81$\pm$3.26 & -- & \\
		       & $\theta_{inc}$ ($^{\circ}$) & 78.09 & 77.91$\pm$1.07 & -- & \\
              & $\theta_{PA}$ ($^{\circ}$) & 337.04 & 337.07$\pm$0.67 & -- & \\
              & $h_{0}$ & 0.05 (fixed) & -- & -- & \\
              & $\alpha_{in}$ & 17.06 & 11.94$\pm$5.37 & -- & \\
              & $\alpha_{out}$ & -3.83 & -3.84$\pm$0.31 & -- & \\
              & $\gamma$ & 1.79 & 2.76$\pm$1.37 & -- & \\
              & $g_{1}$ & 0.97 & 0.90$\pm$0.06 & 0.61 & 0.19$\pm$0.65 \\
              & $g_{2}$ & 0.52 & 0.44$\pm$0.19 & 0.61 & 0.10$\pm$0.66 \\
              & $wg_{1}$ & 0.92 & 0.87$\pm$0.07 & 0.99 & 0.75$\pm$0.17 \\
		\hline
		\hline
	\end{tabular*}
\end{table*}

% \begin{figure*}
%     \centering
%     \includegraphics[width=\textwidth]{figures/twa25_mod.png}
%     \includegraphics[width=\textwidth]{figures/twa25_4um.png}
%     \caption{\label{fig:mod2micron_twa25} Forward modeling results for TWA 25, which are organized similar to Figure \ref{fig:mod2micron}. Again, all images are rotated with North up, and the $+$ represents the star location.} 
% \end{figure*}

\section{Disk Analysis}

\subsection{Surface Brightness Profiles}
In addition to deriving the disk geometry, surface density, and SPF parameters for each disk, we also derive surface brightness profiles in the F200W where the disks are brightest. Because the forward-modeling has shown that the SPF for both disks is very difficult to constrain, we focus on what we can empirically learn from the surface brightness as a function of projected stellar separation. 

\begin{figure}
    \centering
    \includegraphics[width=0.47\textwidth]{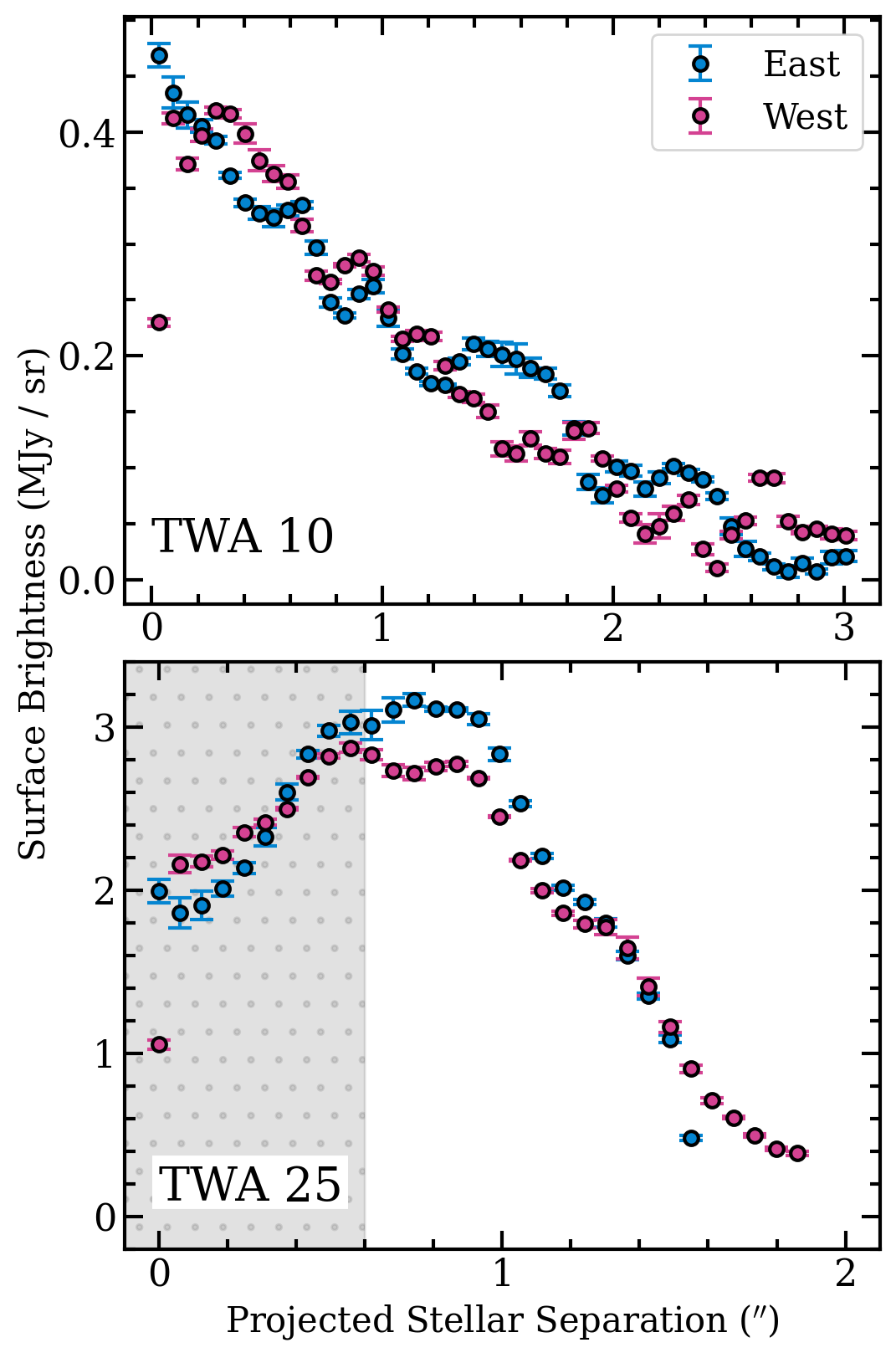}
    \caption{\label{fig:sb_prof} Surface brightness (SB) profiles for the TWA 10 (\textit{top}) and TWA 25 (\textit{bottom}) disks in the F200W as a function of projected separation from the host star. The blue data points represent the East side of the disk and the pink data points represent the West side of the disk. The grey shaded region represents the area where the throughput is $\lesssim$50\% ($\leq 0\farcs6$), as attenuation from the coronagraph affects the surface brightness of the TWA 25 disk at these separations.} 
\end{figure}

We first spatially binned each image into 2x2 bins in order to increase the surface brightness S/N. Additional smoothing is performed by using a Gaussian filter with $\sigma$ = 1 pixel. We then trace the disk spine, i.e. the (x,y) positions of the peak surface brightness along the projected major-axis of the disk. This is done by rotating the data by the derived $PA$, so that the disk's major-axis is horizontal in the image. We then measure the surface brightness along vertical slices perpendicular to the disk's major-axis at various separations from the star (x-position) and fit these profiles with a Gaussian model. The mean value of the best-fit Gaussian is used as the y-position. Next, we place a square 3x3 pixel aperture at each (x,y) position and take the average surface brightness in each aperture. The uncertainty is calculated by measuring the standard deviation within the same apertures placed on the model-subtracted residual maps. We split both surface brightness profiles between the East and West sides of the disk for comparison. The final surface brightness profiles are plotted in Figure \ref{fig:sb_prof}. 

The two disks present varying surface brightness profiles. In the case of TWA 10, the surface brightness profile peaks closest to the star and steadily decreases with increasing projected stellar separation, consistent with a typical debris disk SPF. However, the TWA 25 surface brightness profile differs, increasing between 0$''$ to $\sim 0\farcs8$ from the star before decreasing past 1$''$. This effect is systematic rather than physical. Given the smaller radius of the TWA 25 disk, the forward scattering peak within $\sim 0\farcs8$ becomes attenuated by the coronagraph. The shaded region in the bottom plot of Figure \ref{fig:sb_prof} shows where the coronagraphic throughput is $\lesssim$50\% for comparison.

Most importantly, we can use the surface brightness profiles to search for any asymmetries present. While, the TWA 10 appears fairly axisymmetric, the TWA 25 disk presents a significant brightness asymmetry. Mainly, the East side of the TWA 25 disk is moderately brighter than the West side of the disk between $0\farcs5$ and $1\farcs2$. A brightness asymmetry could be caused by an eccentric disk or potentially a dust density enhancement at these separations. The possibility of these scenarios will be discussed further in Section \ref{sec:TWA25}.

\subsection{Disk Radius Comparison}
Multiple studies of both scattered-light and millimeter data have found tentative trends between debris disk radius and stellar luminosity, where the disk radius increases with stellar luminosity via a power-law (e.g. \citealt{Matra18,Matra25,Esposito20}). Hence, debris disks around M-dwarfs on average are found to be more compact than debris disks around hotter, more luminous stars. We therefore compare the TWA 10 and 25 disk radii to other spatially resolved debris disks to see if they fall within or outside these trends.

\begin{figure}
    \centering
    \includegraphics[width=0.47\textwidth]{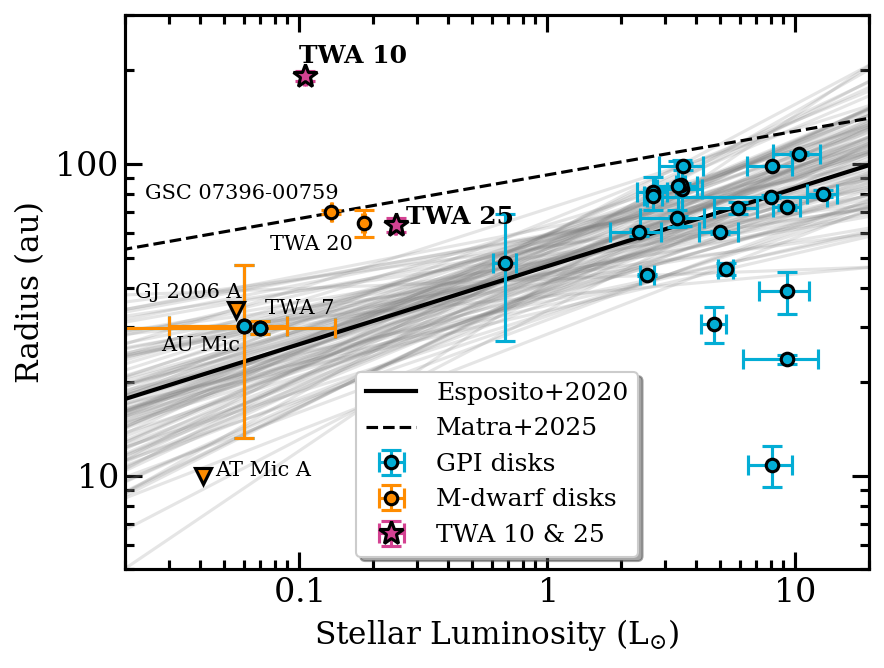}
    \caption{\label{fig:disk_radii} Disk radii ($R_{0}$) vs. stellar luminosity. The blue data points represent debris disks observed with GPI \citep{Esposito20}, while the pink data points represent TWA 10 and 25. The orange data points highlight the other M-dwarfs, where triangles are upper limits. The solid black and light grey lines represents the best-fit power-law with 1$\sigma$ uncertainties derived from \citet{Esposito20} for the GPI sample. The dashed black line represents the best-fit power law derived from \citet{Matra25} for comparison.} 
\end{figure}

Figure \ref{fig:disk_radii} shows the radius versus the stellar luminosity for known debris disks. For comparison, we plot the measured radii and stellar luminosities for debris disks a part of the Gemini Planet Imager (GPI) sample (blue data points; \citealt{Esposito20}). The solid black line and grey lines represent the power-law relationship derived from \citet{Esposito20} using the GPI disk sample, while the dashed black line is the relationship derived from millimeter data via the REASONS survey for comparison \citet{Matra25}. We also highlight the other M-dwarf disks that have measured radii (orange data points). This includes AU Mic and TWA 7, which are a part of the GPI sample, as well as the newly discovered TWA 20 debris disk also observed with NIRCam \citep{Palatnick25} and the GSC 07396-00759 disk \citep{Sissa18}. Additionally, we include two M-dwarf disks that are not fully resolved with ALMA, GJ 2006 A and AT Mic A \citep{Cronin23}, as upper limits on radius. The debris disk Fomalhaut C exists outside the plot on the lower left ($L_{*}\sim0.5$~L$_{\odot}$, $r_{0}$ = 24.5~au; \citealt{Lawson24}). 

While both the majority of resolved M-dwarf disks fall in line with the derived relationships between radius and stellar luminosity, TWA 10 has a significantly larger radius. This discrepancy may have crucial implications for the stellar/disk and dynamical evolution of the TWA 10 system (discussed further in Section \ref{sec:TWA10}). However, the large range in debris disk radii around M-dwarfs, which has become more apparent with the new disk discoveries, also suggests that the trend between disk radius and stellar luminosity does not hold up.

\subsection{Disk SEDs} \label{disk_sed}

\begin{figure}
    \centering
    \includegraphics[width=0.472\textwidth]{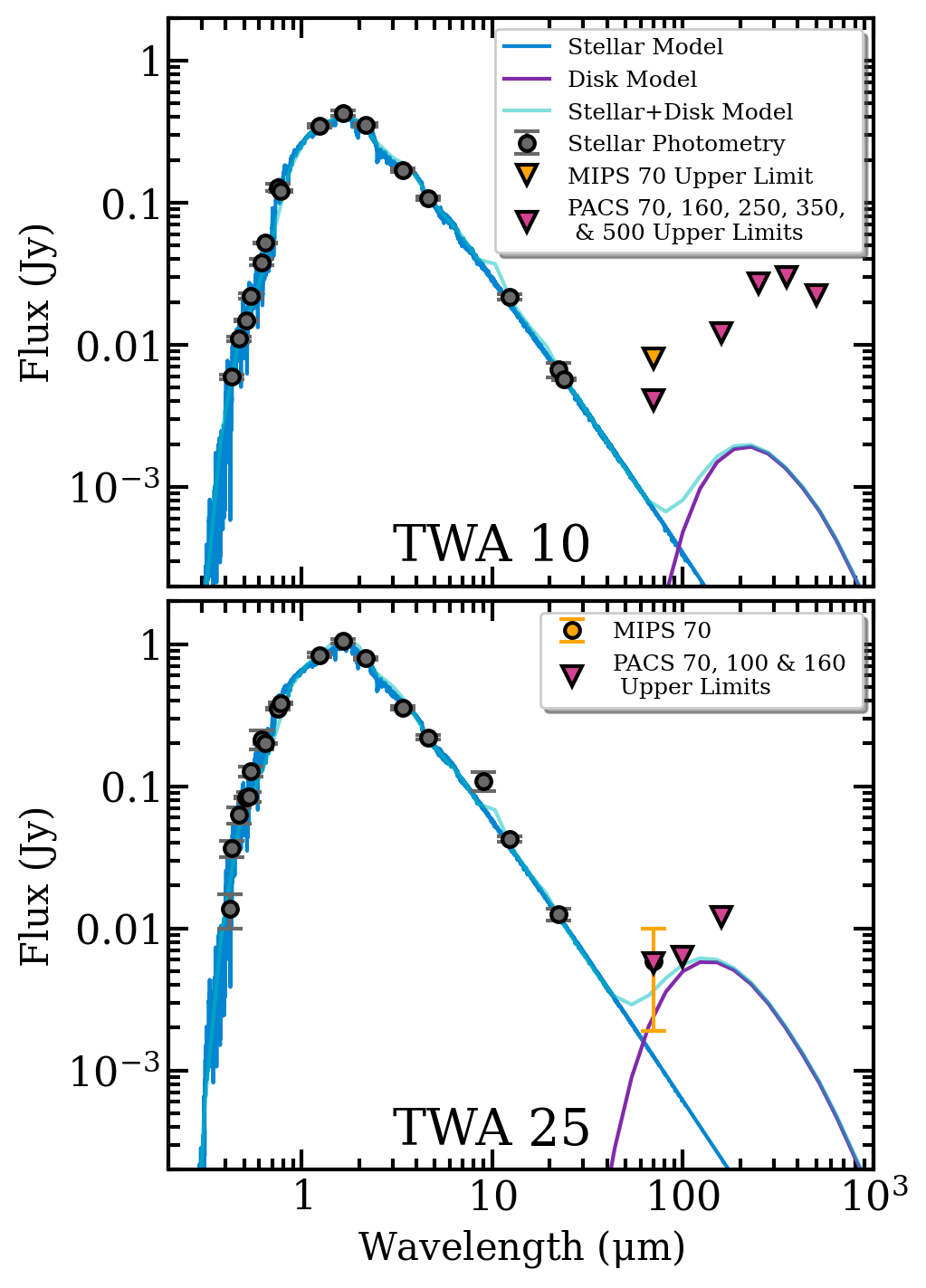}
    \caption{\label{fig:seds} Stellar SEDs of TWA 10 (\textit{top}) and TWA 25 (\textit{bottom}). The grey data points are photometry measurements that are consistent with the stellar SED model (blue line), while the colored data points are measurements from Spitzer/MIPS and Herschel/PACS (triangles are 3$\sigma$ upper limits). The estimated disk SED is shown by the purple line and the stellar+disk SED is represented by the light blue line.} 
\end{figure}

The detection of M-dwarf debris disks has historically been difficult, leading to only a small handful of the brightest M-dwarf disks being spatially resolved. However, the new discovery of M-dwarf debris disks with NIRCam, including TWA 10, may shed light on why this is the case. In Figure \ref{fig:seds}, we plot the stellar SEDs for TWA 10 and TWA 25. Grey data points are measured photometry consistent with the stellar SED model (blue line), which were fit using the SED fitting software from \citet{Yelverton19}. The colored data points represent Spitzer/MIPS (orange) and Herschel/PACS (pink) photometry measurements. The photometry with citations is listed in Table \ref{tab:sed}, located in the Appendix. In both cases, the MIPS and PACS data are either 3$\sigma$ upper limits or have large uncertainties consistent with the stellar SED within 1-2$\sigma$. Although the TWA 25 disk was previously known, there has been no significant IR-excess measured in its SED, similar to TWA 10.

Although we cannot robustly constrain the disk SEDs for these two systems without further observations of the disk emission, we can estimate the SED using the NIRCam data and the limits set by MIPS and PACS. To calculate the disk SED, we utilize the radiative-transfer disk modeling code, \texttt{MCFOST} \citep{Pinte06}. For the disk geometry and dust density parameters ($R_{disk}$, \textit{PA}, \textit{inc}, $\alpha_{in}$, $\alpha_{out}$, $\gamma$), we use the best-fit values from Table \ref{tab:mod_res} and similarly assume a scale height ($h_{0}$) of 0.05. For the dust grain parameters we assume astrosilicate grains with a grain size range of 0.1-1000 $\mu$m (the minimum dust grain size is small for M-dwarfs) and a size distribution of -3.5, which is expected for a collisional cascade \citep{Dohnanyi69}. Finally, the total dust mass is scaled to match the surface brightness of the F200W raw best-fit disk model. The resulting SEDs from the \texttt{MCFOST} models can be seen in Figure \ref{fig:seds} as the purple lines.

Following the described process, we estimate a total dust mass of $\sim$6.7e-3 M$_{\oplus}$ and $\sim$3.3e-3 M$_{\oplus}$ for the TWA 10 and TWA 25 disks, respectively. These are typical dust masses for debris disks, albeit on the lower end for young moving group systems (e.g. \citealt{Matra25}). Additionally, the resulting disk SEDs are consistent with the limits measured with MIPS and PACS. While the TWA 25 disk SED sits only slightly below the PACS 3$\sigma$ upper limits, peaking at a flux of 5.8 mJy, the TWA 10 disk SED is about an order of magnitude fainter than these upper limits, with a peak flux of 1.9 mJy. Dividing the disk SED sum from the stellar SED sum gives a luminosity fraction of $\sim$3.6e-5 for TWA 10 and $\sim$6.9e-5 for TWA 25. For context, cold debris disks (T$<$140 K) with luminosity fractions of $<$1e-4 have been difficult to detect in both scattered-light (e.g. \citealt{Esposito20}) and emission, particularly for M-dwarfs \citep{Matthews14}.

%Additionally, the larger radius of the TWA 10 disk leads to a peak SED flux at a wavelength of 228 $\mu$m, almost twice as far as the TWA 25 SED, which peaks at $\sim$123 $\mu$m.

The main takeaway from this exercise is that the most likely reason the TWA 10 and TWA 25 debris disks (+ potentially other M-dwarf disks) were not detected with \textit{Spitzer} and \textit{Herschel} is because their disk emission is too faint. However, the disk SEDs remain highly uncertain given degeneracies between various dust grain properties and the total dust mass. Thus, future observations at longer wavelengths to detect the disk emission are necessary. 

\section{Discussion} \label{sec:discuss}
In the previous sections, we analyzed the TWA 10 and 25 debris disks in our NIRCam observations by characterizing the disk geometry, surface density, SPF, and surface brightness. In the following sections, we provide further discussion of the implications of our results for both disks.

\subsection{TWA 10} \label{sec:TWA10}
\begin{figure}
    \centering
    \includegraphics[width=0.472\textwidth]{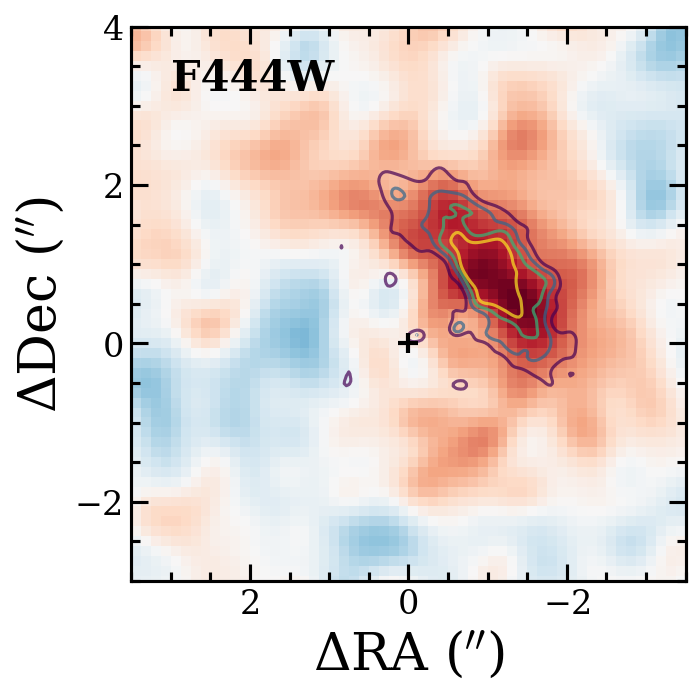}
    \caption{\label{fig:twa10_f444w} Close in look of our convolved MCRDI reduction of TWA 10 in the F444W filter. The data is binned into 3x3 pixel bins and is further smoothed with a gaussian with $\sigma$=2 pixels. Contours represent the surface brightness of the disk in F200W (0.07, 0.10, 0.13, 0.16 MJy sr$^{-1}$). While very faint, a clear signal is observed at the location of the disk as seen in the F200W.} 
\end{figure}

Our observations of TWA 10 provide the discovery and first detection of its debris disk. Previous studies have reported the lack of detected IR-excess in the TWA 10 SED due to a debris disk (e.g. \citealt{Schneider12,RM16}). The SED can be seen in the top plot of Figure \ref{fig:seds} in \ref{sec:seds}, which shows no IR-excess at wavelengths less than 30 $\mu$m and was classified as non-detections by Spitzer/MIPS and Herschel/PACS at 70-500 $\mu$m. Additionally, the disk is relatively faint, with a surface brightness ranging between $\sim$0.1 and 0.4 MJy/sr or $\sim$0.002-0.008 mJy arcsec$^{-2}$ in F200W. Furthermore, the disk is barely detected in F444W, where significant spatial binning and smoothing is required to tease out the disk signal (see Figure \ref{fig:twa10_f444w}). The lack of detected IR-excess, likely due to the faintness of the disk, explains why the TWA 10 disk was unknown until now; it is now the second M-dwarf debris disk to be discovered by NIRCam after TWA 20 \citep{Palatnick25}. Our observations allow us to characterize the morphology of the TWA 10 debris disk for the first time.

Based on our modeling, we can learn basic information about the disk geometry. The disk is moderately inclined, with an inclination of $\sim 63\fdg7$ and a PA of $\sim 220\fdg5$ (measuring from East of North). We derive a disk radius of 191.15$\pm$6.53 au via the emcee method, although the Powell method prefers a slightly smaller radius of $\sim$157 au. In either case, this makes the TWA 10 disk the largest M-dwarf debris disk to be imaged to date, with most M-dwarf debris disks having radii $\lesssim$70 au, as shown in Figure \ref{fig:disk_radii}. In addition to the disk geometry, we find that it is difficult to strongly constrain the surface density profile, although results for $\alpha_{out}$ suggests a relatively steep surface density gradient in the outer edges of the disk. Finally, because only a small range of scattering angles are detected, we find that the SPF is also difficult to constrain, with no constraints made for $g_{2}$ in the F200W or any of the SPF parameters in the F444W. We note that the Powell method shows a preference for $g_{1}$ and $g_{2}$ parameters close to 1, which would suggest that the disk is highly forward scattering. However, higher SNR observations will be required to confirm these results.

Our most interesting result for the TWA 10 disk is its exceptionally large radius, as again, it is by far the largest M-dwarf debris disk to be discovered to date. This discovery challenges our understanding of the formation and evolution of M-dwarfs and their circumstellar environments. For example, observations of class II protoplanetary disks show similar correlations between disk size and spectral type, where disks around low-mass stars tend to be less massive and more compact (e.g. \citealt{Andrews13,marel21,Alvarado25}). While there are also more extended and structured protoplanetary disks observed around low-mass stars, many of these disks still have dust radii less than 100 au (e.g. \citealt{Kurtovic21,Panilla22,Alvarado25}). Thus, it is not clear if the disk originated at its current location (assuming the dust is tracing the planetesimal belt) or was moved out to a larger radius by another mechanism. The latter seems less likely, as radiation pressure \citep{Burns79} is negligible for M-dwarfs due to their low-luminosity. On the other hand, stellar winds, which is a dominant force for M-dwarfs, can export dust both inwards and outwards (e.g. \citealt{Burns79,Pawellek19}). For example, both stellar wind forces have been used to explain the inner regions and extended outer halo for the AU Mic debris disk (e.g. \citealt{Strubbe06,Schuppler15}). However, in the case of AU Mic, the disk surface brightness profile peaks at $\sim$30-40 au, i.e. at the relative location of the planetesimal belt \citep{Augereau06,Strubbe06,Macgregor13,Lawson23}. Because the TWA 10 disk surface brightness peaks at $\sim$191 au, this suggests that the dust observed with NIRCam is tracing the planetesimal belt. Future observations of the TWA 10 disk, especially at longer wavelengths, will be crucial for shedding more light on the disk's origins by constraining the spatial distribution of the large grains, as well as the dust grain size distribution.

% However, TWA 10 is not the first M-dwarf found to have a relatively large disk radius. Other large debris disks around M-dwarfs include GSC 07396-00759 and CP-72 2713 \citep{Sissa18,Moor20}, both located in the $\beta$ Pic moving group. While the GSC 07396-00759 disk is only slightly larger than that of TWA 20 and 25 at $\sim$70 au \citep{Cronin22}, the CP-72 2713 disk is twice as large with a peak radius of $\sim$140 au, although current observations of the disk with \textit{Herschel} and ALMA are not fully resolved \citep{Moor20}. While the TWA 10 disk is still the largest M-dwarf disk, the increasing spread of M-dwarf disk radii may suggest that there is a larger variation in M-dwarf debris disks than previously thought. 

% \begin{figure}
%     \centering
%     \includegraphics[width=0.472\textwidth]{figures/twa10_closeup.png}
%     \caption{\label{fig:twa10} \textbf{Left:} Our convolved MCRDI reduction of TWA 10 in the F200W filter. \textbf{Right:} The best fitting model of the TWA 10 debris disk. The square apertures represent the location of the disk ansae of the model, while the contours in each panel represent the disk flux in the F200W filter.} 
% \end{figure}

\subsection{TWA 25} \label{sec:TWA25}
The TWA 25 debris disk is detected strongly in the F200W and faintly, but still significantly, in the F444W. These observations serve as the fourth detection of the disk (all of which are in scattered-light), where other published observations of the disk include data taken with the NICMOS and STIS instruments on HST \citep{Choquet16,Ren23} and most recently with the SPHERE instrument on the VLT \citep{Engler25}. With NIRCam's improved sensitivity, we are able to better resolve the structure and confirm the morphology of the disk.

We compare our best-fit model values of the disk geometry with those derived from NICMOS and SPHERE. With our modeling, we are able to derive mostly consistent results with the geometrical parameters derived from the NICMOS and SPHERE modeling with additional constraints on the surface density profile. The consistent results include the disk inclination ($77\fdg9 \pm 1\fdg1$ compared to $75^{\circ} \pm 6^{\circ}$ and $78\fdg3 \pm 1\fdg5$ for NICMOS and SPHERE, respectively) and $PA$ ($337\fdg1 \pm 0\fdg7$ compared to $336^{\circ} \pm 8^{\circ}$ and $336\fdg8 \pm 1\fdg8$). However, we derive a smaller radius than previous estimations ($63.8 \pm 3.3$ au compared to $78 \pm 17$ au and $76 \pm 2$). Although the radius is still consistent with the NICMOS estimates within 1$\sigma$ this is not the case for the SPHERE results. This discrepancy may be due to the sensitivity and data reduction techniques of the NICMOS and SPHERE observations. Whereas the NICMOS observations are simply low SNR, the SPHERE observations utilize techniques such as Angular Differential Imaging \citep{Marois06}, which can result in disk self-subtraction (evident by the difference in vertical width between observations) and thus skew the radius estimate. Additionally, the radius measurement from SPHERE is derived from fitting an ellipse to the processed image, which does not take into account potential effects from the PSF subtraction process. Although the backside of the TWA 25 disk is still not detected with NIRCam, the high-sensitivity of the observations, lack of disk self-subtraction at the outer edges, and use of forward modeling suggest that our radius estimation is likely to be closer to reality. Finally, we find that the modeling prefers a steep inner dust density profile ($\alpha_{in}$) of $\sim$12-17, although the uncertainty derived from the emcee method is large. If $\alpha_{in}$ is indeed steep, this may suggest that the disk is being sculpted by a planet within the disk's inner radius. 

For the disk SPF, our modeling prefers a high value for $g_{1}$ of $\gtrsim 0.9$, again suggesting a highly forward scattering disk at 2 $\mu$m (we are unable to constrain the SPF in the F444W). However, because the flux from the disk is significantly removed at the location of the forward scattering peak, due to attenuation from the coronagraph, the results from this analysis should be taken with a grain of salt. For example, conducting the same forward modeling process (via the Powell method) with a single-component HG function leads to a lower preferred $g$ value of 0.79. Thus, future observations, where the flux is not significantly removed from the forward scattering peak, will be necessary to place tighter constraints on the SPF.

% As mentioned previously, the improved sensitivity of observations allow us to probe the backside of the disk (located in the East), shown in Figure \ref{fig:twa25}, which is not present in previous observations. Not only does this give us better constraints on the disk radius and geometry, but it also provides information about the surface density profile. For example, properly modeling the backside of the disk required a steep $\alpha_{in}$, where we derive $\sim$17.7 from our best-fit model. Such a steep inner surface density power-law could suggest possible sculpting by one or multiple planets within the disk's inner radius.

\begin{figure}
    \centering
    \includegraphics[width=0.472\textwidth]{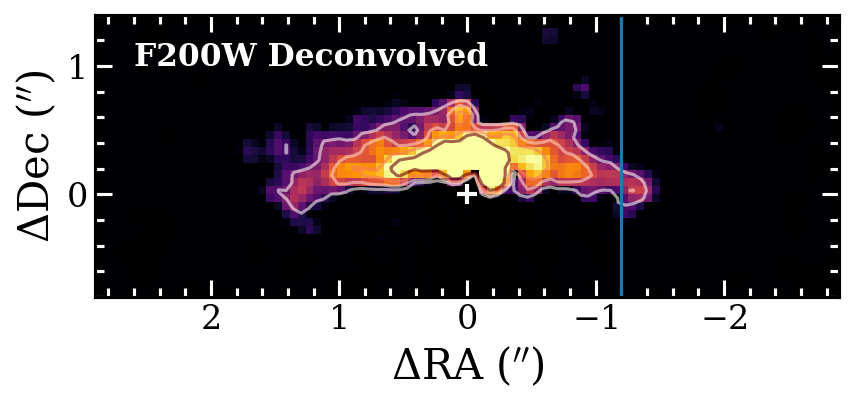}
    \includegraphics[width=0.472\textwidth]{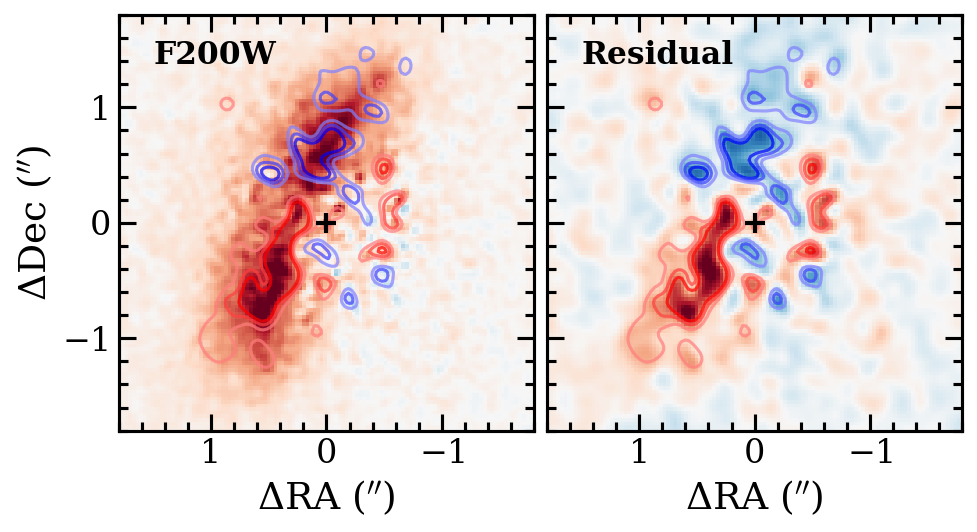}
    \caption{\label{fig:twa25_resid} \textbf{Top:} Our MCRDI reduction in the F200W filter deconvolved with the instrumental PSF. The image is rotated by the disk PA, so that the major-axis is horizontal. Contours are added at 2, 5, 10, and 20 MJy sr$^{-1}$ to highlight the observed brightness asymmetry. \textbf{Bottom Left:} Our MCRDI reduction of TWA 25 in the F200W filter. \textbf{Bottom Right:} The residual map after the best fitting model of the TWA 25 debris disk is subtracted. The red contours represent positive residuals between 0.1 and 0.2 MJy sr$^{-1}$, while the blue contours represent negative residuals between -0.2 and -0.1 MJy sr$^{-1}$.} 
\end{figure}

Additionally, a tentative brightness asymmetry appears to be present at 2 $\mu$m, where the SE side of the disk is moderately brighter than the NW side. This brightness asymmetry can also be seen in the residual profile after subtracting our best-fit model, where a significant positive residual is present in the SE side (see Figure \ref{fig:twa25_resid}). A similar brightness asymmetry also appears to be present in the NICMOS data, which is at a similar wavelength ($\lambda \sim$1.6 $\mu$m), although it is not clear if it is observed in the SPHERE data. We also confirm this brightness asymmetry between a separation of $\sim 0\farcs5$ and $1\farcs2$ in the surface brightness profile (see Figure \ref{fig:sb_prof}). To further quantify the asymmetry, we place two equally sized rectangular apertures on either side of the disk between $0\farcs5$ and $1\farcs2$ in both the F200W and F444W images. We then take the sum and standard deviation of the flux in each aperture and divide the West summed flux from the East summed flux. Doing so, we find that the East side of the disk is roughly 1.10$\pm$0.01 times brighter than the West side at 2 $\mu$m while the 4 $\mu$m data is consistent with being symmetric (1.02$\pm$0.02). However, the actual surface brightness asymmetry may be more extreme than the initial measurements suggest due to artificial asymmetries introduced by the coronagraphic PSF core, which is oval shaped with skewed side-lobes. This effect is evident from the best-fit model to the F200W data, where the coronagraphic PSF has caused the NW side of the disk to appear brighter than the SE side (although the same effect does not appear in the F444W model). Thus, we apply the same measurements to a version of our F200W reduction that is deconvolved from the instrumental PSF for comparison (shown in the top plot of Figure \ref{fig:twa25_resid}). As hypothesized, the asymmetry is even more extreme in the F200W deconvolved image, where we measure the East side of the disk to be 1.25$\pm$0.04 times brighter than the West side.

Brightness asymmetries can occur through several mechanisms. For example, if the disk is eccentric, such as through sculpting from an eccentric planet, the disk pericenter will scatter more light from the star causing a ``pericenter glow" \citep{Wyatt99}. However, if this is the case, we would expect to see a similar brightness asymmetry at both wavelengths. Additionally, we do not detect any significant offset of the disk from the star. One reason the disk may be asymmetric in brightness at different wavelengths is variations in dust grain properties across the disk, such as composition, porosity, or grain size. For example, this can be the result of a recent large collision between two rocky bodies in the system, a hypothesis that has been used to help explain the dust clumps and brightness asymmetries in other debris disks such as $\beta$ Pic and HD 111520 \citep{Dent14,Crotts22,Jones23,Rebollido24}. However, the F200W and F444W observations should trace relatively the same dust grain population given the small difference in wavelength. Because the SNR is lower at 4 $\mu$m, it is possible that we simply do not have the sensitivity to detect the brightness asymmetry in the F444W filter. Future observations at longer wavelengths, such as will ALMA, would be useful to determine the spatial distribution of the larger grains, which can shed more light on the mechanisms shaping the TWA 25 disk.

\section{Searching for Planets}
In addition to analysis of the TWA 10 and 25 debris disks, we also search for potential planetary candidates in these systems which could be dynamically interacting with their disks. We are particularly interested in faint point-sources within $\sim$5$''$ as we can constrain their color between the F200W and F444W filters. We expect that candidates in the planetary regime are very red in color, and therefore should appear in the F444W images but not significantly in F200W images. Point-sources seen outside of 5$''$ may still be potential candidate companions; however, it is less likely given their distances from the host star (higher likelihood of contamination from background objects) and we are unable to constrain their colors. Therefore, we do not consider them in this Section, although we constrain the location and F444W flux of these objects in Section \ref{twa25_bkg} located in the Appendix. We compute the contrast curves and planetary mass sensitivity of our observations, as well as characterize several potential candidate companions found within the TWA 10 and 25 systems.

\subsection{Contrast Curves} \label{sec:contrast}
\begin{figure}
    \centering
    \includegraphics[width=0.472\textwidth]{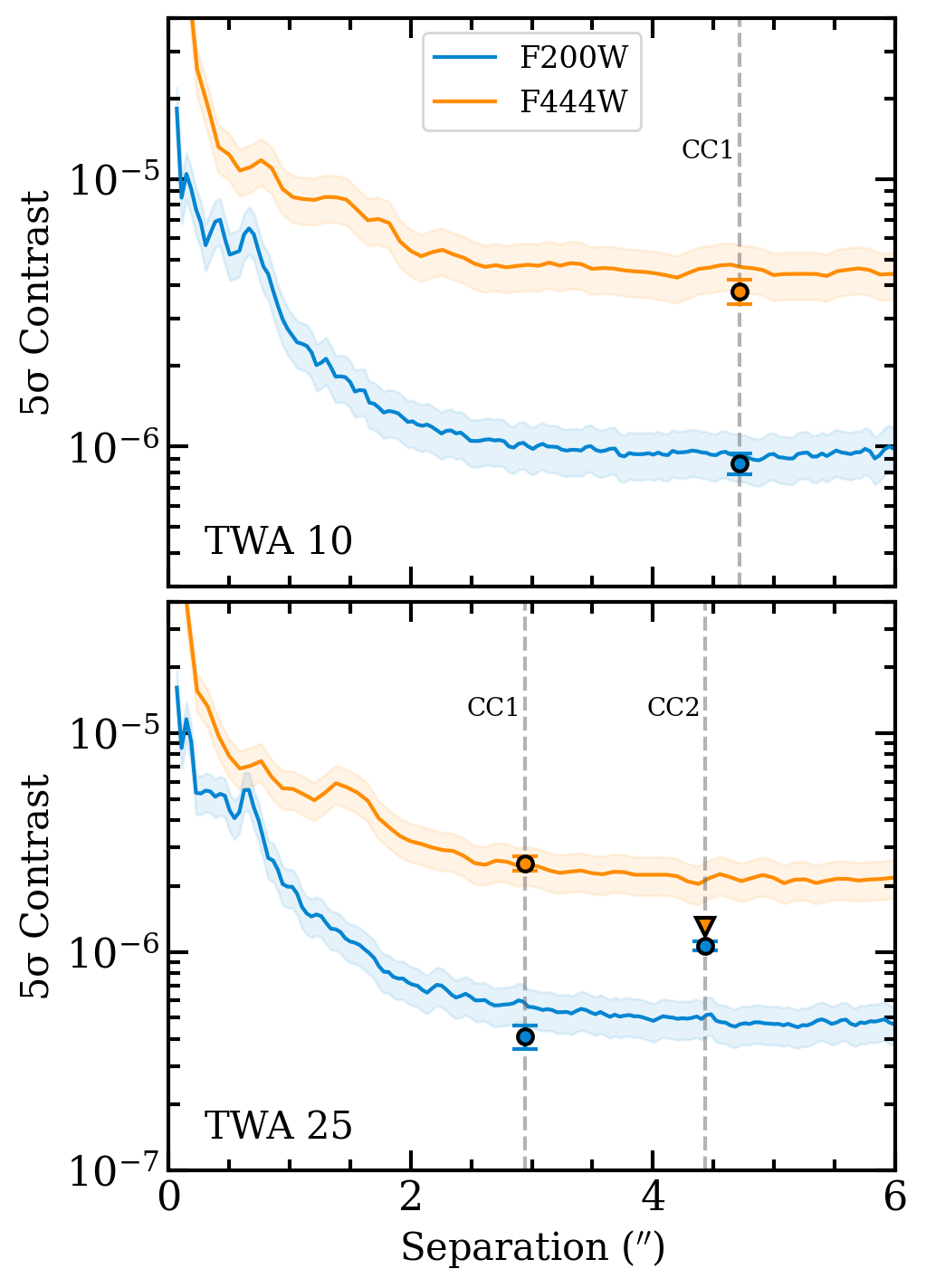}
    \caption{\label{fig:contrast} 5$\sigma$ contrast curves of our disk model subtracted data for TWA 10 (\textit{top}) and TWA 25 (\textit{bottom}). The blue lines are the contrasts for the F200W observations, while the orange lines are the contrasts for the F444W observations. The shaded regions represent $\pm 1 \sigma$ contrasts. The blue data points are the contrasts for potential candidates observed in F200W, while the orange data points are the contrasts for potential candidates observed in the F444W. This includes one candidate in the TWA 10 system and two potential candidates in the TWA 25 system.} 
\end{figure}

To calculate the 5$\sigma$ contrast curves for our observations, we utilize \texttt{spaceKLIP} for which we input the disk model subtracted MCRDI reductions and the stellar SED. The resulting outputs are considered the raw 5$\sigma$ contrast curves, where a second step is usually conducted in \texttt{spaceKLIP} to produce calibrated contrast curves, taking into account extra variables such as the throughput of its KLIP algorithm. However, because the data are reduced with MCRDI instead of KLIP, this step becomes unnecessary as there is no regime where the MCRDI throughput should impact point-source detectability at the 5$\sigma$ level. Thus, the raw contrast curves, which measures the image noise and takes into account coronagraphic throughput, are sufficient to evaluate the planet mass sensitivity of our observations and the contrasts of any potential candidates. Our final contrast curves are plotted in Figure \ref{fig:contrast}. We also include the contrasts for the three candidates identified and analyzed in the later Section \ref{sec:comp_ext}.

\begin{figure*}
    \centering
    \includegraphics[width=0.472\textwidth]{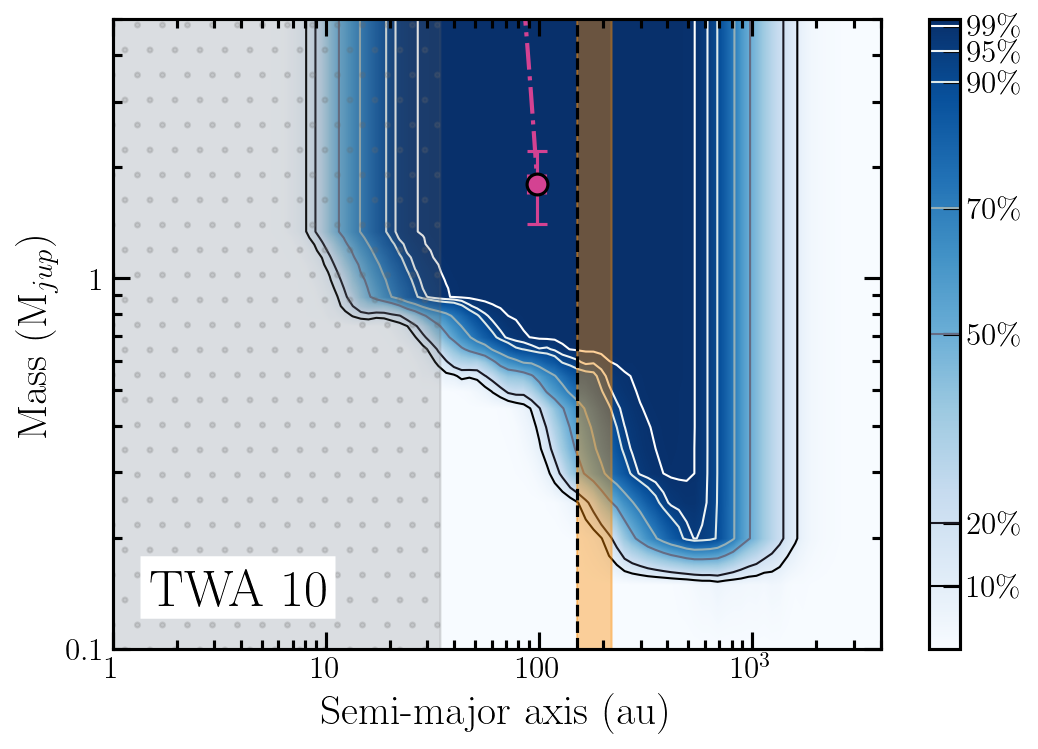}
    \includegraphics[width=0.472\textwidth]{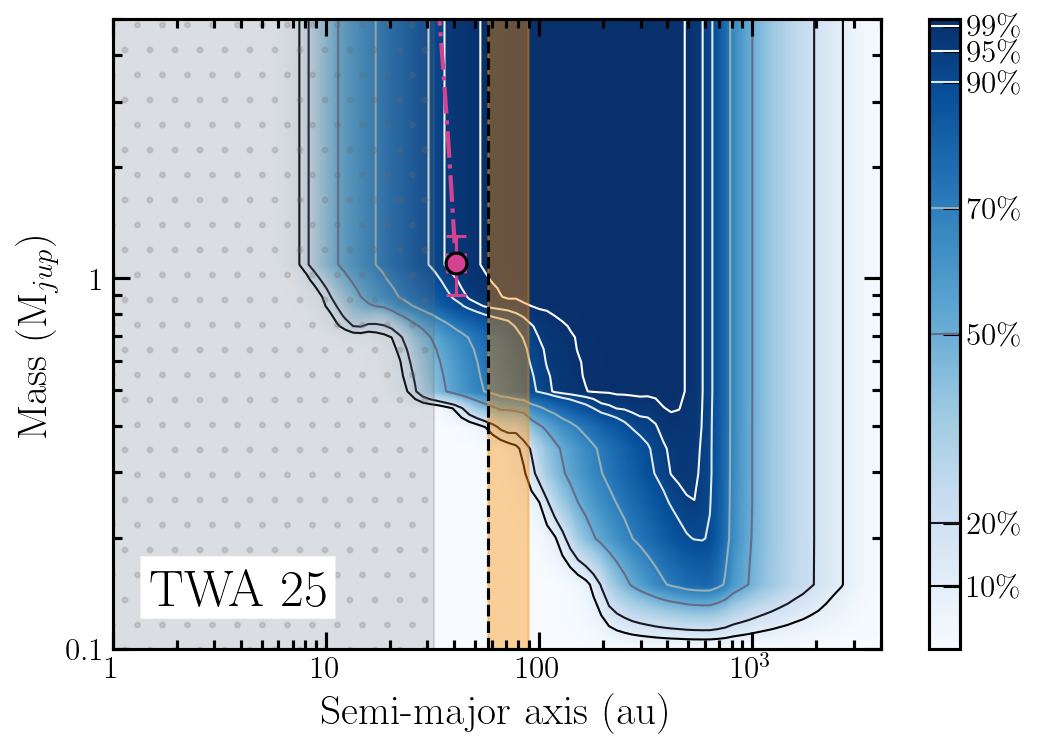}
    \caption{\label{fig:planet_probs} Planet detection probability maps for TWA 10 (\textit{left}) and TWA 25 (\textit{right}) as calculated by \texttt{ExoDMC} using the planet mass sensitivity derived from our 5$\sigma$ contrast curves. The dashed black line represents the estimated disk inner radius, while the orange shaded region represents the disk extent. The grey shaded region represents the area where the throughput is $\le$50\% (e.g. $\le$0.6$''$). The pink data point represents the maximum semi-major axis and minimum planet mass expected for a single planet that is actively sculpting the disk inner edge, while the pink dashed-dot line represents the relationship between planet mass and semi-major axis, both derived using code from \citep{Pearce22}}. %The pink dashed line represents the range of projected stellar separations possible for a planet with the same inclination as the disk.} 
\end{figure*}

We use our F444W 5$\sigma$ contrast curves to also measure the planet mass sensitivity of our observations. To do so, we use the standard ATMO-CEQ planet evolutionary models from \citep{Phillips20} and the cloud-free BEX models with solar metallicity ([M/H]=0) from \citet{Linder19} for a system of age 10 Myr. However, we note that these models represent the most optimistic case for the planet-mass sensitivity as clouds, higher metallicity, and non-equilibrium chemistry can suppress the planet flux (e.g. \citealt{Mang24,Rubin25,Crotts25}). The resulting planet mass sensitivities are then inputted into \texttt{ExoDMC} \citep{Bonavita20}, a rendition of the MESS code \citep{Bonavita12} that calculates exoplanet detection maps by estimating the probability of detection of a synthetic planet population. We assume a gaussian distribution for the eccentricity with a mean of 0 and a $\sigma$ of 0.3. Additionally, we use the measured inclinations of the debris disks in both systems to add further constraints to the detection probability maps. This assumes that any planetary companions present are coplanar with the disks. Coplanar orbits are expected for exoplanets that formed within their protoplanetary disks and have been observed for other directly imaged exoplanets with debris disks (e.g. $\beta$ Pic b: \citealt{Lagrange10,Macintosh14,Nielsen14}, HD 95086 b: \citealt{Rameau16}, HR 8799 b-e: \citealt{Wang18}). The resulting exoplanet detection probability maps can be found in Figure \ref{fig:planet_probs}.

We compare our detection probability maps with the extent of the TWA 10 and 25 debris disks, as well as the expected maximum semi-major axis and minimum mass expected for a single planet actively sculpting the disk. The inner radius and extent of each disk is estimated using the best-fit $r_{0}$, $\alpha_{in}$, and $\alpha_{out}$ values, and are represented by the black dashed lines and orange shaded regions in Figure \ref{fig:planet_probs}. Based on the disk inner radius ($\sim$149.6 and $\sim$57.5 au for TWA 10 and 25, respectively), stellar mass, and system age we estimate the maximum semi-major axis and minimum mass for potential planets in each system. These estimations are based on the assumption that a single planet is actively sculpting its disk near the disk inner edge and is responsible for clearing the material interior to the disk. 

We use the code \texttt{SculptingPlanet}, a model that is used in \citet{Pearce22} and builds upon \citet{PW14}. We assume that the planet has a circular orbit, as our observations do not strongly suggest that either debris disk is eccentric. We also assume a 10\% error on the disk inner radius; observations at longer wavelengths, such as with ALMA, will be crucial for more robustly estimating the disk inner edge, as larger grains more closely trace the planetesimal belt compared to the micron-sized grains probed at NIR wavelengths. This leads to an estimated maximum semi-major axis and minimum planet mass of $98\pm9$ au and $1.8\pm0.4$ M$_{jup}$ for TWA 10 and $41\pm4$ au and $1.1\pm0.2$ M$_{jup}$ for TWA 25, represented by the pink data points in Figure \ref{fig:planet_probs}. For larger planets, the semi-major axis decreases slightly, as the planet does not have to be as close to the disk to sculpt it. The relationship between planet mass and semi-major axis is represented by the pink dashed-dot lines. Our results will be further discussed in Section \ref{sec:implications}.

%Finally, to take into account the planet's inclination, we estimate the range of separations from the star (assuming the minimum planet mass) that such a planet may have along its projected orbit on the sky. Specifically, the estimated sma represents the stellar separation on the sky a planet would have at apocenter/pericenter, while the stellar separation will decrease as it moves along the major-axis of its inclined orbit. Plotting both orbits in \texttt{REBOUND}, an n-body simulation tool, reveals that the predicted minimum mass planets could reach projected stellar separations down to $35\pm7$ au for TWA 10 and $5.5\pm3$ au for TWA 25 (pink dashed lines in Figure \ref{fig:planet_probs}. These projected separations will decrease for increasing planet mass.

\subsection{Companion Extractions \& Colors} \label{sec:comp_ext}
\begin{figure}
    \centering
    \includegraphics[width=0.472\textwidth]{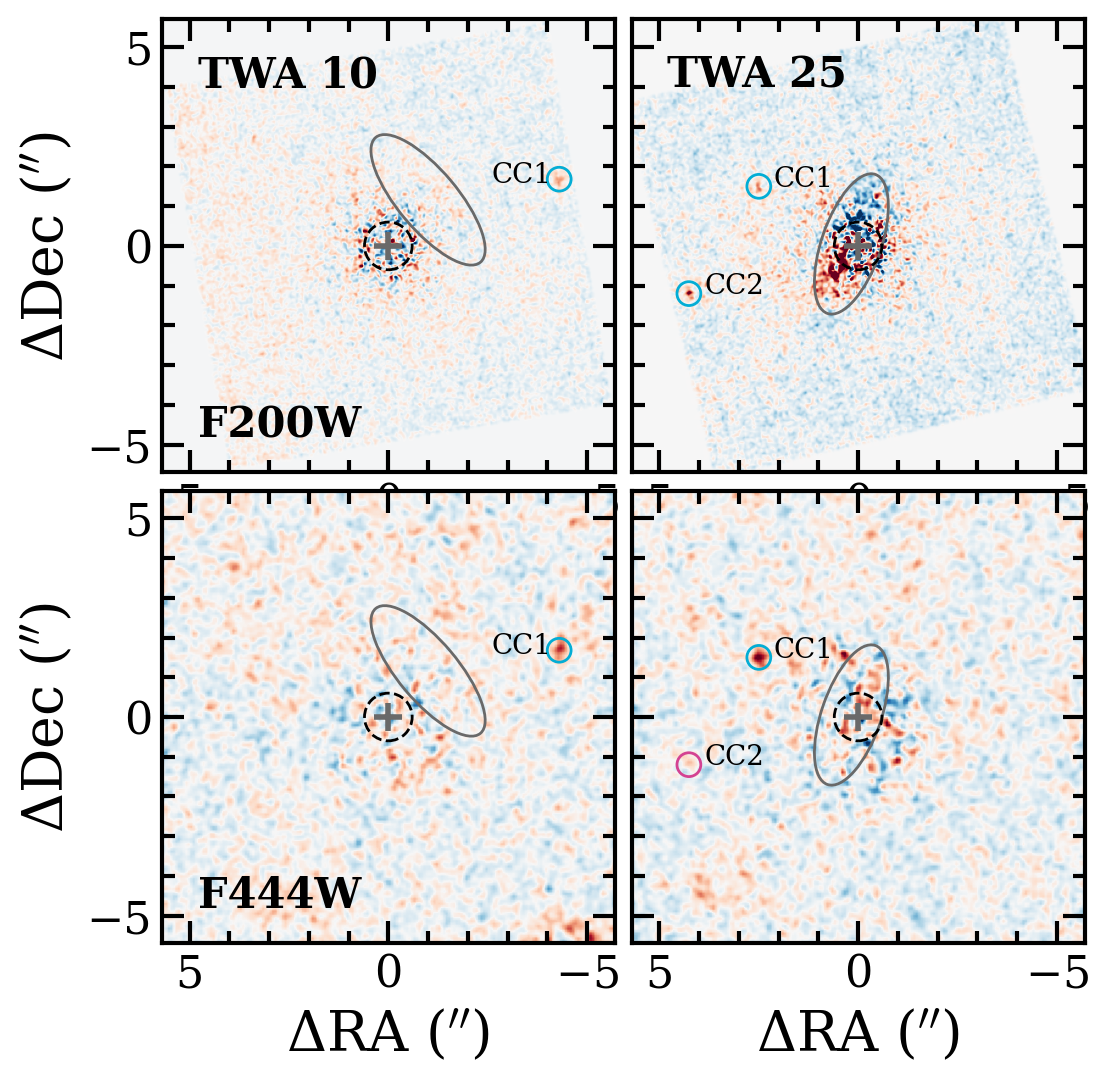}
    \caption{\label{fig:candidates} Model subtracted observations for TWA 10 (\textit{left}) and TWA 25 (\textit{right}), where the F200W data for both systems is shown in the top row and F444W data is shown in the bottom row. The blue circles represent identified candidates that are present in that specific filter, while pink circles represent identified candidates that are only observed in the opposite filter. Each candidate companion is labeled by CC1-CC2. The black dashed circles represents a radius of 0.6$''$ where the throughput reaches 50\%, while the solid grey ellipses highlight the location of the disk.} 
\end{figure}

Analyzing our observations within $\sim$5$''$, we find one significant point-source for TWA 10 and two for TWA 25 that are identifiable by eye above the background noise. One downside to RDI, is that objects observed in reference images may be injected into our final reductions. To ensure that none of the point-sources were injected by one of the reference images, we use the jack-knife test, which removes reference images from our reductions one-by-one. Doing so, we find that all three point-sources pass the jack-knife test, meaning that they are indeed present in our observations of TWA 10 and TWA 25. Thus, we now refer to these sources as candidate companions (CC) and are highlighted in Figure \ref{fig:candidates}. We also note that all the candidates are outside the disk location. Although it is certainly possible for a planet to be orbiting outside the disk (e.g. HD 106906, \citealt{Bailey14}), these candidates are not related to the planets predicted in Section \ref{sec:contrast} near the disk inner edge.

\begin{figure}
    \centering
    \includegraphics[width=0.472\textwidth]{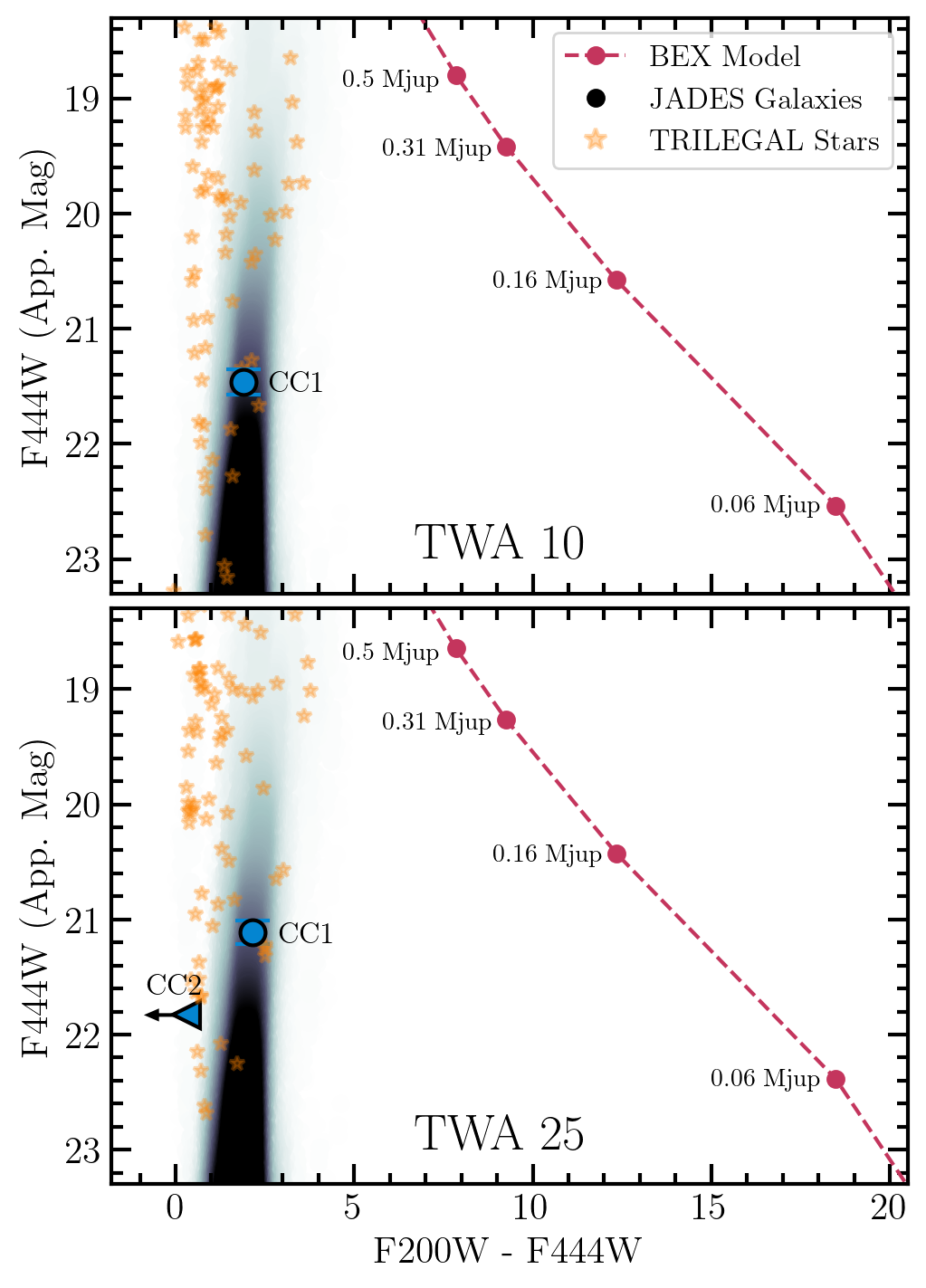}
    \caption{\label{fig:cmd} F200W--F444W colors vs. F444W apparent magnitude for TWA 10 (\textit{top}) and TWA 25 (\textit{bottom}). The black/grey data points represent the sample of galaxies measured as part of the JADES survey, while the orange stars represent the model stars from TRILEGAL. The blue data points represent the candidate companions in the TWA 10 and TWA 25 systems, which we compare to the BEX evolutionary model (pink dashed line).} 
\end{figure}

To estimate their likelihood of being planetary in nature, we first use \texttt{Winnie} to measure their magnitude in each filter. This is done using a point-source model which are forward-modeled and fit to each candidate using \texttt{lmfit.minimize} via the same procedure used to find best-fit disk model. The measured RA/Dec and apparent magnitudes for each candidate can be found in Table \ref{tab:planet_mod} and contrasts are plotted in Figure \ref{fig:contrast}. Because the CC2 candidate is not present in F444W for the TWA 25 system, we set an upper limit on the magnitude equivalent to a 3$\sigma$ contrast at that separation. The measured apparent magnitudes are then used to calculate the F200W-F444W color, which we compare to colors measured for galaxies as part of the JADES survey \citep{Eisenstein23}, as well as stellar models from the TRILEGAL code (TRIdimensional modeL of thE GALaxy; \citealt{Girardi05}). Additionally, we compare to the BEX evolutionary model (assuming a system age of 10 Myr) via a color-magnitude diagram (CMD). The resulting CMD is shown in Figure \ref{fig:cmd}. 

\begin{table*}
	\centering
	\caption{\label{tab:planet_mod}Results from candidate model fitting with \texttt{spaceKLIP} including the measured RA and Dec, as well as apparent magnitudes in both filters. Here, a negative RA is to the right of the star and a positive RA is to the left of the star.}
	\begin{tabular*}{\textwidth}{c @{\extracolsep{\fill}} ccccc}
	    \hline
	    \hline
		Name & Candidate & RA ($''$) & Dec ($''$) & F200W (App. Mag) & F444W (App. Mag) \\
		\hline
        TWA 10 & CC1 & -4.32$\pm$0.01 & 1.70$\pm$0.01 & 23.37$\pm$0.10 & 21.46$\pm$0.11 \\
		  TWA 25 & CC1 & 2.51$\pm$0.01 & 1.50$\pm$0.01 & 23.26$\pm$0.14 & 21.11$\pm$0.08 \\
		  ... & CC2 & 4.26$\pm$0.01 & -1.19$\pm$0.01 & 22.23$\pm$0.05 & $>$21.83 \\
		\hline
		\hline
	\end{tabular*}
\end{table*}

We find that all three candidates across both systems are likely too blue in color to be potential planet companions, especially CC2, which only has an upper limit on color. Additionally, all three candidates lie within the same parameter space as the JADES galaxies and TRILEGAL stars, further suggesting that they are likely background objects. Although we do not yet fully understand the full extent of possible colors and evolutionary tracks for the demographic of low-mass, wide-orbit planets, we do not further consider these three sources as planet candidates in this work.

\subsection{Implications of Exoplanet Non-Detections} \label{sec:implications}
In the previous two sections, we rule out the three sources found in our observations within 5$''$ to be exoplanet companions, as well as infer the possible planet semi-major axis and masses based on the disk inner radius. Given these analyses, we can say something about the implications of both the lack of detection of potential planet candidates in our observations and where we expected to find such candidates based on the debris disk morphologies. 

For both the TWA 10 and 25 systems, the predicted planets have a significant probability of being detected in the F444W filter based on the disk inner edge, stellar mass, and age. While the probability is slightly lower for TWA 25 ($\sim$90-95\% depending on planet mass) given the higher inclination and smaller disk extent, the predicted planet based on the TWA 10 disk would have a 99\% probability of being detected. Even in the case of a larger planet mass and thus decreased required semi-major axis, the planet would still have a 99\% probability of being detected in the F444W. Additionally, we create the same detection probability maps using the F200W planet mass sensitivity curves for comparison, which shows that a planet of mass $\gtrsim$2.5 M$_{jup}$ would also have a 99\% probability of being detected in either system. In summary, we should have detected a planet in both the TWA 10 and 25 systems, assuming a single planet sculpting the disk inner edge. In the case of a multi-Jupiter mass planet, we would have also likely detected it in the F200W filter.

In the case of TWA 25, our analysis suggests that there is a $\lesssim$10\% chance that a disk sculpting planet of $\lesssim$2.5 M$_{jup}$ (again, planets with masses $\gtrsim$2.5 M$_{jup}$ would have been detected in the F200W filter) may have eluded detection given its small semi-major axis. TWA 10 presents an even more perplexing situation. Regardless of planet mass, if a single planet is responsible for sculpting the disk, we should have detected it with at least 5$\sigma$ significance. However, we do not detect any significant point sources in either filter outside of CC1, which we have already determined is most likely a background object. This conundrum raises the question as to why we are not detecting any planet companions in either system when their debris disks suggest that there should be one.

There are several potential scenarios that could explain the lack of a planet detection in the TWA 10 and TWA 25 systems; 

\begin{enumerate}
\item The atmospheric evolutionary models used over-predict the expected planet mass sensitivity, given the lack of observed sub-Jupiter mass planets that help inform such models. For example, water clouds,  disequilibrium chemistry, and increased metallicity can cause suppression in the F444W band (e.g. \citealt{Rubin25}), which is not accounted for in the \citet{Linder19} models. A similar conclusion was reached in the case of TWA 7 b, where the same BEX models predicted a planet mass closer to 0.06 M$_{jup}$ \citep{Crotts25}. However, combined photometry of NIRCam and MIRI observations suggested a warm, Saturn-mass planet with a disequilibrium chemistry and above solar metallicity atmosphere \citep{Lagrange25,Crotts25}. Therefore, it is possible that the atmospheric chemistry of sub-Jupiter mass planets present in these systems differ from the chemistry assumed in the BEX models, making them more difficult to detect than expected. A detailed investigation regarding the effects of clouds and chemistry on the estimated sensitivity from the NIRCam surveys (including GO 4050) are currently being conducted in a separate effort \citep{Strampelli26}.

\item Multiple smaller planets are sculpting the debris disks rather than one single larger planet, making them out of reach of direct detection, even in the case of the BEX evolutionary model. This scenario is also explored in \citet{Pearce22}, assuming multiple planets of equal mass, where they found that the planets in the multi-planet scenario were on average 40 times less massive than a single planet. This scenario is also supported for the TWA 25 system, where the sharp inner-edge of the disk ($\alpha_{in}=11.94\pm5.37$) suggests sculpting by multiple smaller planets, as more massive planets excite dust eccentricities and cause the inner edge to smooth out \citep{Pearce24}.

\item Planets may have started closer to the disk early on and then migrated inwards over time, beyond what we can detect given the IWA of NIRCam. This hypothesis has been suggested in recent studies, where \citet{Pearce24} found that, on average, the inner edges of debris disks were too steep to be caused solely by collisional activity, but not steep enough to be solely due to planet sculpting. In summary, the process of inward migration would cause debris caught in resonance by the planet to become excited on increasingly high eccentricities \citep{Wyatt03,Reche08,Friebe22}, resulting in a more shallow inner-edge profile. %Although this may be a plausible scenario for the TWA 10 system, which has a disk with a more shallow inner surface density power-law, albeit with a large uncertainty ($\alpha_{in} = 4.53\pm3.36$), the sharp inner surface density power-law for TWA 25 suggests that it is actively being sculpted by one or more inner planets. 

\item The location of the debris disk inner edge is unrelated to planets, but is rather set by processes during the protoplanetary disk phase. For example, planetesimals may naturally form in distinct radial zones via streaming-instability \citep{Carrera17} or snow-lines (e.g. \citealt{Ida16,drazkowska17,Schoonenberg17,Schoonenberg18,Izidoro22,Morbidelli22}). Other mechanisms can also stir planetesimals in the disk to produce a collisional cascade, such as self-stirring (if the disk is massive enough; \citealt{Kenyon01}) or stellar flybys \citep{Ida00,KB02}.
\end{enumerate}

\section{Summary}
We presented new scattered-light observations of the M-dwarf debris disks around TWA 10 and TWA 25 with NIRCam, taken in the F200W and F444W filters. While the TWA 25 debris disk has been previously imaged with HST and SPHERE, NIRCam has provided the most sensitive observations of the disk to date, probing the disk substructure on a much deeper level. Additionally, our observations have led to the discovery of the TWA 10 debris disk, providing the opportunity to characterize the disk for the very first time.

Through modeling and analysis of each disk, we find the following results;

\begin{itemize}
    \item The TWA 10 disk is by far the largest M-dwarf debris disk detected to date, with a radius of $\sim$191 au. This finding suggests that M-dwarf debris disks may have a larger range of radii than previously thought, bringing into question the possible evolutionary tracks for these disks and the relationship between disk radius and stellar luminosity.
    \item We are able to probe the TWA 25 disk morphology with higher sensitivity than other scattered-light observations, allowing for greater characterization of the disk. We find that the disk radius is smaller than previous estimations, where we derive a radius of $\sim$63.8 au compared to $\sim$76-78 au. Additionally, we find that the disk has a very steep inner dust-density slope of 11.94$\pm$5.37, suggesting that the disk inner radius may be actively sculpted by planets. Measuring the surface brightness reveals a small but significant brightness asymmetry between the SE and NW sides of the disk in the F200W, while the disk st{is} appears axisymmetric in F444W. 
    \item Analysis of the potential disk SED for both systems, based on our NIRCam data and limits set by MIPS and PACS, suggests that the IR-excess was not robustly detected in the MIR-FIR likely due to low luminosity fractions. However, future observations of the disk emission, such as with ALMA, will be required to place more robust constraints the disk SEDs.
\end{itemize}

In addition to analyzing the disk morphology, we also use our observations to search for potential planetary candidates. We find one potential candidate in the TWA 10 system and two in the TWA 25 system within 5$''$. Color analysis of these candidates reveal that they are likely background objects as they are too blue and overlap with both galaxies observed in the JADES survey and stellar models from the TRILEGAL code.

Finally, we create planet detection probability maps using the measured 5$\sigma$ contrast curves and ATMO/BEX evolutionary models, as well as calculate the expected minimum planet mass and maximum semi-major axis assuming a single planet actively sculpting the disk inner edge. We find that such a planet should have been detected in our observations with a high probability. Thus, the non-detection of any planet candidates within the disk inner radius suggests that either the planet atmospheric evolutionary models assumed are incorrect, multiple smaller planets are sculpting the disk rather than a single giant planet, planets are migrating inwards over time, the disk inner-edge is not related to planets, or a combination of two or more.

In summary, the NIRCam instrument is an amazing tool for the discovery and analysis of M-dwarf debris disks, both due to its wavelength range and sensitivity. This has led to new insights that will be important to better understanding the architectures and evolution of such systems compared to those around other spectral-types.

\begin{acknowledgements}
The authors wish to thank the anonymous referee for helpful suggestions that improved this manuscript. This work is based on observations with the NASA/ESA/CSA JWST, obtained at the Space Telescope Science Institute, which is operated by AURA, Inc., under NASA contract NAS 5-03127. These observations are associated with the JWST program 4050 (PI: A.Carter). The JWST data presented in this article were obtained from the Mikulski Archive for Space Telescopes (MAST) at the Space Telescope Science Institute. The specific observations analyzed can be accessed via \dataset[doi: 10.17909/haeg-5t51]{https://doi.org/10.17909/haeg-5t51}. Support for program 4050 was provided by NASA through a grant from the Space Telescope Science Institute, which is operated by the Association of Universities for Research in Astronomy, Inc., under NASA contract NAS 5-03127. This work benefited from the 2024 and 2025 Exoplanet Summer Program in the Other Worlds Laboratory (OWL) at the University of California, Santa Cruz, a program funded by the Heising-Simons Foundation. This material is based upon work supported by the National Science Foundation Astronomy \& Astrophysics Postdoctoral Fellowship Award No. 2401654 for author BLL. Any opinions, findings, and conclusions or recommendations expressed in this material are those of the authors(s) and do not necessarily reflect the views of the National Science Foundation. B.J.S. acknowledges funding by the UK Science and Technology Facilities Council (STFC) grant nos. ST/V000594/1 and UKRI1196.
\end{acknowledgements}

\facilities{JWST}

\software{spaceKLIP (\citealt{Kammerer22,Carter23}, \url{https://spaceklip.readthedocs.io/en/latest/index.html}),
Winnie (\citealt{Lawson22}, \url{https://github.com/kdlawson/Winnie}),
lmfit (\citealt{lmfit25}, \url{https://lmfit.github.io/lmfit-py/}),
matplotlib (\citealt{Hunter07}), 
iPython (\citealt{Perez07}), 
Astropy (\citealt{astropy13,astropy18,astropy22}),
NumPy (\citealt{Harris20}; \url{https://numpy.org}),
SciPy (\citealt{Virtanen20}; \url{http://www.scipy.org/})}

\appendix

\section{SED Photometry} \label{sec:seds}

In Section \ref{disk_sed}, we compiled photometry to build the SED for both TWA 10 and TWA 25, in order to understand why no significant IR-excess has been detected for these systems. The photometry for each filter, as well as the appropriate references, can be found below in Table \ref{tab:sed}.

\begin{table}
	\centering
	\caption{\label{tab:sed} Photometry for the TWA 10 and TWA 25 SEDs plotted in Figure \ref{fig:seds}. References: (1) \citealt{Henden16}, (2) \citealt{gaia18}, (3) \citealt{Cutri03}, (4) \citealt{Marocco21}, (5) \citealt{Wright10}, (6) \citealt{Spitzer21}, (7) \citealt{Low05}, (8) \citealt{RM13}, (9) \citealt{Hog00}, (10) \citealt{Ishihara10}, (11) \citealt{Cieza13}.}
	\begin{tabular}{cccc}
	    \hline
	    \hline
		Name & Filter & Flux (Jy) & Reference \\
		\hline
        TWA 10 & BAPASS & 5.9e-3$\pm$2.3e-4 & (1) \\
               & GAPASS & 1.1e-2$\pm$4.6e-4 & (1) \\
               & GAIA.BP & 1.5e-2$\pm$2.9e-4 & (2) \\
               & VAPASS & 2.2e-5$\pm$9.9e-4 & (1) \\
               & RAPASS & 3.8e-2$\pm$1.8e-3 & (1) \\
               & GAIA.G & 5.2e-2$\pm$5.e-4 & (2) \\
               & IAPASS & 1.3e-1$\pm$8.0e-3 & (1) \\
               & GAIA.RP & 1.2e-1$\pm$1.5e-3 & (2) \\
               & 2MASS J & 3.5e-1$\pm$8.3e-3 & (3) \\
               & 2MASS H & 4.3e-1$\pm$1.8e-2 & (3) \\
               & 2MASS Ks & 3.5e-2$\pm$9.9e-3 & (3) \\
               & WISE3 & 1.7e-1$\pm$4.2e-3 & (4) \\
               & WISE4 & 1.1e-1$\pm$2.9e-3 & (4) \\
               & WISE12 & 2.2e-2$\pm$9.6e-4 & (5) \\
               & WISE22 & 6.7e-3$\pm$8.2e-4 & (5) \\
               & MIPS24 & 5.8e-3$\pm$3.0e-5 & (6) \\
               & MIPS70 & $<$8.0e-3 & (7) \\
               & PACS70 & $<$4.1e-3 & (8) \\
               & PACS160 & $<$1.2e-2 & (8) \\
               & PACS250 & $<$2.70e-2 & (11) \\
               & PACS350 & $<$2.70e-2 & (11) \\
               & PACS500 & $<$2.25e-2 & (11) \\
        \hline
        TWA 25 & BT & 1.4e-2$\pm$3.7e-3 & (9) \\
               & BAPASS & 3.7e-2$\pm$4.7e-3 & (1) \\
               & GAPASS & 6.3e-2$\pm$8.4e-3 & (1) \\
               & GAIA.BP & 8.4e-2$\pm$1.2e-3 & (2) \\
               & VT & 8.4e-2$\pm$7.3e-3 & (9) \\
               & VAPASS & 1.3e-1$\pm$9.9e-3 & (1) \\
               & RAPASS & 2.1e-1$\pm$3.2e-2 & (1) \\
               & GAIA.G & 2.0e-1$\pm$1.9e-3 & (2) \\
               & IAPASS & 3.5e-1$\pm$6.4e-3 & (1) \\
               & GAIA.RP & 3.8e-1$\pm$4.4e-3 & (2) \\
               & 2MASS J & 8.4e-1$\pm$2.7e-2 & (3) \\
               & 2MASS H & 1.1e0$\pm$4.2e-2 & (3) \\
               & 2MASS Ks & 7.9e-1$\pm$1.6e-2 & (3) \\
               & WISE3 & 3.6e-1$\pm$9.1e-3 & (4) \\
               & WISE4 & 2.2e-1$\pm$7.0e-3 & (4) \\
               & AKARI9 & 1.1e-1$\pm$1.6e-2 & (10) \\
               & WISE12 & 4.2e-2$\pm$1.9e-3 & (5) \\
               & WISE22 & 1.3e-2$\pm$1.1e-3 & (5) \\
               & MIPS24 & 1.1e-2$\pm$2.3e-4 & (6) \\
               & MIPS 70 & 5.9e-3$\pm$4.0e-3 & (6) \\
               & PACS70 & $<$5.7e-3 & (8) \\
               & PACS100 & $<$6.3e-3 & (8) \\
               & PACS160 & $<$1.2e-2 & (8) \\
		\hline
		\hline
	\end{tabular}
\end{table}

\section{Additional TWA 25 F444W Objects} \label{twa25_bkg}
\begin{figure*}
    \centering
    \includegraphics[width=0.9\textwidth]{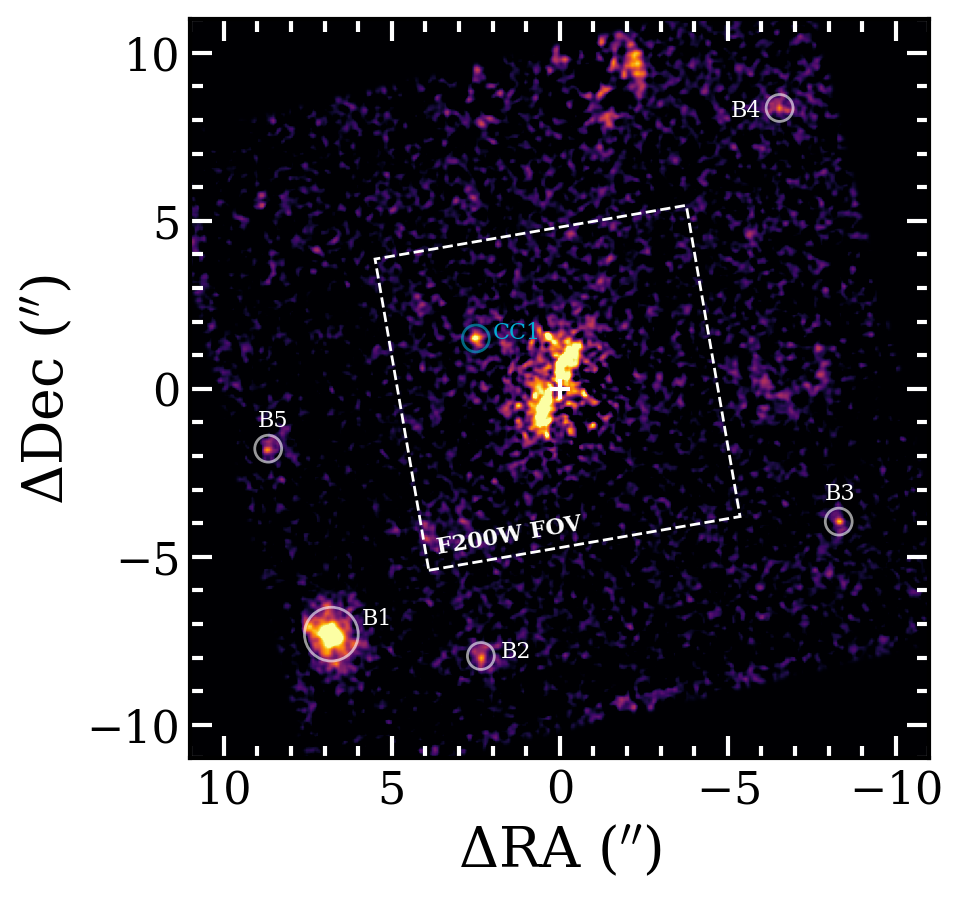}
    \caption{\label{fig:twa25_bg} TWA 25 F444W data zoomed out to show the additional sources, which are circled in whie and labeled B1-B5. The white dashed box represents the FOV of the F200W observations. Additionally, CC1 is circled in blue.} 
\end{figure*}

Although the TWA 25 system only has two candidates within both the F200W and F444W FOVs, there are 5 objects in the F444W data located outside the F200W FOV ($\gtrsim$5$''$). These objects are labeled B1-B5 in Figure \ref{fig:twa25_bg}. While B2-B5 appear point-source like, B1 is clearly extended and therefore is very likely a background galaxy. For B2-B5, we cannot make any claims on whether or not these objects are bound companions or background objects, as we cannot measure their colors, although it is less likely that these are bound companions given their distance from the star. Proper motion followup would be required to more confidently determine that these are indeed background objects. Similar to the three candidates in Section \ref{sec:comp_ext}, we use \texttt{Winnie} and \texttt{lmfit.minimize} to fit each point-source with a forward-model in order to estimate their RA/Dec and apparent magnitude. The results from this forward modeling can be found in Table \ref{tab:twa25_bg}.

\begin{table}
	\centering
	\caption{\label{tab:twa25_bg}Results from candidate model fitting with \texttt{spaceKLIP} for the TWA 25 objects outside of the F200W FOV including the measured RA and Dec, as well as apparent magnitude in the F444W filter. A negative RA is to the right of the star and a positive RA is to the left of the star.}
	\begin{tabular}{cccc}
	    \hline
	    \hline
		Candidate & RA ($''$) & Dec ($''$) & F444W (App. Mag) \\
		\hline
        B1 & 6.79$\pm$0.01 & -7.37$\pm$0.01 & 20.00$\pm$0.08 \\
		  B2 & 2.33$\pm$0.01 & -8.02$\pm$0.01 & 21.28$\pm$0.09 \\
        B3 & -8.31$\pm$0.01 & -3.97$\pm$0.01 & 21.31$\pm$0.09 \\
        B4 & -6.80$\pm$0.02 & 8.25$\pm$0.01 & 21.53$\pm$0.13 \\
        B5 & 8.69$\pm$0.01 & -1.84$\pm$0.01 &  21.54$\pm$0.12 \\
		\hline
		\hline
	\end{tabular}
\end{table}


\clearpage

\bibliography{sample631}{}

@ARTICLE{Augereau06,
       author = {{Augereau}, J.-C. and {Beust}, H.},
        title = "{On the AU Microscopii debris disk. Density profiles, grain properties, and dust dynamics}",
      journal = {\aap},
         year = 2006,
        month = sep,
       volume = {455},
       number = {3},
        pages = {987-999},
          doi = {10.1051/0004-6361:20054250},
archivePrefix = {arXiv},
       eprint = {astro-ph/0604313},
 primaryClass = {astro-ph},
       adsurl = {https://ui.adsabs.harvard.edu/abs/2006A&A...455..987A}
}

@ARTICLE{Macgregor13,
       author = {{MacGregor}, Meredith A. and {Wilner}, David J. and {Rosenfeld}, Katherine A. and {Andrews}, Sean M. and {Matthews}, Brenda and {Hughes}, A. Meredith and {Booth}, Mark and {Chiang}, Eugene and {Graham}, James R. and {Kalas}, Paul and {Kennedy}, Grant and {Sibthorpe}, Bruce},
        title = "{Millimeter Emission Structure in the First ALMA Image of the AU Mic Debris Disk}",
      journal = {\apjl},
         year = 2013,
        month = jan,
       volume = {762},
       number = {2},
          eid = {L21},
        pages = {L21},
          doi = {10.1088/2041-8205/762/2/L21},
archivePrefix = {arXiv},
       eprint = {1211.5148},
 primaryClass = {astro-ph.EP},
       adsurl = {https://ui.adsabs.harvard.edu/abs/2013ApJ...762L..21M}
}

@ARTICLE{Strubbe06,
       author = {{Strubbe}, Linda E. and {Chiang}, Eugene I.},
        title = "{Dust Dynamics, Surface Brightness Profiles, and Thermal Spectra of Debris Disks: The Case of AU Microscopii}",
      journal = {\apj},
         year = 2006,
        month = sep,
       volume = {648},
       number = {1},
        pages = {652-665},
          doi = {10.1086/505736},
archivePrefix = {arXiv},
       eprint = {astro-ph/0510527},
 primaryClass = {astro-ph},
       adsurl = {https://ui.adsabs.harvard.edu/abs/2006ApJ...648..652S}
}

@ARTICLE{Strampelli26,
       author = {{Strampelli}, G.~M.},
         year = {2026, in prep}
}

@ARTICLE{Schuppler15,
       author = {{Sch{\"u}ppler}, Ch. and {L{\"o}hne}, T. and {Krivov}, A.~V. and {Ertel}, S. and {Marshall}, J.~P. and {Wolf}, S. and {Wyatt}, M.~C. and {Augereau}, J.-C. and {Metchev}, S.~A.},
        title = "{Collisional modelling of the AU Microscopii debris disc}",
      journal = {\aap},
         year = 2015,
        month = sep,
       volume = {581},
          eid = {A97},
        pages = {A97},
          doi = {10.1051/0004-6361/201525664},
archivePrefix = {arXiv},
       eprint = {1506.04564},
 primaryClass = {astro-ph.EP},
       adsurl = {https://ui.adsabs.harvard.edu/abs/2015A&A...581A..97S}
}

@ARTICLE{Yelverton19,
       author = {{Yelverton}, Ben and {Kennedy}, Grant M. and {Su}, Kate Y.~L. and {Wyatt}, Mark C.},
        title = "{A statistically significant lack of debris discs in medium separation binary systems}",
      journal = {\mnras},
         year = 2019,
        month = sep,
       volume = {488},
       number = {3},
        pages = {3588-3606},
          doi = {10.1093/mnras/stz1927},
archivePrefix = {arXiv},
       eprint = {1907.04800},
 primaryClass = {astro-ph.EP},
       adsurl = {https://ui.adsabs.harvard.edu/abs/2019MNRAS.488.3588Y}
}

@ARTICLE{Wyatt99,
       author = {{Wyatt}, M.~C. and {Dermott}, S.~F. and {Telesco}, C.~M. and {Fisher}, R.~S. and {Grogan}, K. and {Holmes}, E.~K. and {Pi{\~n}a}, R.~K.},
        title = "{How Observations of Circumstellar Disk Asymmetries Can Reveal Hidden Planets: Pericenter Glow and Its Application to the HR 4796 Disk}",
      journal = {\apj},
         year = 1999,
        month = dec,
       volume = {527},
       number = {2},
        pages = {918-944},
          doi = {10.1086/308093},
archivePrefix = {arXiv},
       eprint = {astro-ph/9908267},
 primaryClass = {astro-ph},
       adsurl = {https://ui.adsabs.harvard.edu/abs/1999ApJ...527..918W}
}

@ARTICLE{Cieza13,
       author = {{Cieza}, Lucas A. and {Olofsson}, Johan and {Harvey}, Paul M. and {Evans}, II, Neal J. and {Najita}, Joan and {Henning}, Thomas and {Mer{\'\i}n}, Bruno and {Liebhart}, Armin and {G{\"u}del}, Manuel and {Augereau}, Jean-Charles and {Pinte}, Christophe},
        title = "{The Herschel DIGIT Survey of Weak-line T Tauri Stars: Implications for Disk Evolution and Dissipation}",
      journal = {\apj},
         year = 2013,
        month = jan,
       volume = {762},
       number = {2},
          eid = {100},
        pages = {100},
          doi = {10.1088/0004-637X/762/2/100},
archivePrefix = {arXiv},
       eprint = {1211.4510},
 primaryClass = {astro-ph.GA},
       adsurl = {https://ui.adsabs.harvard.edu/abs/2013ApJ...762..100C}
}

@ARTICLE{astropy22,
       author = {{Astropy Collaboration} and {Price-Whelan}, Adrian M. and {Lim}, Pey Lian and {Earl}, Nicholas and {Starkman}, Nathaniel and {Bradley}, Larry and {Shupe}, David L. and {Patil}, Aarya A. and {Corrales}, Lia and {Brasseur}, C.~E. and {N{\"o}the}, Maximilian and {Donath}, Axel and {Tollerud}, Erik and {Morris}, Brett M. and {Ginsburg}, Adam and {Vaher}, Eero and {Weaver}, Benjamin A. and {Tocknell}, James and {Jamieson}, William and {van Kerkwijk}, Marten H. and {Robitaille}, Thomas P. and {Merry}, Bruce and {Bachetti}, Matteo and {G{\"u}nther}, H. Moritz and {Aldcroft}, Thomas L. and {Alvarado-Montes}, Jaime A. and {Archibald}, Anne M. and {B{\'o}di}, Attila and {Bapat}, Shreyas and {Barentsen}, Geert and {Baz{\'a}n}, Juanjo and {Biswas}, Manish and {Boquien}, M{\'e}d{\'e}ric and {Burke}, D.~J. and {Cara}, Daria and {Cara}, Mihai and {Conroy}, Kyle E. and {Conseil}, Simon and {Craig}, Matthew W. and {Cross}, Robert M. and {Cruz}, Kelle L. and {D'Eugenio}, Francesco and {Dencheva}, Nadia and {Devillepoix}, Hadrien A.~R. and {Dietrich}, J{\"o}rg P. and {Eigenbrot}, Arthur Davis and {Erben}, Thomas and {Ferreira}, Leonardo and {Foreman-Mackey}, Daniel and {Fox}, Ryan and {Freij}, Nabil and {Garg}, Suyog and {Geda}, Robel and {Glattly}, Lauren and {Gondhalekar}, Yash and {Gordon}, Karl D. and {Grant}, David and {Greenfield}, Perry and {Groener}, Austen M. and {Guest}, Steve and {Gurovich}, Sebastian and {Handberg}, Rasmus and {Hart}, Akeem and {Hatfield-Dodds}, Zac and {Homeier}, Derek and {Hosseinzadeh}, Griffin and {Jenness}, Tim and {Jones}, Craig K. and {Joseph}, Prajwel and {Kalmbach}, J. Bryce and {Karamehmetoglu}, Emir and {Ka{\l}uszy{\'n}ski}, Miko{\l}aj and {Kelley}, Michael S.~P. and {Kern}, Nicholas and {Kerzendorf}, Wolfgang E. and {Koch}, Eric W. and {Kulumani}, Shankar and {Lee}, Antony and {Ly}, Chun and {Ma}, Zhiyuan and {MacBride}, Conor and {Maljaars}, Jakob M. and {Muna}, Demitri and {Murphy}, N.~A. and {Norman}, Henrik and {O'Steen}, Richard and {Oman}, Kyle A. and {Pacifici}, Camilla and {Pascual}, Sergio and {Pascual-Granado}, J. and {Patil}, Rohit R. and {Perren}, Gabriel I. and {Pickering}, Timothy E. and {Rastogi}, Tanuj and {Roulston}, Benjamin R. and {Ryan}, Daniel F. and {Rykoff}, Eli S. and {Sabater}, Jose and {Sakurikar}, Parikshit and {Salgado}, Jes{\'u}s and {Sanghi}, Aniket and {Saunders}, Nicholas and {Savchenko}, Volodymyr and {Schwardt}, Ludwig and {Seifert-Eckert}, Michael and {Shih}, Albert Y. and {Jain}, Anany Shrey and {Shukla}, Gyanendra and {Sick}, Jonathan and {Simpson}, Chris and {Singanamalla}, Sudheesh and {Singer}, Leo P. and {Singhal}, Jaladh and {Sinha}, Manodeep and {Sip{\H{o}}cz}, Brigitta M. and {Spitler}, Lee R. and {Stansby}, David and {Streicher}, Ole and {{\v{S}}umak}, Jani and {Swinbank}, John D. and {Taranu}, Dan S. and {Tewary}, Nikita and {Tremblay}, Grant R. and {de Val-Borro}, Miguel and {Van Kooten}, Samuel J. and {Vasovi{\'c}}, Zlatan and {Verma}, Shresth and {de Miranda Cardoso}, Jos{\'e} Vin{\'\i}cius and {Williams}, Peter K.~G. and {Wilson}, Tom J. and {Winkel}, Benjamin and {Wood-Vasey}, W.~M. and {Xue}, Rui and {Yoachim}, Peter and {Zhang}, Chen and {Zonca}, Andrea and {Astropy Project Contributors}},
        title = "{The Astropy Project: Sustaining and Growing a Community-oriented Open-source Project and the Latest Major Release (v5.0) of the Core Package}",
      journal = {\apj},
         year = 2022,
        month = aug,
       volume = {935},
       number = {2},
          eid = {167},
        pages = {167},
          doi = {10.3847/1538-4357/ac7c74},
archivePrefix = {arXiv},
       eprint = {2206.14220},
 primaryClass = {astro-ph.IM},
       adsurl = {https://ui.adsabs.harvard.edu/abs/2022ApJ...935..167A}
}

@ARTICLE{astropy13,
       author = {{Astropy Collaboration} and {Robitaille}, Thomas P. and {Tollerud}, Erik J. and {Greenfield}, Perry and {Droettboom}, Michael and {Bray}, Erik and {Aldcroft}, Tom and {Davis}, Matt and {Ginsburg}, Adam and {Price-Whelan}, Adrian M. and {Kerzendorf}, Wolfgang E. and {Conley}, Alexander and {Crighton}, Neil and {Barbary}, Kyle and {Muna}, Demitri and {Ferguson}, Henry and {Grollier}, Fr{\'e}d{\'e}ric and {Parikh}, Madhura M. and {Nair}, Prasanth H. and {Unther}, Hans M. and {Deil}, Christoph and {Woillez}, Julien and {Conseil}, Simon and {Kramer}, Roban and {Turner}, James E.~H. and {Singer}, Leo and {Fox}, Ryan and {Weaver}, Benjamin A. and {Zabalza}, Victor and {Edwards}, Zachary I. and {Azalee Bostroem}, K. and {Burke}, D.~J. and {Casey}, Andrew R. and {Crawford}, Steven M. and {Dencheva}, Nadia and {Ely}, Justin and {Jenness}, Tim and {Labrie}, Kathleen and {Lim}, Pey Lian and {Pierfederici}, Francesco and {Pontzen}, Andrew and {Ptak}, Andy and {Refsdal}, Brian and {Servillat}, Mathieu and {Streicher}, Ole},
        title = "{Astropy: A community Python package for astronomy}",
      journal = {\aap},
         year = 2013,
        month = oct,
       volume = {558},
          eid = {A33},
        pages = {A33},
          doi = {10.1051/0004-6361/201322068},
archivePrefix = {arXiv},
       eprint = {1307.6212},
 primaryClass = {astro-ph.IM},
       adsurl = {https://ui.adsabs.harvard.edu/abs/2013A&A...558A..33A}
}

@ARTICLE{Morbidelli22,
       author = {{Morbidelli}, A. and {Bailli{\'e}}, K. and {Batygin}, K. and {Charnoz}, S. and {Guillot}, T. and {Rubie}, D.~C. and {Kleine}, T.},
        title = "{Contemporary formation of early Solar System planetesimals at two distinct radial locations}",
      journal = {Nature Astronomy},
         year = 2022,
        month = jan,
       volume = {6},
        pages = {72-79},
          doi = {10.1038/s41550-021-01517-7},
archivePrefix = {arXiv},
       eprint = {2112.15413},
 primaryClass = {astro-ph.EP},
       adsurl = {https://ui.adsabs.harvard.edu/abs/2022NatAs...6...72M}
}

@ARTICLE{Izidoro22,
       author = {{Izidoro}, Andre and {Dasgupta}, Rajdeep and {Raymond}, Sean N. and {Deienno}, Rogerio and {Bitsch}, Bertram and {Isella}, Andrea},
        title = "{Planetesimal rings as the cause of the Solar System's planetary architecture}",
      journal = {Nature Astronomy},
         year = 2022,
        month = mar,
       volume = {6},
        pages = {357-366},
          doi = {10.1038/s41550-021-01557-z},
archivePrefix = {arXiv},
       eprint = {2112.15558},
 primaryClass = {astro-ph.EP},
       adsurl = {https://ui.adsabs.harvard.edu/abs/2022NatAs...6..357I}
}

@ARTICLE{Schoonenberg18,
       author = {{Schoonenberg}, Djoeke and {Ormel}, Chris W. and {Krijt}, Sebastiaan},
        title = "{A Lagrangian model for dust evolution in protoplanetary disks: formation of wet and dry planetesimals at different stellar masses}",
      journal = {\aap},
         year = 2018,
        month = dec,
       volume = {620},
          eid = {A134},
        pages = {A134},
          doi = {10.1051/0004-6361/201834047},
archivePrefix = {arXiv},
       eprint = {1810.02370},
 primaryClass = {astro-ph.EP},
       adsurl = {https://ui.adsabs.harvard.edu/abs/2018A&A...620A.134S}
}

@ARTICLE{Schoonenberg17,
       author = {{Schoonenberg}, Djoeke and {Ormel}, Chris W.},
        title = "{Planetesimal formation near the snowline: in or out?}",
      journal = {\aap},
         year = 2017,
        month = jun,
       volume = {602},
          eid = {A21},
        pages = {A21},
          doi = {10.1051/0004-6361/201630013},
archivePrefix = {arXiv},
       eprint = {1702.02151},
 primaryClass = {astro-ph.EP},
       adsurl = {https://ui.adsabs.harvard.edu/abs/2017A&A...602A..21S}
}

@ARTICLE{drazkowska17,
       author = {{Dra{\.z}kowska}, J. and {Alibert}, Y.},
        title = "{Planetesimal formation starts at the snow line}",
      journal = {\aap},
         year = 2017,
        month = dec,
       volume = {608},
          eid = {A92},
        pages = {A92},
          doi = {10.1051/0004-6361/201731491},
archivePrefix = {arXiv},
       eprint = {1710.00009},
 primaryClass = {astro-ph.EP},
       adsurl = {https://ui.adsabs.harvard.edu/abs/2017A&A...608A..92D}
}

@ARTICLE{Ida16,
       author = {{Ida}, S. and {Guillot}, T.},
        title = "{Formation of dust-rich planetesimals from sublimated pebbles inside of the snow line}",
      journal = {\aap},
         year = 2016,
        month = nov,
       volume = {596},
          eid = {L3},
        pages = {L3},
          doi = {10.1051/0004-6361/201629680},
archivePrefix = {arXiv},
       eprint = {1610.09643},
 primaryClass = {astro-ph.EP},
       adsurl = {https://ui.adsabs.harvard.edu/abs/2016A&A...596L...3I}
}

@ARTICLE{Carrera17,
       author = {{Carrera}, Daniel and {Gorti}, Uma and {Johansen}, Anders and {Davies}, Melvyn B.},
        title = "{Planetesimal Formation by the Streaming Instability in a Photoevaporating Disk}",
      journal = {\apj},
         year = 2017,
        month = apr,
       volume = {839},
       number = {1},
          eid = {16},
        pages = {16},
          doi = {10.3847/1538-4357/aa6932},
archivePrefix = {arXiv},
       eprint = {1703.07895},
 primaryClass = {astro-ph.EP},
       adsurl = {https://ui.adsabs.harvard.edu/abs/2017ApJ...839...16C}
}

@ARTICLE{Girardi05,
       author = {{Girardi}, L. and {Groenewegen}, M.~A.~T. and {Hatziminaoglou}, E. and {da Costa}, L.},
        title = "{Star counts in the Galaxy. Simulating from very deep to very shallow photometric surveys with the TRILEGAL code}",
      journal = {\aap},
         year = 2005,
        month = jun,
       volume = {436},
       number = {3},
        pages = {895-915},
          doi = {10.1051/0004-6361:20042352},
archivePrefix = {arXiv},
       eprint = {astro-ph/0504047},
 primaryClass = {astro-ph},
       adsurl = {https://ui.adsabs.harvard.edu/abs/2005A&A...436..895G}
}

@ARTICLE{Rubin25,
       author = {{Bowens-Rubin}, Rachel and {Mang}, James and {Limbach}, Mary Anne and {Carter}, Aarynn L. and {Stevenson}, Kevin B. and {Wagner}, Kevin and {Strampelli}, Giovanni and {Morley}, Caroline V. and {Kennedy}, Grant and {Matthews}, Elisabeth and {Vanderburg}, Andrew and {Salama}, Ma{\"\i}ssa},
        title = "{NIRCam Yells at Cloud: JWST MIRI Imaging Can Directly Detect Exoplanets of the Same Temperature, Mass, Age, and Orbital Separation as Saturn and Jupiter}",
      journal = {\apjl},
         year = 2025,
        month = jun,
       volume = {986},
       number = {2},
          eid = {L26},
        pages = {L26},
          doi = {10.3847/2041-8213/addbde},
archivePrefix = {arXiv},
       eprint = {2505.15995},
 primaryClass = {astro-ph.EP},
       adsurl = {https://ui.adsabs.harvard.edu/abs/2025ApJ...986L..26B}
}

@ARTICLE{Mang24,
       author = {{Mang}, James and {Morley}, Caroline V. and {Robinson}, Tyler D. and {Gao}, Peter},
        title = "{Microphysical Prescriptions for Parameterized Water Cloud Formation on Ultra-cool Substellar Objects}",
      journal = {\apj},
         year = 2024,
        month = oct,
       volume = {974},
       number = {2},
          eid = {190},
        pages = {190},
          doi = {10.3847/1538-4357/ad6c4c},
archivePrefix = {arXiv},
       eprint = {2408.08958},
 primaryClass = {astro-ph.EP},
       adsurl = {https://ui.adsabs.harvard.edu/abs/2024ApJ...974..190M}
}

@ARTICLE{Phillips20,
       author = {{Phillips}, M.~W. and {Tremblin}, P. and {Baraffe}, I. and {Chabrier}, G. and {Allard}, N.~F. and {Spiegelman}, F. and {Goyal}, J.~M. and {Drummond}, B. and {H{\'e}brard}, E.},
        title = "{A new set of atmosphere and evolution models for cool T-Y brown dwarfs and giant exoplanets}",
      journal = {\aap},
         year = 2020,
        month = may,
       volume = {637},
          eid = {A38},
        pages = {A38},
          doi = {10.1051/0004-6361/201937381},
archivePrefix = {arXiv},
       eprint = {2003.13717},
 primaryClass = {astro-ph.SR},
       adsurl = {https://ui.adsabs.harvard.edu/abs/2020A&A...637A..38P}
}

@ARTICLE{Rebollido24,
       author = {{Rebollido}, Isabel and {Stark}, Christopher C. and {Kammerer}, Jens and {Perrin}, Marshall D. and {Lawson}, Kellen and {Pueyo}, Laurent and {Chen}, Christine and {Hines}, Dean and {Girard}, Julien H. and {Worthen}, Kadin and {Ingerbretsen}, Carl and {Betti}, Sarah and {Clampin}, Mark and {Golimowski}, David and {Hoch}, Kielan and {Lewis}, Nikole K. and {Lu}, Cicero X. and {van der Marel}, Roeland P. and {Rickman}, Emily and {Seager}, Sara and {Soummer}, R{\'e}mi and {Valenti}, Jeff A. and {Ward-Duong}, Kimberly and {Mountain}, C. Matt},
        title = "{JWST-TST High Contrast: Asymmetries, Dust Populations, and Hints of a Collision in the {\ensuremath{\beta}} Pictoris Disk with NIRCam and MIRI}",
      journal = {\aj},
         year = 2024,
        month = feb,
       volume = {167},
       number = {2},
          eid = {69},
        pages = {69},
          doi = {10.3847/1538-3881/ad1759},
archivePrefix = {arXiv},
       eprint = {2401.05271},
 primaryClass = {astro-ph.EP},
       adsurl = {https://ui.adsabs.harvard.edu/abs/2024AJ....167...69R}
}

@ARTICLE{Dohnanyi69,
       author = {{Dohnanyi}, J.~S.},
        title = "{Collisional Model of Asteroids and Their Debris}",
      journal = {JGR},
         year = 1969,
        month = may,
       volume = {74},
        pages = {2531-2554},
          doi = {10.1029/JB074i010p02531},
       adsurl = {https://ui.adsabs.harvard.edu/abs/1969JGR....74.2531D}
}

@ARTICLE{Pinte06,
       author = {{Pinte}, C. and {M{\'e}nard}, F. and {Duch{\^e}ne}, G. and {Bastien}, P.},
        title = "{Monte Carlo radiative transfer in protoplanetary disks}",
      journal = {\aap},
         year = 2006,
        month = dec,
       volume = {459},
       number = {3},
        pages = {797-804},
          doi = {10.1051/0004-6361:20053275},
archivePrefix = {arXiv},
       eprint = {astro-ph/0606550},
 primaryClass = {astro-ph},
       adsurl = {https://ui.adsabs.harvard.edu/abs/2006A&A...459..797P}
}

@ARTICLE{Pawellek19,
       author = {{Pawellek}, Nicole and {Mo{\'o}r}, Attila and {Pascucci}, Ilaria and {Krivov}, Alexander V.},
        title = "{Dust spreading in debris discs: do small grains cling on to their birth environment?}",
      journal = {\mnras},
         year = 2019,
        month = aug,
       volume = {487},
       number = {4},
        pages = {5874-5888},
          doi = {10.1093/mnras/stz1682},
archivePrefix = {arXiv},
       eprint = {1906.06953},
 primaryClass = {astro-ph.EP},
       adsurl = {https://ui.adsabs.harvard.edu/abs/2019MNRAS.487.5874P}
}

@ARTICLE{Panilla22,
       author = {{Pinilla}, Paola},
        title = "{First steps of planet formation around very low mass stars and brown dwarfs}",
      journal = {European Physical Journal Plus},
         year = 2022,
        month = nov,
       volume = {137},
       number = {11},
          eid = {1206},
        pages = {1206},
          doi = {10.1140/epjp/s13360-022-03384-1},
archivePrefix = {arXiv},
       eprint = {2210.06560},
 primaryClass = {astro-ph.EP},
       adsurl = {https://ui.adsabs.harvard.edu/abs/2022EPJP..137.1206P}
}

@ARTICLE{Kurtovic21,
       author = {{Kurtovic}, N.~T. and {Pinilla}, P. and {Long}, F. and {Benisty}, M. and {Manara}, C.~F. and {Natta}, A. and {Pascucci}, I. and {Ricci}, L. and {Scholz}, A. and {Testi}, L.},
        title = "{Size and structures of disks around very low mass stars in the Taurus star-forming region}",
      journal = {\aap},
         year = 2021,
        month = jan,
       volume = {645},
          eid = {A139},
        pages = {A139},
          doi = {10.1051/0004-6361/202038983},
archivePrefix = {arXiv},
       eprint = {2012.02225},
 primaryClass = {astro-ph.EP},
       adsurl = {https://ui.adsabs.harvard.edu/abs/2021A&A...645A.139K}
}

@ARTICLE{Andrews13,
       author = {{Andrews}, Sean M. and {Rosenfeld}, Katherine A. and {Kraus}, Adam L. and {Wilner}, David J.},
        title = "{The Mass Dependence between Protoplanetary Disks and their Stellar Hosts}",
      journal = {\apj},
         year = 2013,
        month = jul,
       volume = {771},
       number = {2},
          eid = {129},
        pages = {129},
          doi = {10.1088/0004-637X/771/2/129},
archivePrefix = {arXiv},
       eprint = {1305.5262},
 primaryClass = {astro-ph.SR},
       adsurl = {https://ui.adsabs.harvard.edu/abs/2013ApJ...771..129A}
}

@ARTICLE{Alvarado25,
       author = {{Guerra-Alvarado}, Osmar M. and {van der Marel}, Nienke and {Williams}, Jonathan P. and {Pinilla}, Paola and {Mulders}, Gijs D. and {Lambrechts}, Michiel and {Sanchez}, Mariana},
        title = "{A high-resolution survey of protoplanetary disks in Lupus and the nature of compact disks}",
      journal = {\aap},
         year = 2025,
        month = apr,
       volume = {696},
          eid = {A232},
        pages = {A232},
          doi = {10.1051/0004-6361/202453338},
archivePrefix = {arXiv},
       eprint = {2503.19504},
 primaryClass = {astro-ph.EP},
       adsurl = {https://ui.adsabs.harvard.edu/abs/2025A&A...696A.232G}
}

@ARTICLE{Matra18,
       author = {{Matr{\`a}}, L. and {Marino}, S. and {Kennedy}, G.~M. and {Wyatt}, M.~C. and {{\"O}berg}, K.~I. and {Wilner}, D.~J.},
        title = "{An Empirical Planetesimal Belt Radius-Stellar Luminosity Relation}",
      journal = {\apj},
         year = 2018,
        month = may,
       volume = {859},
       number = {1},
          eid = {72},
        pages = {72},
          doi = {10.3847/1538-4357/aabcc4},
archivePrefix = {arXiv},
       eprint = {1804.01094},
 primaryClass = {astro-ph.EP},
       adsurl = {https://ui.adsabs.harvard.edu/abs/2018ApJ...859...72M}
}

@ARTICLE{Lawson24,
       author = {{Lawson}, Kellen and {Schlieder}, Joshua E. and {Leisenring}, Jarron M. and {Bogat}, Ell and {Beichman}, Charles A. and {Bryden}, Geoffrey and {G{\'a}sp{\'a}r}, Andr{\'a}s and {Groff}, Tyler D. and {McElwain}, Michael W. and {Meyer}, Michael R. and {Barclay}, Thomas and {Calissendorff}, Per and {De Furio}, Matthew and {Li}, Yiting and {Rieke}, Marcia J. and {Ygouf}, Marie and {Greene}, Thomas P. and {Girard}, Julien H. and {Gennaro}, Mario and {Kammerer}, Jens and {Rest}, Armin and {Roellig}, Thomas L. and {Sunnquist}, Ben},
        title = "{JWST/NIRCam Detection of the Fomalhaut C Debris Disk in Scattered Light}",
      journal = {\apjl},
         year = 2024,
        month = may,
       volume = {967},
       number = {1},
          eid = {L8},
        pages = {L8},
          doi = {10.3847/2041-8213/ad4496},
archivePrefix = {arXiv},
       eprint = {2405.00573},
 primaryClass = {astro-ph.EP},
       adsurl = {https://ui.adsabs.harvard.edu/abs/2024ApJ...967L...8L}
}

@ARTICLE{Matra25,
       author = {{Matr{\`a}}, L. and {Marino}, S. and {Wilner}, D.~J. and {Kennedy}, G.~M. and {Booth}, M. and {Krivov}, A.~V. and {Williams}, J.~P. and {Hughes}, A.~M. and {del Burgo}, C. and {Carpenter}, J. and {Davies}, C.~L. and {Ertel}, S. and {Kral}, Q. and {Lestrade}, J.-F. and {Marshall}, J.~P. and {Milli}, J. and {{\"O}berg}, K.~I. and {Pawellek}, N. and {Sepulveda}, A.~G. and {Wyatt}, M.~C. and {Matthews}, B.~C. and {MacGregor}, M.},
        title = "{REsolved ALMA and SMA Observations of Nearby Stars (REASONS): A population of 74 resolved planetesimal belts at millimetre wavelengths}",
      journal = {\aap},
         year = 2025,
        month = jan,
       volume = {693},
          eid = {A151},
        pages = {A151},
          doi = {10.1051/0004-6361/202451397},
archivePrefix = {arXiv},
       eprint = {2501.09058},
 primaryClass = {astro-ph.EP},
       adsurl = {https://ui.adsabs.harvard.edu/abs/2025A&A...693A.151M}
}

@ARTICLE{emcee13,
       author = {{Foreman-Mackey}, Daniel and {Hogg}, David W. and {Lang}, Dustin and {Goodman}, Jonathan},
        title = "{emcee: The MCMC Hammer}",
      journal = {\pasp},
         year = 2013,
        month = mar,
       volume = {125},
       number = {925},
        pages = {306},
          doi = {10.1086/670067},
archivePrefix = {arXiv},
       eprint = {1202.3665},
 primaryClass = {astro-ph.IM},
       adsurl = {https://ui.adsabs.harvard.edu/abs/2013PASP..125..306F}
}

@article{Powell65,
    author = {Powell, M. J. D.},
    title = {An efficient method for finding the minimum of a function of several variables without calculating derivatives},
    journal = {The Computer Journal},
    volume = {7},
    number = {2},
    pages = {155-162},
    year = {1964},
    month = {01},
    issn = {0010-4620},
    doi = {10.1093/comjnl/7.2.155},
    url = {https://doi.org/10.1093/comjnl/7.2.155},
    eprint = {https://academic.oup.com/comjnl/article-pdf/7/2/155/959784/070155.pdf},
}

@ARTICLE{Lawson23,
       author = {{Lawson}, Kellen and {Schlieder}, Joshua E. and {Leisenring}, Jarron M. and {Bogat}, Ell and {Beichman}, Charles A. and {Bryden}, Geoffrey and {G{\'a}sp{\'a}r}, Andr{\'a}s and {Groff}, Tyler D. and {McElwain}, Michael W. and {Meyer}, Michael R. and {Barclay}, Thomas and {Calissendorff}, Per and {De Furio}, Matthew and {Ygouf}, Marie and {Boccaletti}, Anthony and {Greene}, Thomas P. and {Krist}, John and {Plavchan}, Peter and {Rieke}, Marcia J. and {Roellig}, Thomas L. and {Stansberry}, John and {Wisniewski}, John P. and {Young}, Erick T.},
        title = "{JWST/NIRCam Coronagraphy of the Young Planet-hosting Debris Disk AU Microscopii}",
      journal = {\aj},
         year = 2023,
        month = oct,
       volume = {166},
       number = {4},
          eid = {150},
        pages = {150},
          doi = {10.3847/1538-3881/aced08},
archivePrefix = {arXiv},
       eprint = {2308.02486},
 primaryClass = {astro-ph.EP},
       adsurl = {https://ui.adsabs.harvard.edu/abs/2023AJ....166..150L}
}

@ARTICLE{Herczeg14,
       author = {{Herczeg}, Gregory J. and {Hillenbrand}, Lynne A.},
        title = "{An Optical Spectroscopic Study of T Tauri Stars. I. Photospheric Properties}",
      journal = {\apj},
         year = 2014,
        month = may,
       volume = {786},
       number = {2},
          eid = {97},
        pages = {97},
          doi = {10.1088/0004-637X/786/2/97},
archivePrefix = {arXiv},
       eprint = {1403.1675},
 primaryClass = {astro-ph.SR},
       adsurl = {https://ui.adsabs.harvard.edu/abs/2014ApJ...786...97H}
}

@ARTICLE{Torres06,
       author = {{Torres}, C.~A.~O. and {Quast}, G.~R. and {da Silva}, L. and {de La Reza}, R. and {Melo}, C.~H.~F. and {Sterzik}, M.},
        title = "{Search for associations containing young stars (SACY). I. Sample and searching method}",
      journal = {\aap},
         year = 2006,
        month = dec,
       volume = {460},
       number = {3},
        pages = {695-708},
          doi = {10.1051/0004-6361:20065602},
archivePrefix = {arXiv},
       eprint = {astro-ph/0609258},
 primaryClass = {astro-ph},
       adsurl = {https://ui.adsabs.harvard.edu/abs/2006A&A...460..695T}
}

@ARTICLE{Miret25,
       author = {{Miret-Roig}, N. and {Alves}, J. and {Ratzenb{\"o}ck}, S. and {Galli}, P.~A.~B. and {Bouy}, H. and {Figueras}, F. and {Gro{\ss}schedl}, J. and {Meingast}, S. and {Posch}, L. and {Rottensteiner}, A. and {Swiggum}, C. and {Wagner}, N.},
        title = "{The TW Hydrae Association is a cluster chain of Sco-Cen}",
      journal = {\aap},
         year = 2025,
        month = feb,
       volume = {694},
          eid = {A60},
        pages = {A60},
          doi = {10.1051/0004-6361/202452558},
archivePrefix = {arXiv},
       eprint = {2501.11716},
 primaryClass = {astro-ph.SR},
       adsurl = {https://ui.adsabs.harvard.edu/abs/2025A&A...694A..60M}
}

@ARTICLE{Luppe20,
       author = {{Luppe}, Patricia and {Krivov}, Alexander V. and {Booth}, Mark and {Lestrade}, Jean-Fran{\c{c}}ois},
        title = "{Observability of dusty debris discs around M-stars}",
      journal = {\mnras},
         year = 2020,
        month = dec,
       volume = {499},
       number = {3},
        pages = {3932-3942},
          doi = {10.1093/mnras/staa2608},
archivePrefix = {arXiv},
       eprint = {1910.13142},
 primaryClass = {astro-ph.EP},
       adsurl = {https://ui.adsabs.harvard.edu/abs/2020MNRAS.499.3932L}
}

@ARTICLE{Cronin23,
       author = {{Cronin-Coltsmann}, Patrick F. and {Kennedy}, Grant M. and {Kral}, Quentin and {Lestrade}, Jean-Fran{\c{c}}ois and {Marino}, Sebastian and {Matr{\`a}}, Luca and {Wyatt}, Mark C.},
        title = "{An ALMA Survey of M-dwarfs in the Beta Pictoris Moving Group with two new debris disc detections}",
      journal = {\mnras},
         year = 2023,
        month = dec,
       volume = {526},
       number = {4},
        pages = {5401-5417},
          doi = {10.1093/mnras/stad3083},
archivePrefix = {arXiv},
       eprint = {2310.15255},
 primaryClass = {astro-ph.SR},
       adsurl = {https://ui.adsabs.harvard.edu/abs/2023MNRAS.526.5401C}
}

@ARTICLE{Patel14,
       author = {{Patel}, Rahul I. and {Metchev}, Stanimir A. and {Heinze}, Aren},
        title = "{A Sensitive Identification of Warm Debris Disks in the Solar Neighborhood through Precise Calibration of Saturated WISE Photometry}",
      journal = {\apjs},
         year = 2014,
        month = may,
       volume = {212},
       number = {1},
          eid = {10},
        pages = {10},
          doi = {10.1088/0067-0049/212/1/10},
archivePrefix = {arXiv},
       eprint = {1403.3435},
 primaryClass = {astro-ph.SR},
       adsurl = {https://ui.adsabs.harvard.edu/abs/2014ApJS..212...10P}
}

@ARTICLE{Ida00,
       author = {{Ida}, Shigeru and {Larwood}, John and {Burkert}, Andreas},
        title = "{Evidence for Early Stellar Encounters in the Orbital Distribution of Edgeworth-Kuiper Belt Objects}",
      journal = {\apj},
         year = 2000,
        month = jan,
       volume = {528},
       number = {1},
        pages = {351-356},
          doi = {10.1086/308179},
archivePrefix = {arXiv},
       eprint = {astro-ph/9907217},
 primaryClass = {astro-ph},
       adsurl = {https://ui.adsabs.harvard.edu/abs/2000ApJ...528..351I}
}

@ARTICLE{Kenyon01,
       author = {{Kenyon}, Scott J. and {Bromley}, Benjamin C.},
        title = "{Gravitational Stirring in Planetary Debris Disks}",
      journal = {\aj},
         year = 2001,
        month = jan,
       volume = {121},
       number = {1},
        pages = {538-551},
          doi = {10.1086/318019},
archivePrefix = {arXiv},
       eprint = {astro-ph/0009185},
 primaryClass = {astro-ph},
       adsurl = {https://ui.adsabs.harvard.edu/abs/2001AJ....121..538K}
}

@ARTICLE{Mustill09,
       author = {{Mustill}, Alexander J. and {Wyatt}, Mark C.},
        title = "{Debris disc stirring by secular perturbations from giant planets}",
      journal = {\mnras},
         year = 2009,
        month = nov,
       volume = {399},
       number = {3},
        pages = {1403-1414},
          doi = {10.1111/j.1365-2966.2009.15360.x},
archivePrefix = {arXiv},
       eprint = {0907.1389},
 primaryClass = {astro-ph.EP},
       adsurl = {https://ui.adsabs.harvard.edu/abs/2009MNRAS.399.1403M}
}

@ARTICLE{Hughes18,
       author = {{Hughes}, A. Meredith and {Duch{\^e}ne}, Gaspard and {Matthews}, Brenda C.},
        title = "{Debris Disks: Structure, Composition, and Variability}",
      journal = {\araa},
         year = 2018,
        month = sep,
       volume = {56},
        pages = {541-591},
          doi = {10.1146/annurev-astro-081817-052035},
archivePrefix = {arXiv},
       eprint = {1802.04313},
 primaryClass = {astro-ph.EP},
       adsurl = {https://ui.adsabs.harvard.edu/abs/2018ARA&A..56..541H}
}

@INPROCEEDINGS{Matthews14,
       author = {{Matthews}, B.~C. and {Krivov}, A.~V. and {Wyatt}, M.~C. and {Bryden}, G. and {Eiroa}, C.},
        title = "{Observations, Modeling, and Theory of Debris Disks}",
    booktitle = {Protostars and Planets VI},
         year = 2014,
       editor = {{Beuther}, Henrik and {Klessen}, Ralf S. and {Dullemond}, Cornelis P. and {Henning}, Thomas},
        month = jan,
        pages = {521-544},
          doi = {10.2458/azu_uapress_9780816531240-ch023},
archivePrefix = {arXiv},
       eprint = {1401.0743},
 primaryClass = {astro-ph.SR},
       adsurl = {https://ui.adsabs.harvard.edu/abs/2014prpl.conf..521M}
}

@ARTICLE{Wyatt08,
       author = {{Wyatt}, M.~C.},
        title = "{Evolution of debris disks.}",
      journal = {\araa},
         year = 2008,
        month = sep,
       volume = {46},
        pages = {339-383},
          doi = {10.1146/annurev.astro.45.051806.110525},
       adsurl = {https://ui.adsabs.harvard.edu/abs/2008ARA&A..46..339W}
}

@ARTICLE{Ishihara10,
       author = {{Ishihara}, D. and {Onaka}, T. and {Kataza}, H. and {Salama}, A. and {Alfageme}, C. and {Cassatella}, A. and {Cox}, N. and {Garc{\'\i}a-Lario}, P. and {Stephenson}, C. and {Cohen}, M. and {Fujishiro}, N. and {Fujiwara}, H. and {Hasegawa}, S. and {Ita}, Y. and {Kim}, W. and {Matsuhara}, H. and {Murakami}, H. and {M{\"u}ller}, T.~G. and {Nakagawa}, T. and {Ohyama}, Y. and {Oyabu}, S. and {Pyo}, J. and {Sakon}, I. and {Shibai}, H. and {Takita}, S. and {Tanab{\'e}}, T. and {Uemizu}, K. and {Ueno}, M. and {Usui}, F. and {Wada}, T. and {Watarai}, H. and {Yamamura}, I. and {Yamauchi}, C.},
        title = "{The AKARI/IRC mid-infrared all-sky survey}",
      journal = {\aap},
         year = 2010,
        month = may,
       volume = {514},
          eid = {A1},
        pages = {A1},
          doi = {10.1051/0004-6361/200913811},
archivePrefix = {arXiv},
       eprint = {1003.0270},
 primaryClass = {astro-ph.IM},
       adsurl = {https://ui.adsabs.harvard.edu/abs/2010A&A...514A...1I}
}

@MISC{CarterJWSTprop2023,
       author = {{Carter}, Aarynn and {Balmer}, William and {Biller}, Beth and {Bogat}, Ell and {Bonavita}, Mariangela and {Bowler}, Brendan and {Calissendorff}, Per and {Fontanive}, Clemence and {Franson}, Kyle and {Gagne}, Jonathan and {Girard}, Julien and {Hinkley}, Sasha and {Hoch}, Kielan K.~W. and {Kammerer}, Jens and {Kennedy}, Grant and {Leisenring}, Jarron Michael and {Macintosh}, Bruce A. and {Matthews}, Elisabeth C. and {Meyer}, Michael R. and {Millar-Blanchaer}, Maxwell Andrew and {Morley}, Caroline and {Perrin}, Marshall and {Pueyo}, Laurent and {Ray}, Shrishmoy and {Rebollido}, Isabel and {Rickman}, Emily and {Skemer}, Andrew and {Wang}, Jason J.},
        title = "{Uncharted Worlds: Towards a Legacy of Direct Imaging of Sub-Jupiter Mass Exoplanets}",
 howpublished = {JWST Proposal. Cycle 2, ID. \#4050},
         year = 2023,
        month = may,
        pages = {4050},
       adsurl = {https://ui.adsabs.harvard.edu/abs/2023jwst.prop.4050C}
}

@ARTICLE{Hog00,
       author = {{H{\o}g}, E. and {Fabricius}, C. and {Makarov}, V.~V. and {Urban}, S. and {Corbin}, T. and {Wycoff}, G. and {Bastian}, U. and {Schwekendiek}, P. and {Wicenec}, A.},
        title = "{The Tycho-2 catalogue of the 2.5 million brightest stars}",
      journal = {\aap},
         year = 2000,
        month = mar,
       volume = {355},
        pages = {L27-L30},
       adsurl = {https://ui.adsabs.harvard.edu/abs/2000A&A...355L..27H}
}

@ARTICLE{RM13,
       author = {{Riviere-Marichalar}, P. and {Pinte}, C. and {Barrado}, D. and {Thi}, W.~F. and {Eiroa}, C. and {Kamp}, I. and {Montesinos}, B. and {Donaldson}, J. and {Augereau}, J.~C. and {Hu{\'e}lamo}, N. and {Roberge}, A. and {Ardila}, D. and {Sandell}, G. and {Williams}, J.~P. and {Dent}, W.~R.~F. and {Menard}, F. and {Lillo-Box}, J. and {Duch{\^e}ne}, G.},
        title = "{Gas and dust in the TW Hydrae association as seen by the Herschel Space Observatory}",
      journal = {\aap},
         year = 2013,
        month = jul,
       volume = {555},
          eid = {A67},
        pages = {A67},
          doi = {10.1051/0004-6361/201321506},
archivePrefix = {arXiv},
       eprint = {1306.0328},
 primaryClass = {astro-ph.SR},
       adsurl = {https://ui.adsabs.harvard.edu/abs/2013A&A...555A..67R}
}

@ARTICLE{Low05,
       author = {{Low}, Frank J. and {Smith}, Paul S. and {Werner}, Michael and {Chen}, Christine and {Krause}, Vanessa and {Jura}, Michael and {Hines}, Dean C.},
        title = "{Exploring Terrestrial Planet Formation in the TW Hydrae Association}",
      journal = {\apj},
         year = 2005,
        month = oct,
       volume = {631},
       number = {2},
        pages = {1170-1179},
          doi = {10.1086/432640},
archivePrefix = {arXiv},
       eprint = {astro-ph/0506291},
 primaryClass = {astro-ph},
       adsurl = {https://ui.adsabs.harvard.edu/abs/2005ApJ...631.1170L}
}

@dataset{Spitzer21,
       author = {{SSC} and {IRSA}},
        title = "{VizieR Online Data Catalog: The Spitzer (SEIP) source list (SSTSL2) (Spitzer Science Center, 2021)}",
 howpublished = {VizieR On-line Data Catalog: II/368.  Originally published in: Spitzer Science Center (SSC), IRSA (2021)},
         year = 2021,
        month = mar,
          eid = {II/368},
       adsurl = {https://ui.adsabs.harvard.edu/abs/2021yCat.2368....0S}
}

@ARTICLE{Wright10,
       author = {{Wright}, Edward L. and {Eisenhardt}, Peter R.~M. and {Mainzer}, Amy K. and {Ressler}, Michael E. and {Cutri}, Roc M. and {Jarrett}, Thomas and {Kirkpatrick}, J. Davy and {Padgett}, Deborah and {McMillan}, Robert S. and {Skrutskie}, Michael and {Stanford}, S.~A. and {Cohen}, Martin and {Walker}, Russell G. and {Mather}, John C. and {Leisawitz}, David and {Gautier}, III, Thomas N. and {McLean}, Ian and {Benford}, Dominic and {Lonsdale}, Carol J. and {Blain}, Andrew and {Mendez}, Bryan and {Irace}, William R. and {Duval}, Valerie and {Liu}, Fengchuan and {Royer}, Don and {Heinrichsen}, Ingolf and {Howard}, Joan and {Shannon}, Mark and {Kendall}, Martha and {Walsh}, Amy L. and {Larsen}, Mark and {Cardon}, Joel G. and {Schick}, Scott and {Schwalm}, Mark and {Abid}, Mohamed and {Fabinsky}, Beth and {Naes}, Larry and {Tsai}, Chao-Wei},
        title = "{The Wide-field Infrared Survey Explorer (WISE): Mission Description and Initial On-orbit Performance}",
      journal = {\aj},
         year = 2010,
        month = dec,
       volume = {140},
       number = {6},
        pages = {1868-1881},
          doi = {10.1088/0004-6256/140/6/1868},
archivePrefix = {arXiv},
       eprint = {1008.0031},
 primaryClass = {astro-ph.IM},
       adsurl = {https://ui.adsabs.harvard.edu/abs/2010AJ....140.1868W}
}

@ARTICLE{Marocco21,
       author = {{Marocco}, Federico and {Eisenhardt}, Peter R.~M. and {Fowler}, John W. and {Kirkpatrick}, J. Davy and {Meisner}, Aaron M. and {Schlafly}, Edward F. and {Stanford}, S.~A. and {Garcia}, Nelson and {Caselden}, Dan and {Cushing}, Michael C. and {Cutri}, Roc M. and {Faherty}, Jacqueline K. and {Gelino}, Christopher R. and {Gonzalez}, Anthony H. and {Jarrett}, Thomas H. and {Koontz}, Renata and {Mainzer}, Amanda and {Marchese}, Elijah J. and {Mobasher}, Bahram and {Schlegel}, David J. and {Stern}, Daniel and {Teplitz}, Harry I. and {Wright}, Edward L.},
        title = "{The CatWISE2020 Catalog}",
      journal = {\apjs},
         year = 2021,
        month = mar,
       volume = {253},
       number = {1},
          eid = {8},
        pages = {8},
          doi = {10.3847/1538-4365/abd805},
archivePrefix = {arXiv},
       eprint = {2012.13084},
 primaryClass = {astro-ph.IM},
       adsurl = {https://ui.adsabs.harvard.edu/abs/2021ApJS..253....8M}
}

@BOOK{Cutri03,
       author = {{Cutri}, R.~M. and {Skrutskie}, M.~F. and {van Dyk}, S. and {Beichman}, C.~A. and {Carpenter}, J.~M. and {Chester}, T. and {Cambresy}, L. and {Evans}, T. and {Fowler}, J. and {Gizis}, J. and {Howard}, E. and {Huchra}, J. and {Jarrett}, T. and {Kopan}, E.~L. and {Kirkpatrick}, J.~D. and {Light}, R.~M. and {Marsh}, K.~A. and {McCallon}, H. and {Schneider}, S. and {Stiening}, R. and {Sykes}, M. and {Weinberg}, M. and {Wheaton}, W.~A. and {Wheelock}, S. and {Zacarias}, N.},
        title = "{2MASS All Sky Catalog of point sources.}",
         year = 2003,
       adsurl = {https://ui.adsabs.harvard.edu/abs/2003tmc..book.....C}
}

@dataset{gaia18,
       author = {{Gaia Collaboration}},
        title = "{VizieR Online Data Catalog: Gaia DR2 (Gaia Collaboration, 2018)}",
 howpublished = {VizieR On-line Data Catalog: I/345.  Originally published in: 2018A\&A...616A...1G},
         year = 2018,
        month = apr,
          eid = {I/345},
          doi = {10.26093/cds/vizier.1345},
       adsurl = {https://ui.adsabs.harvard.edu/abs/2018yCat.1345....0G}
}

@dataset{Henden16,
       author = {{Henden}, A.~A. and {Templeton}, M. and {Terrell}, D. and {Smith}, T.~C. and {Levine}, S. and {Welch}, D.},
        title = "{VizieR Online Data Catalog: AAVSO Photometric All Sky Survey (APASS) DR9 (Henden+, 2016)}",
 howpublished = {VizieR On-line Data Catalog: II/336.  Originally published in: 2015AAS...22533616H},
         year = 2016,
        month = jan,
          eid = {II/336},
       adsurl = {https://ui.adsabs.harvard.edu/abs/2016yCat.2336....0H}
}

@ARTICLE{Virtanen20,
       author = {{Virtanen}, Pauli and {Gommers}, Ralf and {Oliphant}, Travis E. and {Haberland}, Matt and {Reddy}, Tyler and {Cournapeau}, David and {Burovski}, Evgeni and {Peterson}, Pearu and {Weckesser}, Warren and {Bright}, Jonathan and {van der Walt}, St{\'e}fan J. and {Brett}, Matthew and {Wilson}, Joshua and {Millman}, K. Jarrod and {Mayorov}, Nikolay and {Nelson}, Andrew R.~J. and {Jones}, Eric and {Kern}, Robert and {Larson}, Eric and {Carey}, C.~J. and {Polat}, {\.I}lhan and {Feng}, Yu and {Moore}, Eric W. and {VanderPlas}, Jake and {Laxalde}, Denis and {Perktold}, Josef and {Cimrman}, Robert and {Henriksen}, Ian and {Quintero}, E.~A. and {Harris}, Charles R. and {Archibald}, Anne M. and {Ribeiro}, Ant{\^o}nio H. and {Pedregosa}, Fabian and {van Mulbregt}, Paul and {SciPy 1. 0 Contributors}},
        title = "{SciPy 1.0: fundamental algorithms for scientific computing in Python}",
      journal = {Nature Methods},
         year = 2020,
        month = feb,
       volume = {17},
        pages = {261-272},
          doi = {10.1038/s41592-019-0686-2},
archivePrefix = {arXiv},
       eprint = {1907.10121},
 primaryClass = {cs.MS},
       adsurl = {https://ui.adsabs.harvard.edu/abs/2020NatMe..17..261V}
}

@ARTICLE{Harris20,
       author = {{Harris}, Charles R. and {Millman}, K. Jarrod and {van der Walt}, St{\'e}fan J. and {Gommers}, Ralf and {Virtanen}, Pauli and {Cournapeau}, David and {Wieser}, Eric and {Taylor}, Julian and {Berg}, Sebastian and {Smith}, Nathaniel J. and {Kern}, Robert and {Picus}, Matti and {Hoyer}, Stephan and {van Kerkwijk}, Marten H. and {Brett}, Matthew and {Haldane}, Allan and {del R{\'\i}o}, Jaime Fern{\'a}ndez and {Wiebe}, Mark and {Peterson}, Pearu and {G{\'e}rard-Marchant}, Pierre and {Sheppard}, Kevin and {Reddy}, Tyler and {Weckesser}, Warren and {Abbasi}, Hameer and {Gohlke}, Christoph and {Oliphant}, Travis E.},
        title = "{Array programming with NumPy}",
      journal = {\nat},
         year = 2020,
        month = sep,
       volume = {585},
       number = {7825},
        pages = {357-362},
          doi = {10.1038/s41586-020-2649-2},
archivePrefix = {arXiv},
       eprint = {2006.10256},
 primaryClass = {cs.MS},
       adsurl = {https://ui.adsabs.harvard.edu/abs/2020Natur.585..357H}
}

@ARTICLE{astropy18,
       author = {{Astropy Collaboration} and {Price-Whelan}, A.~M. and {Sip{\H{o}}cz}, B.~M. and {G{\"u}nther}, H.~M. and {Lim}, P.~L. and {Crawford}, S.~M. and {Conseil}, S. and {Shupe}, D.~L. and {Craig}, M.~W. and {Dencheva}, N. and {Ginsburg}, A. and {VanderPlas}, J.~T. and {Bradley}, L.~D. and {P{\'e}rez-Su{\'a}rez}, D. and {de Val-Borro}, M. and {Aldcroft}, T.~L. and {Cruz}, K.~L. and {Robitaille}, T.~P. and {Tollerud}, E.~J. and {Ardelean}, C. and {Babej}, T. and {Bach}, Y.~P. and {Bachetti}, M. and {Bakanov}, A.~V. and {Bamford}, S.~P. and {Barentsen}, G. and {Barmby}, P. and {Baumbach}, A. and {Berry}, K.~L. and {Biscani}, F. and {Boquien}, M. and {Bostroem}, K.~A. and {Bouma}, L.~G. and {Brammer}, G.~B. and {Bray}, E.~M. and {Breytenbach}, H. and {Buddelmeijer}, H. and {Burke}, D.~J. and {Calderone}, G. and {Cano Rodr{\'\i}guez}, J.~L. and {Cara}, M. and {Cardoso}, J.~V.~M. and {Cheedella}, S. and {Copin}, Y. and {Corrales}, L. and {Crichton}, D. and {D'Avella}, D. and {Deil}, C. and {Depagne}, {\'E}. and {Dietrich}, J.~P. and {Donath}, A. and {Droettboom}, M. and {Earl}, N. and {Erben}, T. and {Fabbro}, S. and {Ferreira}, L.~A. and {Finethy}, T. and {Fox}, R.~T. and {Garrison}, L.~H. and {Gibbons}, S.~L.~J. and {Goldstein}, D.~A. and {Gommers}, R. and {Greco}, J.~P. and {Greenfield}, P. and {Groener}, A.~M. and {Grollier}, F. and {Hagen}, A. and {Hirst}, P. and {Homeier}, D. and {Horton}, A.~J. and {Hosseinzadeh}, G. and {Hu}, L. and {Hunkeler}, J.~S. and {Ivezi{\'c}}, {\v{Z}}. and {Jain}, A. and {Jenness}, T. and {Kanarek}, G. and {Kendrew}, S. and {Kern}, N.~S. and {Kerzendorf}, W.~E. and {Khvalko}, A. and {King}, J. and {Kirkby}, D. and {Kulkarni}, A.~M. and {Kumar}, A. and {Lee}, A. and {Lenz}, D. and {Littlefair}, S.~P. and {Ma}, Z. and {Macleod}, D.~M. and {Mastropietro}, M. and {McCully}, C. and {Montagnac}, S. and {Morris}, B.~M. and {Mueller}, M. and {Mumford}, S.~J. and {Muna}, D. and {Murphy}, N.~A. and {Nelson}, S. and {Nguyen}, G.~H. and {Ninan}, J.~P. and {N{\"o}the}, M. and {Ogaz}, S. and {Oh}, S. and {Parejko}, J.~K. and {Parley}, N. and {Pascual}, S. and {Patil}, R. and {Patil}, A.~A. and {Plunkett}, A.~L. and {Prochaska}, J.~X. and {Rastogi}, T. and {Reddy Janga}, V. and {Sabater}, J. and {Sakurikar}, P. and {Seifert}, M. and {Sherbert}, L.~E. and {Sherwood-Taylor}, H. and {Shih}, A.~Y. and {Sick}, J. and {Silbiger}, M.~T. and {Singanamalla}, S. and {Singer}, L.~P. and {Sladen}, P.~H. and {Sooley}, K.~A. and {Sornarajah}, S. and {Streicher}, O. and {Teuben}, P. and {Thomas}, S.~W. and {Tremblay}, G.~R. and {Turner}, J.~E.~H. and {Terr{\'o}n}, V. and {van Kerkwijk}, M.~H. and {de la Vega}, A. and {Watkins}, L.~L. and {Weaver}, B.~A. and {Whitmore}, J.~B. and {Woillez}, J. and {Zabalza}, V. and {Astropy Contributors}},
        title = "{The Astropy Project: Building an Open-science Project and Status of the v2.0 Core Package}",
      journal = {\aj},
         year = 2018,
        month = sep,
       volume = {156},
       number = {3},
          eid = {123},
        pages = {123},
          doi = {10.3847/1538-3881/aabc4f},
archivePrefix = {arXiv},
       eprint = {1801.02634},
 primaryClass = {astro-ph.IM},
       adsurl = {https://ui.adsabs.harvard.edu/abs/2018AJ....156..123A}
}

@ARTICLE{Perez07,
       author = {{Perez}, Fernando and {Granger}, Brian E.},
        title = "{IPython: A System for Interactive Scientific Computing}",
      journal = {Computing in Science and Engineering},
         year = 2007,
        month = jan,
       volume = {9},
       number = {3},
        pages = {21-29},
          doi = {10.1109/MCSE.2007.53},
       adsurl = {https://ui.adsabs.harvard.edu/abs/2007CSE.....9c..21P}}

@article{Hunter07,
	Author = {J. D. Hunter},
	Journal = {CSE},
        volume = {9}, 
        pages = {90-95},
	Number = {3},
	Title = {Matplotlib: A 2D Graphics Environment},
	Year = {2007}}

@software{lmfit25,
  author       = {Newville, Matthew and
                  Otten, Renee and
                  Nelson, Andrew and
                  Stensitzki, Till and
                  Ingargiola, Antonino and
                  Allan, Daniel and
                  Fox, Austin and
                  Carter, Faustin and
                  Rawlik, Michal},
  title        = {LMFIT: Non-Linear Least-Squares Minimization and
                   Curve-Fitting for Python
                  },
  month        = mar,
  year         = 2025,
  publisher    = {Zenodo},
  version      = {1.3.3},
  doi          = {10.5281/zenodo.15014437},
  url          = {https://doi.org/10.5281/zenodo.15014437},
  swhid        = {swh:1:dir:6dbda1361832412880d315cbf608ba498c177d82
                   ;origin=https://doi.org/10.5281/zenodo.598352;visi
                   t=swh:1:snp:0d236d4d8ff7f7248600297b145d047734607d
                   14;anchor=swh:1:rel:5f2a70f7c390d9b78cb661b349b044
                   753cf7e89e;path=lmfit-lmfit-py-f97ddf7
                  },
}

@ARTICLE{Bailey14,
       author = {{Bailey}, Vanessa and {Meshkat}, Tiffany and {Reiter}, Megan and {Morzinski}, Katie and {Males}, Jared and {Su}, Kate Y.~L. and {Hinz}, Philip M. and {Kenworthy}, Matthew and {Stark}, Daniel and {Mamajek}, Eric and {Briguglio}, Runa and {Close}, Laird M. and {Follette}, Katherine B. and {Puglisi}, Alfio and {Rodigas}, Timothy and {Weinberger}, Alycia J. and {Xompero}, Marco},
        title = "{HD 106906 b: A Planetary-mass Companion Outside a Massive Debris Disk}",
      journal = {\apjl},
         year = 2014,
        month = jan,
       volume = {780},
       number = {1},
          eid = {L4},
        pages = {L4},
          doi = {10.1088/2041-8205/780/1/L4},
archivePrefix = {arXiv},
       eprint = {1312.1265},
 primaryClass = {astro-ph.EP},
       adsurl = {https://ui.adsabs.harvard.edu/abs/2014ApJ...780L...4B}
}

@ARTICLE{Jones23,
       author = {{Jones}, Joshua W. and {Chiang}, Eugene and {Duch{\^e}ne}, Gaspard and {Kalas}, Paul and {Esposito}, Thomas M.},
        title = "{Giant Impacts and Debris Disk Morphology}",
      journal = {\apj},
         year = 2023,
        month = may,
       volume = {948},
       number = {2},
          eid = {102},
        pages = {102},
          doi = {10.3847/1538-4357/acc466},
archivePrefix = {arXiv},
       eprint = {2303.10189},
 primaryClass = {astro-ph.EP},
       adsurl = {https://ui.adsabs.harvard.edu/abs/2023ApJ...948..102J}
}

@ARTICLE{Crotts22,
       author = {{Crotts}, Katie A. and {Draper}, Zachary H. and {Matthews}, Brenda C. and {Duch{\^e}ne}, Gaspard and {Esposito}, Thomas M. and {Wilner}, David and {Mazoyer}, Johan and {Padgett}, Deborah and {Kalas}, Paul and {Stapelfeldt}, Karl},
        title = "{A Multiwavelength Study of the Highly Asymmetrical Debris Disk around HD 111520}",
      journal = {\apj},
         year = 2022,
        month = jun,
       volume = {932},
       number = {1},
          eid = {23},
        pages = {23},
          doi = {10.3847/1538-4357/ac6c86},
archivePrefix = {arXiv},
       eprint = {2204.11759},
 primaryClass = {astro-ph.EP},
       adsurl = {https://ui.adsabs.harvard.edu/abs/2022ApJ...932...23C}
}

@ARTICLE{Dent14,
       author = {{Dent}, W.~R.~F. and {Wyatt}, M.~C. and {Roberge}, A. and {Augereau}, J.-C. and {Casassus}, S. and {Corder}, S. and {Greaves}, J.~S. and {de Gregorio-Monsalvo}, I. and {Hales}, A. and {Jackson}, A.~P. and {Hughes}, A. Meredith and {Lagrange}, A.-M. and {Matthews}, B. and {Wilner}, D.},
        title = "{Molecular Gas Clumps from the Destruction of Icy Bodies in the {\ensuremath{\beta}} Pictoris Debris Disk}",
      journal = {Science},
         year = 2014,
        month = mar,
       volume = {343},
       number = {6178},
        pages = {1490-1492},
          doi = {10.1126/science.1248726},
archivePrefix = {arXiv},
       eprint = {1404.1380},
 primaryClass = {astro-ph.SR},
       adsurl = {https://ui.adsabs.harvard.edu/abs/2014Sci...343.1490D}
}

@ARTICLE{Marois06,
       author = {{Marois}, Christian and {Lafreni{\`e}re}, David and {Doyon}, Ren{\'e} and {Macintosh}, Bruce and {Nadeau}, Daniel},
        title = "{Angular Differential Imaging: A Powerful High-Contrast Imaging Technique}",
      journal = {\apj},
         year = 2006,
        month = apr,
       volume = {641},
       number = {1},
        pages = {556-564},
          doi = {10.1086/500401},
archivePrefix = {arXiv},
       eprint = {astro-ph/0512335},
 primaryClass = {astro-ph},
       adsurl = {https://ui.adsabs.harvard.edu/abs/2006ApJ...641..556M}
}

@ARTICLE{Engler25,
       author = {{Engler}, N. and {Milli}, J. and {Pawellek}, N. and {Gratton}, R. and {Th{\'e}bault}, P. and {Lazzoni}, C. and {Olofsson}, J. and {Schmid}, H.~M. and {Ulmer-Moll}, S. and {Perrot}, C. and {Augereau}, J.-C. and {Desidera}, S. and {Chauvin}, G. and {Janson}, M. and {Xie}, C. and {Henning}, Th. and {Boccaletti}, A. and {Brown-Sevilla}, S.~B. and {Choquet}, E. and {Dominik}, C. and {Ginski}, C. and {Zurlo}, A. and {Feldt}, M. and {Fusco}, T. and {Girard}, J.~H. and {Gisler}, D. and {van Holstein}, R.~G. and {Langlois}, M. and {Maire}, A.-L. and {Mesa}, D. and {Rabou}, P. and {Rodet}, L. and {Samland}, M. and {Schmidt}, T. and {Vigan}, A.},
        title = "{Characterization of debris disks observed with SPHERE}",
      journal = {\aap},
         year = 2025,
        month = dec,
       volume = {704},
          eid = {A21},
        pages = {A21},
          doi = {10.1051/0004-6361/202554953},
archivePrefix = {arXiv},
       eprint = {2512.03128},
 primaryClass = {astro-ph.EP},
       adsurl = {https://ui.adsabs.harvard.edu/abs/2025A&A...704A..21E}
}

@ARTICLE{Sissa18,
       author = {{Sissa}, E. and {Olofsson}, J. and {Vigan}, A. and {Augereau}, J.~C. and {D'Orazi}, V. and {Desidera}, S. and {Gratton}, R. and {Langlois}, M. and {Rigliaco}, E. and {Boccaletti}, A. and {Kral}, Q. and {Lazzoni}, C. and {Mesa}, D. and {Messina}, S. and {Sezestre}, E. and {Th{\'e}bault}, P. and {Zurlo}, A. and {Bhowmik}, T. and {Bonnefoy}, M. and {Chauvin}, G. and {Feldt}, M. and {Hagelberg}, J. and {Lagrange}, A.-M. and {Janson}, M. and {Maire}, A.-L. and {M{\'e}nard}, F. and {Schlieder}, J. and {Schmidt}, T. and {Szul{\'a}gyi}, J. and {Stadler}, E. and {Maurel}, D. and {Delboulb{\'e}}, A. and {Feautrier}, P. and {Ramos}, J. and {Rigal}, F.},
        title = "{New disk discovered with VLT/SPHERE around the M star GSC 07396-00759}",
      journal = {\aap},
         year = 2018,
        month = may,
       volume = {613},
          eid = {L6},
        pages = {L6},
          doi = {10.1051/0004-6361/201832740},
archivePrefix = {arXiv},
       eprint = {1804.02882},
 primaryClass = {astro-ph.EP},
       adsurl = {https://ui.adsabs.harvard.edu/abs/2018A&A...613L...6S}
}

@ARTICLE{HG41,
       author = {{Henyey}, L.~G. and {Greenstein}, J.~L.},
        title = "{Diffuse radiation in the Galaxy.}",
      journal = {\apj},
         year = 1941,
        month = jan,
       volume = {93},
        pages = {70-83},
          doi = {10.1086/144246},
       adsurl = {https://ui.adsabs.harvard.edu/abs/1941ApJ....93...70H}
}

@ARTICLE{Augereau99,
       author = {{Augereau}, J.~C. and {Lagrange}, A.~M. and {Mouillet}, D. and {Papaloizou}, J.~C.~B. and {Grorod}, P.~A.},
        title = "{On the HR 4796 A circumstellar disk}",
      journal = {\aap},
         year = 1999,
        month = aug,
       volume = {348},
        pages = {557-569},
          doi = {10.48550/arXiv.astro-ph/9906429},
archivePrefix = {arXiv},
       eprint = {astro-ph/9906429},
 primaryClass = {astro-ph},
       adsurl = {https://ui.adsabs.harvard.edu/abs/1999A&A...348..557A}
}

@ARTICLE{Gaia21,
       author = {{Gaia Collaboration} and {Brown}, A.~G.~A. and {Vallenari}, A. and {Prusti}, T. and {de Bruijne}, J.~H.~J. and {Babusiaux}, C. and {Biermann}, M. and {Creevey}, O.~L. and {Evans}, D.~W. and {Eyer}, L. and {Hutton}, A. and {Jansen}, F. and {Jordi}, C. and {Klioner}, S.~A. and {Lammers}, U. and {Lindegren}, L. and {Luri}, X. and {Mignard}, F. and {Panem}, C. and {Pourbaix}, D. and {Randich}, S. and {Sartoretti}, P. and {Soubiran}, C. and {Walton}, N.~A. and {Arenou}, F. and {Bailer-Jones}, C.~A.~L. and {Bastian}, U. and {Cropper}, M. and {Drimmel}, R. and {Katz}, D. and {Lattanzi}, M.~G. and {van Leeuwen}, F. and {Bakker}, J. and {Cacciari}, C. and {Casta{\~n}eda}, J. and {De Angeli}, F. and {Ducourant}, C. and {Fabricius}, C. and {Fouesneau}, M. and {Fr{\'e}mat}, Y. and {Guerra}, R. and {Guerrier}, A. and {Guiraud}, J. and {Jean-Antoine Piccolo}, A. and {Masana}, E. and {Messineo}, R. and {Mowlavi}, N. and {Nicolas}, C. and {Nienartowicz}, K. and {Pailler}, F. and {Panuzzo}, P. and {Riclet}, F. and {Roux}, W. and {Seabroke}, G.~M. and {Sordo}, R. and {Tanga}, P. and {Th{\'e}venin}, F. and {Gracia-Abril}, G. and {Portell}, J. and {Teyssier}, D. and {Altmann}, M. and {Andrae}, R. and {Bellas-Velidis}, I. and {Benson}, K. and {Berthier}, J. and {Blomme}, R. and {Brugaletta}, E. and {Burgess}, P.~W. and {Busso}, G. and {Carry}, B. and {Cellino}, A. and {Cheek}, N. and {Clementini}, G. and {Damerdji}, Y. and {Davidson}, M. and {Delchambre}, L. and {Dell'Oro}, A. and {Fern{\'a}ndez-Hern{\'a}ndez}, J. and {Galluccio}, L. and {Garc{\'\i}a-Lario}, P. and {Garcia-Reinaldos}, M. and {Gonz{\'a}lez-N{\'u}{\~n}ez}, J. and {Gosset}, E. and {Haigron}, R. and {Halbwachs}, J.-L. and {Hambly}, N.~C. and {Harrison}, D.~L. and {Hatzidimitriou}, D. and {Heiter}, U. and {Hern{\'a}ndez}, J. and {Hestroffer}, D. and {Hodgkin}, S.~T. and {Holl}, B. and {Jan{\ss}en}, K. and {Jevardat de Fombelle}, G. and {Jordan}, S. and {Krone-Martins}, A. and {Lanzafame}, A.~C. and {L{\"o}ffler}, W. and {Lorca}, A. and {Manteiga}, M. and {Marchal}, O. and {Marrese}, P.~M. and {Moitinho}, A. and {Mora}, A. and {Muinonen}, K. and {Osborne}, P. and {Pancino}, E. and {Pauwels}, T. and {Petit}, J.-M. and {Recio-Blanco}, A. and {Richards}, P.~J. and {Riello}, M. and {Rimoldini}, L. and {Robin}, A.~C. and {Roegiers}, T. and {Rybizki}, J. and {Sarro}, L.~M. and {Siopis}, C. and {Smith}, M. and {Sozzetti}, A. and {Ulla}, A. and {Utrilla}, E. and {van Leeuwen}, M. and {van Reeven}, W. and {Abbas}, U. and {Abreu Aramburu}, A. and {Accart}, S. and {Aerts}, C. and {Aguado}, J.~J. and {Ajaj}, M. and {Altavilla}, G. and {{\'A}lvarez}, M.~A. and {{\'A}lvarez Cid-Fuentes}, J. and {Alves}, J. and {Anderson}, R.~I. and {Anglada Varela}, E. and {Antoja}, T. and {Audard}, M. and {Baines}, D. and {Baker}, S.~G. and {Balaguer-N{\'u}{\~n}ez}, L. and {Balbinot}, E. and {Balog}, Z. and {Barache}, C. and {Barbato}, D. and {Barros}, M. and {Barstow}, M.~A. and {Bartolom{\'e}}, S. and {Bassilana}, J.-L. and {Bauchet}, N. and {Baudesson-Stella}, A. and {Becciani}, U. and {Bellazzini}, M. and {Bernet}, M. and {Bertone}, S. and {Bianchi}, L. and {Blanco-Cuaresma}, S. and {Boch}, T. and {Bombrun}, A. and {Bossini}, D. and {Bouquillon}, S. and {Bragaglia}, A. and {Bramante}, L. and {Breedt}, E. and {Bressan}, A. and {Brouillet}, N. and {Bucciarelli}, B. and {Burlacu}, A. and {Busonero}, D. and {Butkevich}, A.~G. and {Buzzi}, R. and {Caffau}, E. and {Cancelliere}, R. and {C{\'a}novas}, H. and {Cantat-Gaudin}, T. and {Carballo}, R. and {Carlucci}, T. and {Carnerero}, M.~I. and {Carrasco}, J.~M. and {Casamiquela}, L. and {Castellani}, M. and {Castro-Ginard}, A. and {Castro Sampol}, P. and {Chaoul}, L. and {Charlot}, P. and {Chemin}, L. and {Chiavassa}, A. and {Cioni}, M.-R.~L. and {Comoretto}, G. and {Cooper}, W.~J. and {Cornez}, T. and {Cowell}, S. and {Crifo}, F. and {Crosta}, M. and {Crowley}, C. and {Dafonte}, C. and {Dapergolas}, A. and {David}, M. and {David}, P.},
        title = "{Gaia Early Data Release 3. Summary of the contents and survey properties}",
      journal = {\aap},
         year = 2021,
        month = may,
       volume = {649},
          eid = {A1},
        pages = {A1},
          doi = {10.1051/0004-6361/202039657},
archivePrefix = {arXiv},
       eprint = {2012.01533},
 primaryClass = {astro-ph.GA},
       adsurl = {https://ui.adsabs.harvard.edu/abs/2021A&A...649A...1G}
}

@ARTICLE{Lawson22,
       author = {{Lawson}, Kellen and {Currie}, Thayne and {Wisniewski}, John P. and {Groff}, Tyler D. and {McElwain}, Michael W. and {Schlieder}, Joshua E.},
        title = "{Constrained Reference Star Differential Imaging: Enabling High-fidelity Imagery of Highly Structured Circumstellar Disks}",
      journal = {\apjl},
         year = 2022,
        month = aug,
       volume = {935},
       number = {2},
          eid = {L25},
        pages = {L25},
          doi = {10.3847/2041-8213/ac853b},
archivePrefix = {arXiv},
       eprint = {2208.01606},
 primaryClass = {astro-ph.EP},
       adsurl = {https://ui.adsabs.harvard.edu/abs/2022ApJ...935L..25L}
}

@INPROCEEDINGS{Kammerer22,
       author = {{Kammerer}, Jens and {Girard}, Julien and {Carter}, Aarynn L. and {Perrin}, Marshall D. and {Cooper}, Rachel and {Thatte}, Deepashri and {Vandal}, Thomas and {Leisenring}, Jarron and {Wang}, Jason and {Balmer}, William O. and {Sivaramakrishnan}, Anand and {Pueyo}, Laurent and {Ward-Duong}, Kimberly and {Sunnquist}, Ben and {Adams Redai}, J{\'e}a.},
        title = "{Performance of near-infrared high-contrast imaging methods with JWST from commissioning}",
    booktitle = {Space Telescopes and Instrumentation 2022: Optical, Infrared, and Millimeter Wave},
         year = 2022,
       editor = {{Coyle}, Laura E. and {Matsuura}, Shuji and {Perrin}, Marshall D.},
       series = {Society of Photo-Optical Instrumentation Engineers (SPIE) Conference Series},
       volume = {12180},
        month = aug,
          eid = {121803N},
        pages = {121803N},
          doi = {10.1117/12.2628865},
archivePrefix = {arXiv},
       eprint = {2208.00996},
 primaryClass = {astro-ph.EP},
       adsurl = {https://ui.adsabs.harvard.edu/abs/2022SPIE12180E..3NK}
}

@ARTICLE{Carter23,
       author = {{Carter}, Aarynn L. and {Hinkley}, Sasha and {Kammerer}, Jens and {Skemer}, Andrew and {Biller}, Beth A. and {Leisenring}, Jarron M. and {Millar-Blanchaer}, Maxwell A. and {Petrus}, Simon and {Stone}, Jordan M. and {Ward-Duong}, Kimberly and {Wang}, Jason J. and {Girard}, Julien H. and {Hines}, Dean C. and {Perrin}, Marshall D. and {Pueyo}, Laurent and {Balmer}, William O. and {Bonavita}, Mariangela and {Bonnefoy}, Mickael and {Chauvin}, Gael and {Choquet}, Elodie and {Christiaens}, Valentin and {Danielski}, Camilla and {Kennedy}, Grant M. and {Matthews}, Elisabeth C. and {Miles}, Brittany E. and {Patapis}, Polychronis and {Ray}, Shrishmoy and {Rickman}, Emily and {Sallum}, Steph and {Stapelfeldt}, Karl R. and {Whiteford}, Niall and {Zhou}, Yifan and {Absil}, Olivier and {Boccaletti}, Anthony and {Booth}, Mark and {Bowler}, Brendan P. and {Chen}, Christine H. and {Currie}, Thayne and {Fortney}, Jonathan J. and {Grady}, Carol A. and {Greebaum}, Alexandra Z. and {Henning}, Thomas and {Hoch}, Kielan K.~W. and {Janson}, Markus and {Kalas}, Paul and {Kenworthy}, Matthew A. and {Kervella}, Pierre and {Kraus}, Adam L. and {Lagage}, Pierre-Olivier and {Liu}, Michael C. and {Macintosh}, Bruce and {Marino}, Sebastian and {Marley}, Mark S. and {Marois}, Christian and {Matthews}, Brenda C. and {Mawet}, Dimitri and {McElwain}, Michael W. and {Metchev}, Stanimir and {Meyer}, Michael R. and {Molliere}, Paul and {Moran}, Sarah E. and {Morley}, Caroline V. and {Mukherjee}, Sagnick and {Pantin}, Eric and {Quirrenbach}, Andreas and {Rebollido}, Isabel and {Ren}, Bin B. and {Schneider}, Glenn and {Vasist}, Malavika and {Worthen}, Kadin and {Wyatt}, Mark C. and {Briesemeister}, Zackery W. and {Bryan}, Marta L. and {Calissendorff}, Per and {Cantalloube}, Faustine and {Cugno}, Gabriele and {De Furio}, Matthew and {Dupuy}, Trent J. and {Factor}, Samuel M. and {Faherty}, Jacqueline K. and {Fitzgerald}, Michael P. and {Franson}, Kyle and {Gonzales}, Eileen C. and {Hood}, Callie E. and {Howe}, Alex R. and {Kuzuhara}, Masayuki and {Lagrange}, Anne-Marie and {Lawson}, Kellen and {Lazzoni}, Cecilia and {Lew}, Ben W.~P. and {Liu}, Pengyu and {Llop-Sayson}, Jorge and {Lloyd}, James P. and {Martinez}, Raquel A. and {Mazoyer}, Johan and {Palma-Bifani}, Paulina and {Quanz}, Sascha P. and {Redai}, Jea Adams and {Samland}, Matthias and {Schlieder}, Joshua E. and {Tamura}, Motohide and {Tan}, Xianyu and {Uyama}, Taichi and {Vigan}, Arthur and {Vos}, Johanna M. and {Wagner}, Kevin and {Wolff}, Schuyler G. and {Ygouf}, Marie and {Zhang}, Xi and {Zhang}, Keming and {Zhang}, Zhoujian},
        title = "{The JWST Early Release Science Program for Direct Observations of Exoplanetary Systems I: High-contrast Imaging of the Exoplanet HIP 65426 b from 2 to 16 {\ensuremath{\mu}}m}",
      journal = {\apjl},
         year = 2023,
        month = jul,
       volume = {951},
       number = {1},
          eid = {L20},
        pages = {L20},
          doi = {10.3847/2041-8213/acd93e},
archivePrefix = {arXiv},
       eprint = {2208.14990},
 primaryClass = {astro-ph.EP},
       adsurl = {https://ui.adsabs.harvard.edu/abs/2023ApJ...951L..20C}
}

@INPROCEEDINGS{Girard22,
       author = {{Girard}, Julien H. and {Leisenring}, Jarron and {Kammerer}, Jens and {Gennaro}, Mario and {Rieke}, Marcia and {Stansberry}, John and {Rest}, Armin and {Egami}, Eiichi and {Sunnquist}, Ben and {Boyer}, Martha and {Canipe}, Alicia and {Correnti}, Matteo and {Hilbert}, Bryan and {Perrin}, Marshall D. and {Pueyo}, Laurent and {Soummer}, Remi and {Allen}, Marsha and {Bushouse}, Howard and {Aguilar}, Jonathan and {Brooks}, Brian and {Coe}, Dan and {DiFelice}, Audrey and {Golimowski}, David and {Hartig}, George and {Hines}, Dean C. and {Koekemoer}, Anton and {Nickson}, Bryony and {Nikolov}, Nikolay and {Kozhurina-Platais}, Vera and {Pirzkal}, Nor and {Robberto}, Massimo and {Sivaramakrishnan}, Anand and {Sohn}, Sangmo Tony and {Telfer}, Randal and {Wu}, Chi Rai and {Beatty}, Thomas and {Florian}, Michael and {Hainline}, Kevin and {Kelly}, Doug and {Misselt}, Karl and {Schlawin}, Everett and {Sun}, Fengwu and {Williams}, Christina and {Willmer}, Christopher and {Stark}, Christopher and {Ygouf}, Marie and {Carter}, Aarynn and {Beichman}, Charles and {Greene}, Thomas P. and {Roellig}, Thomas and {Krist}, John and {Adams Redai}, J{\'e}a. and {Wang}, Jason and {Clark}, Charles R. and {Lewis}, Dan and {Ferry}, Malcolm},
        title = "{JWST/NIRCam coronagraphy: commissioning and first on-sky results}",
    booktitle = {Space Telescopes and Instrumentation 2022: Optical, Infrared, and Millimeter Wave},
         year = 2022,
       editor = {{Coyle}, Laura E. and {Matsuura}, Shuji and {Perrin}, Marshall D.},
       series = {Society of Photo-Optical Instrumentation Engineers (SPIE) Conference Series},
       volume = {12180},
        month = aug,
          eid = {121803Q},
        pages = {121803Q},
          doi = {10.1117/12.2629636},
archivePrefix = {arXiv},
       eprint = {2208.00998},
 primaryClass = {astro-ph.IM},
       adsurl = {https://ui.adsabs.harvard.edu/abs/2022SPIE12180E..3QG}
}

@ARTICLE{Kalas04,
       author = {{Kalas}, Paul and {Liu}, Michael C. and {Matthews}, Brenda C.},
        title = "{Discovery of a Large Dust Disk Around the Nearby Star AU Microscopii}",
      journal = {Science},
         year = 2004,
        month = mar,
       volume = {303},
       number = {5666},
        pages = {1990-1992},
          doi = {10.1126/science.1093420},
archivePrefix = {arXiv},
       eprint = {astro-ph/0403132},
 primaryClass = {astro-ph},
       adsurl = {https://ui.adsabs.harvard.edu/abs/2004Sci...303.1990K}
}

@ARTICLE{Gaidos17,
       author = {{Gaidos}, E.},
        title = "{A minimum mass nebula for M dwarfs}",
      journal = {\mnras},
         year = 2017,
        month = sep,
       volume = {470},
       number = {1},
        pages = {L1-L5},
          doi = {10.1093/mnrasl/slx063},
archivePrefix = {arXiv},
       eprint = {1704.03265},
 primaryClass = {astro-ph.EP},
       adsurl = {https://ui.adsabs.harvard.edu/abs/2017MNRAS.470L...1G}
}

@ARTICLE{ML14,
       author = {{Morey}, {\'E}tienne and {Lestrade}, Jean-Fran{\c{c}}ois},
        title = "{On the steady state collisional evolution of debris disks around M dwarfs}",
      journal = {\aap},
         year = 2014,
        month = may,
       volume = {565},
          eid = {A58},
        pages = {A58},
          doi = {10.1051/0004-6361/201322567},
archivePrefix = {arXiv},
       eprint = {1404.1954},
 primaryClass = {astro-ph.EP},
       adsurl = {https://ui.adsabs.harvard.edu/abs/2014A&A...565A..58M}
}

@ARTICLE{KB02,
       author = {{Kenyon}, Scott J. and {Bromley}, Benjamin C.},
        title = "{Collisional Cascades in Planetesimal Disks. I. Stellar Flybys}",
      journal = {\aj},
         year = 2002,
        month = mar,
       volume = {123},
       number = {3},
        pages = {1757-1775},
          doi = {10.1086/338850},
archivePrefix = {arXiv},
       eprint = {astro-ph/0111384},
 primaryClass = {astro-ph},
       adsurl = {https://ui.adsabs.harvard.edu/abs/2002AJ....123.1757K}
}

@ARTICLE{Trilling08,
       author = {{Trilling}, D.~E. and {Bryden}, G. and {Beichman}, C.~A. and {Rieke}, G.~H. and {Su}, K.~Y.~L. and {Stansberry}, J.~A. and {Blaylock}, M. and {Stapelfeldt}, K.~R. and {Beeman}, J.~W. and {Haller}, E.~E.},
        title = "{Debris Disks around Sun-like Stars}",
      journal = {\apj},
         year = 2008,
        month = feb,
       volume = {674},
       number = {2},
        pages = {1086-1105},
          doi = {10.1086/525514},
archivePrefix = {arXiv},
       eprint = {0710.5498},
 primaryClass = {astro-ph},
       adsurl = {https://ui.adsabs.harvard.edu/abs/2008ApJ...674.1086T}
}

@ARTICLE{Montesinos16,
       author = {{Montesinos}, B. and {Eiroa}, C. and {Krivov}, A.~V. and {Marshall}, J.~P. and {Pilbratt}, G.~L. and {Liseau}, R. and {Mora}, A. and {Maldonado}, J. and {Wolf}, S. and {Ertel}, S. and {Bayo}, A. and {Augereau}, J.-C. and {Heras}, A.~M. and {Fridlund}, M. and {Danchi}, W.~C. and {Solano}, E. and {Kirchschlager}, F. and {del Burgo}, C. and {Montes}, D.},
        title = "{Incidence of debris discs around FGK stars in the solar neighbourhood}",
      journal = {\aap},
         year = 2016,
        month = sep,
       volume = {593},
          eid = {A51},
        pages = {A51},
          doi = {10.1051/0004-6361/201628329},
archivePrefix = {arXiv},
       eprint = {1605.05837},
 primaryClass = {astro-ph.SR},
       adsurl = {https://ui.adsabs.harvard.edu/abs/2016A&A...593A..51M}
}

@ARTICLE{Sibthorpe18,
       author = {{Sibthorpe}, B. and {Kennedy}, G.~M. and {Wyatt}, M.~C. and {Lestrade}, J.-F. and {Greaves}, J.~S. and {Matthews}, B.~C. and {Duch{\^e}ne}, G.},
        title = "{Analysis of the Herschel DEBRIS Sun-like star sample}",
      journal = {\mnras},
         year = 2018,
        month = apr,
       volume = {475},
       number = {3},
        pages = {3046-3064},
          doi = {10.1093/mnras/stx3188},
archivePrefix = {arXiv},
       eprint = {1803.00072},
 primaryClass = {astro-ph.EP},
       adsurl = {https://ui.adsabs.harvard.edu/abs/2018MNRAS.475.3046S}
}

@ARTICLE{Friebe22,
       author = {{Friebe}, Marc F. and {Pearce}, Tim D. and {L{\"o}hne}, Torsten},
        title = "{Gap carving by a migrating planet embedded in a massive debris disc}",
      journal = {\mnras},
         year = 2022,
        month = may,
       volume = {512},
       number = {3},
        pages = {4441-4454},
          doi = {10.1093/mnras/stac664},
archivePrefix = {arXiv},
       eprint = {2203.03611},
 primaryClass = {astro-ph.EP},
       adsurl = {https://ui.adsabs.harvard.edu/abs/2022MNRAS.512.4441F}
}

@ARTICLE{Reche08,
       author = {{Reche}, R. and {Beust}, H. and {Augereau}, J.-C. and {Absil}, O.},
        title = "{On the observability of resonant structures in planetesimal disks due to planetary migration}",
      journal = {\aap},
         year = 2008,
        month = mar,
       volume = {480},
       number = {2},
        pages = {551-561},
          doi = {10.1051/0004-6361:20077934},
archivePrefix = {arXiv},
       eprint = {0801.2691},
 primaryClass = {astro-ph},
       adsurl = {https://ui.adsabs.harvard.edu/abs/2008A&A...480..551R}
}

@ARTICLE{Wyatt03,
       author = {{Wyatt}, M.~C.},
        title = "{Resonant Trapping of Planetesimals by Planet Migration: Debris Disk Clumps and Vega's Similarity to the Solar System}",
      journal = {\apj},
         year = 2003,
        month = dec,
       volume = {598},
       number = {2},
        pages = {1321-1340},
          doi = {10.1086/379064},
archivePrefix = {arXiv},
       eprint = {astro-ph/0308253},
 primaryClass = {astro-ph},
       adsurl = {https://ui.adsabs.harvard.edu/abs/2003ApJ...598.1321W}
}

@ARTICLE{Pearce24,
       author = {{Pearce}, Tim D. and {Krivov}, Alexander V. and {Sefilian}, Antranik A. and {Jankovic}, Marija R. and {L{\"o}hne}, Torsten and {Morgner}, Tobias and {Wyatt}, Mark C. and {Booth}, Mark and {Marino}, Sebastian},
        title = "{The effect of sculpting planets on the steepness of debris-disc inner edges}",
      journal = {\mnras},
         year = 2024,
        month = jan,
       volume = {527},
       number = {2},
        pages = {3876-3899},
          doi = {10.1093/mnras/stad3462},
archivePrefix = {arXiv},
       eprint = {2311.04265},
 primaryClass = {astro-ph.EP},
       adsurl = {https://ui.adsabs.harvard.edu/abs/2024MNRAS.527.3876P}
}

@ARTICLE{Lagrange25,
       author = {{Lagrange}, A.-M. and {Wilkinson}, C. and {M{\^a}lin}, M. and {Boccaletti}, A. and {Perrot}, C. and {Matr{\`a}}, L. and {Combes}, F. and {Beust}, H. and {Rouan}, D. and {Chomez}, A. and {Milli}, J. and {Charnay}, B. and {Mazevet}, S. and {Flasseur}, O. and {Olofsson}, J. and {Bayo}, A. and {Kral}, Q. and {Carter}, A. and {Crotts}, K.~A. and {Delorme}, P. and {Chauvin}, G. and {Thebault}, P. and {Rubini}, P. and {Kiefer}, F. and {Radcliffe}, A. and {Mazoyer}, J. and {Bodrito}, T. and {Stasevic}, S. and {Langlois}, M.},
        title = "{Evidence for a sub-Jovian planet in the young TWA 7 disk}",
      journal = {\nat},
         year = 2025,
        month = jun,
       volume = {642},
       number = {8069},
        pages = {905-908},
          doi = {10.1038/s41586-025-09150-4},
archivePrefix = {arXiv},
       eprint = {2502.15081},
 primaryClass = {astro-ph.EP},
       adsurl = {https://ui.adsabs.harvard.edu/abs/2025Natur.642..905L}
}

@ARTICLE{Crotts25,
       author = {{Crotts}, Katie A. and {Carter}, Aarynn L. and {Lawson}, Kellen and {Mang}, James and {Biller}, Beth and {Booth}, Mark and {Ferrer-Chavez}, Rodrigo and {Girard}, Julien H. and {Lagrange}, Anne-Marie and {Liu}, Michael C. and {Marino}, Sebastian and {Millar-Blanchaer}, Maxwell A. and {Skemer}, Andy and {Strampelli}, Giovanni M. and {Wang}, Jason and {Absil}, Olivier and {Balmer}, William O. and {Bendahan-West}, Rapha{\"e}l and {Bogat}, Ellis and {Bowens-Rubin}, Rachel and {Chauvin}, Ga{\"e}l and {Fontanive}, Cl{\'e}mence and {Franson}, Kyle and {Kammerer}, Jens and {Leisenring}, Jarron and {Morley}, Caroline V. and {Rebollido}, Isabel and {Skaf}, Nour and {Sutlieff}, Ben J. and {Bruinsma}, Evelyn L. and {Hinkley}, Sasha and {Hoch}, Kielan and {James}, Andrew D. and {Kane}, Rohan and {Mawet}, Dimitri and {Meyer}, Michael R. and {Palatnick}, Skyler and {Perrin}, Marshall D. and {Ray}, Shrishmoy and {Rickman}, Emily and {Sanghi}, Aniket and {Stephenson}, Klaus Subbotina},
        title = "{Follow-up Exploration of the TWA 7 Planet{\textendash}Disk System with JWST NIRCam}",
      journal = {\apjl},
         year = 2025,
        month = jul,
       volume = {987},
       number = {2},
          eid = {L41},
        pages = {L41},
          doi = {10.3847/2041-8213/ade798},
archivePrefix = {arXiv},
       eprint = {2506.19932},
 primaryClass = {astro-ph.EP},
       adsurl = {https://ui.adsabs.harvard.edu/abs/2025ApJ...987L..41C}
}

@ARTICLE{Eisenstein23,
       author = {{Eisenstein}, Daniel J. and {Willott}, Chris and {Alberts}, Stacey and {Arribas}, Santiago and {Bonaventura}, Nina and {Bunker}, Andrew J. and {Cameron}, Alex J. and {Carniani}, Stefano and {Charlot}, Stephane and {Curtis-Lake}, Emma and {D'Eugenio}, Francesco and {Endsley}, Ryan and {Ferruit}, Pierre and {Giardino}, Giovanna and {Hainline}, Kevin and {Hausen}, Ryan and {Jakobsen}, Peter and {Johnson}, Benjamin D. and {Maiolino}, Roberto and {Rieke}, Marcia and {Rieke}, George and {Rix}, Hans-Walter and {Robertson}, Brant and {Stark}, Daniel P. and {Tacchella}, Sandro and {Williams}, Christina C. and {Willmer}, Christopher N.~A. and {Baker}, William M. and {Baum}, Stefi and {Bhatawdekar}, Rachana and {Boyett}, Kristan and {Chen}, Zuyi and {Chevallard}, Jacopo and {Circosta}, Chiara and {Curti}, Mirko and {Danhaive}, A. Lola and {DeCoursey}, Christa and {de Graaff}, Anna and {Dressler}, Alan and {Egami}, Eiichi and {Helton}, Jakob M. and {Hviding}, Raphael E. and {Ji}, Zhiyuan and {Jones}, Gareth C. and {Kumari}, Nimisha and {L{\"u}tzgendorf}, Nora and {Laseter}, Isaac and {Looser}, Tobias J. and {Lyu}, Jianwei and {Maseda}, Michael V. and {Nelson}, Erica and {Parlanti}, Eleonora and {Perna}, Michele and {Pusk{\'a}s}, D{\'a}vid and {Rawle}, Tim and {Rodr{\'\i}guez Del Pino}, Bruno and {Sandles}, Lester and {Saxena}, Aayush and {Scholtz}, Jan and {Sharpe}, Katherine and {Shivaei}, Irene and {Silcock}, Maddie S. and {Simmonds}, Charlotte and {Skarbinski}, Maya and {Smit}, Renske and {Stone}, Meredith and {Suess}, Katherine A. and {Sun}, Fengwu and {Tang}, Mengtao and {Topping}, Michael W. and {{\"U}bler}, Hannah and {Villanueva}, Natalia C. and {Wallace}, Imaan E.~B. and {Whitler}, Lily and {Witstok}, Joris and {Woodrum}, Charity},
        title = "{Overview of the JWST Advanced Deep Extragalactic Survey (JADES)}",
      journal = {arXiv e-prints},
         year = 2023,
        month = jun,
          eid = {arXiv:2306.02465},
        pages = {arXiv:2306.02465},
          doi = {10.48550/arXiv.2306.02465},
archivePrefix = {arXiv},
       eprint = {2306.02465},
 primaryClass = {astro-ph.GA},
       adsurl = {https://ui.adsabs.harvard.edu/abs/2023arXiv230602465E}
}

@ARTICLE{PW14,
       author = {{Pearce}, Tim D. and {Wyatt}, Mark C.},
        title = "{Dynamical evolution of an eccentric planet and a less massive debris disc}",
      journal = {\mnras},
         year = 2014,
        month = sep,
       volume = {443},
       number = {3},
        pages = {2541-2560},
          doi = {10.1093/mnras/stu1302},
archivePrefix = {arXiv},
       eprint = {1406.7294},
 primaryClass = {astro-ph.EP},
       adsurl = {https://ui.adsabs.harvard.edu/abs/2014MNRAS.443.2541P}
}

@ARTICLE{Pearce22,
       author = {{Pearce}, Tim D. and {Launhardt}, Ralf and {Ostermann}, Robert and {Kennedy}, Grant M. and {Gennaro}, Mario and {Booth}, Mark and {Krivov}, Alexander V. and {Cugno}, Gabriele and {Henning}, Thomas K. and {Quirrenbach}, Andreas and {Barcucci}, Arianna Musso and {Matthews}, Elisabeth C. and {Ruh}, Henrik L. and {Stone}, Jordan M.},
        title = "{Planet populations inferred from debris discs. Insights from 178 debris systems in the ISPY, LEECH, and LIStEN planet-hunting surveys}",
      journal = {\aap},
         year = 2022,
        month = mar,
       volume = {659},
          eid = {A135},
        pages = {A135},
          doi = {10.1051/0004-6361/202142720},
archivePrefix = {arXiv},
       eprint = {2201.08369},
 primaryClass = {astro-ph.EP},
       adsurl = {https://ui.adsabs.harvard.edu/abs/2022A&A...659A.135P}
}

@ARTICLE{Wang18,
       author = {{Wang}, Jason J. and {Graham}, James R. and {Dawson}, Rebekah and {Fabrycky}, Daniel and {De Rosa}, Robert J. and {Pueyo}, Laurent and {Konopacky}, Quinn and {Macintosh}, Bruce and {Marois}, Christian and {Chiang}, Eugene and {Ammons}, S. Mark and {Arriaga}, Pauline and {Bailey}, Vanessa P. and {Barman}, Travis and {Bulger}, Joanna and {Chilcote}, Jeffrey and {Cotten}, Tara and {Doyon}, Rene and {Duch{\^e}ne}, Gaspard and {Esposito}, Thomas M. and {Fitzgerald}, Michael P. and {Follette}, Katherine B. and {Gerard}, Benjamin L. and {Goodsell}, Stephen J. and {Greenbaum}, Alexandra Z. and {Hibon}, Pascale and {Hung}, Li-Wei and {Ingraham}, Patrick and {Kalas}, Paul and {Larkin}, James E. and {Maire}, J{\'e}r{\^o}me and {Marchis}, Franck and {Marley}, Mark S. and {Metchev}, Stanimir and {Millar-Blanchaer}, Maxwell A. and {Nielsen}, Eric L. and {Oppenheimer}, Rebecca and {Palmer}, David and {Patience}, Jennifer and {Perrin}, Marshall and {Poyneer}, Lisa and {Rajan}, Abhijith and {Rameau}, Julien and {Rantakyr{\"o}}, Fredrik T. and {Ruffio}, Jean-Baptiste and {Savransky}, Dmitry and {Schneider}, Adam C. and {Sivaramakrishnan}, Anand and {Song}, Inseok and {Soummer}, Remi and {Thomas}, Sandrine and {Wallace}, J. Kent and {Ward-Duong}, Kimberly and {Wiktorowicz}, Sloane and {Wolff}, Schuyler},
        title = "{Dynamical Constraints on the HR 8799 Planets with GPI}",
      journal = {\aj},
         year = 2018,
        month = nov,
       volume = {156},
       number = {5},
          eid = {192},
        pages = {192},
          doi = {10.3847/1538-3881/aae150},
archivePrefix = {arXiv},
       eprint = {1809.04107},
 primaryClass = {astro-ph.EP},
       adsurl = {https://ui.adsabs.harvard.edu/abs/2018AJ....156..192W}
}

@ARTICLE{Nielsen14,
       author = {{Nielsen}, Eric L. and {Liu}, Michael C. and {Wahhaj}, Zahed and {Biller}, Beth A. and {Hayward}, Thomas L. and {Males}, Jared R. and {Close}, Laird M. and {Morzinski}, Katie M. and {Skemer}, Andrew J. and {Kuchner}, Marc J. and {Rodigas}, Timothy J. and {Hinz}, Philip M. and {Chun}, Mark and {Ftaclas}, Christ and {Toomey}, Douglas W.},
        title = "{The Gemini NICI Planet-Finding Campaign: The Orbit of the Young Exoplanet {\ensuremath{\beta}} Pictoris b}",
      journal = {\apj},
         year = 2014,
        month = oct,
       volume = {794},
       number = {2},
          eid = {158},
        pages = {158},
          doi = {10.1088/0004-637X/794/2/158},
archivePrefix = {arXiv},
       eprint = {1403.7195},
 primaryClass = {astro-ph.EP},
       adsurl = {https://ui.adsabs.harvard.edu/abs/2014ApJ...794..158N}
}
\bibliographystyle{aasjournal}



\end{document}